\documentclass{jfm}

\usepackage{graphicx}
\usepackage{newtxtext}
\usepackage{newtxmath}
\usepackage{natbib}
\usepackage{hyperref}
\hypersetup{
    colorlinks = true,
    urlcolor   = blue,
    citecolor  = black,
}

\newcommand{\RomanNumeralCaps}[1]

\usepackage{enumitem}
\newtheorem{example}{Example}
\newcommand{\bD}{\mathbf{D}}
\newcommand{\bF}{\mathbf{F}}
\newcommand{\bi}{\mathbf{i}}
\newcommand{\bj}{\mathbf{j}}
\newcommand{\bk}{\mathbf{k}}
\newcommand{\bn}{\mathbf{n}}
\newcommand{\bN}{\mathbf{N}}
\newcommand{\br}{\mathbf{r}}
\newcommand{\bR}{\mathbf{R}}

\newcommand{\bu}{\mathbf{u}}
\newcommand{\bv}{\mathbf{v}}
\newcommand{\bx}{\mathbf{x}}
\newcommand{\bzero}{\mathbf{0}}
\newcommand{\hbn}{\hat{\bn}}
\newcommand{\hbN}{\hat{\bN}}
\newcommand{\rd}{\mathrm{d}}

\newcommand{\R}{\mathbb{R}}
\newcommand{\Z}{\mathbb{Z}}

\title{A mathematical model for wind-generated particle-fluid flow fields with an application to the helicopter cloud problem}

\author{D. J. Needham\aff{1}
  \and S. Langdon\aff{2}\corresp{\email{stephen.langdon@brunel.ac.uk}}}

\affiliation{\aff{1}School of Mathematics, University of Birmingham, Birmingham, B15 2TT, United Kingdom.
\aff{2}Department of Mathematics, Brunel University London, Uxbridge, UB8 3PH, United Kingdom.}

\begin{document}
\maketitle

\begin{abstract}
We develop a model for the interaction of a fluid flowing above an otherwise static particle bed, with generally the particles being entrained or detrained into the fluid from the upper surface of the particle bed, and thereby forming a fully two phase  fluidized cloud above the particle bed. The flow in this large scale fluidized region is treated as a two-phase flow, whilst the key processes of entrainment and detrainment from the particle bed are treated by examining the local dynamical force balances on the particles in a thin transition layer at the interface between the fully fluidized region and the static particle bed. This detailed consideration leads to the formation of an additional macroscopic boundary condition at this interface, which closes the two phase flow problem in the bulk fluidized region above. We then introduce an elementary model of the well known helicopter brownout problem, and use the theory developed in the first part of the paper to fully analyse this model, both analytically and numerically.
\end{abstract}




\section{Introduction}
\label{sec:1}
In this paper we develop a rational (that is, based directly on the fundamental Newtonian laws of mechanics) physically based mathematical model for the generation of wind-driven particle flow fields from a static particle bed. The physical model is composed of three structured regions: the fluidized region in which the particles lifted from the surface of the static particle bed are in suspension in the fluid, forming a fully developed two-phase flow; the key interfacial layer in which the interaction mechanisms between the local near surface fluid flow and the particle bed enables surface particles to be entrained by the local flow from the static bed of particles into the fluidized region and/or detrained from the local flow in the fluidized region into the static bed of particles; the static bed region where the particles are stored.

A particular example of such wind-driven particle flow fields is the scenario in which a helicopter may descend towards a bed of sand, the \emph{helicopter cloud problem} (often refered to as the \emph{`brownout problem'} in the aero-engineering literature).  In this case, the local interaction between the down-draft and swirling flow, generated by the helicopter rotor, and the upper surface of the otherwise static sand bed, entrains sand particles from a thin interfacial layer into the fluidized region, which then flow in the form  of a high velocity particle laden cloud around the helicopter body, which we refer to as the \emph{helicopter cloud}, and a consequence of which is generally a significant deterioration in visibility for the helicopter pilot - when this becomes too severe, it is referred to as \emph{brownout}.  This problem has received much attention in the engineering literature, but most work to date has primarily been driven by observation and experimentation.  Attempts to model the physics of brownout include those by \citet{WaWhKeMcGiDo08,PhBr08,PhKiBr11,GLG,PoMiPaPeVe20,TaGaBaLiZhHu21,TGZWCW,TaYoHeYuWa22,LSXW,LXJHL}.  In addition a very recent thesis by \citet{thesis} reviews particle tracking simulations together with experiments and high speed photographic evidence related to the helicopter cloud problem, and provides an excellent survey of the engineering literature in this area. These modelling approaches broadly split into two classes, which we now delineate.

The first approach is purely computational, and treats each particle within the fluid flow as an individual entity, with no effort being made to address the entrainment or detrainment of particles into the particle laden flow via local interaction of the flow with the upper surface of the otherwise static particle bed. Indeed, the particle bed is removed from consideration, and instead the particles are simply placed into the flow in a random spatial distribution at an initial reference time. Subsequently each particle is tracked as it moves within the fluid flow according to its own dynamical equation of motion under the action of the locally induced fluid interaction forces and gravity (the following, amongst many others, are papers adopting this principal strategy as a key modelling component: \citet{GLG,PoMiPaPeVe20,TaGaBaLiZhHu21,TGZWCW,TaYoHeYuWa22,LSXW,LXJHL}). Although this approach does give rational information about particulate behaviour \emph{once the particles have made their way into the fluidized flow}, the lack of consideration of the entrainment and detrainment process at the interface between the fluidized region and the static particle bed is a very severe drawback for this modelling approach - it is this interaction process that is the fundamental and key process in initiating and driving the whole phenomenon of the particle cloud. In addition, given that particle clouds in the fluidized region are generally concentrated rather than dilute, with a low fluid voidage (high particle concentration), the approach of treating each particle in the fluidized region as an individual entity is exceptionally inefficient in computational time, and a two-phase flow approach is much more natural and efficient.

The second modelling approach is to introduce a continuum particle density field in the fluidized region (measuring particle volume per unit spatial volume of the two phase flow), and then to postulate that this field satisfies a suitable advection-diffusion PDE throughout the flow field.
This approach does go some way to address the entrainment and detrainment of particles from the static particle bed. Specifically, this is done in a phenomenological way by the introduction of a localised particle mass source term into the advection-diffusion PDE. This source term is localised in space, so that it acts only in a thin neighbourhood of the interface between the fluidized region and the static particle bed, and is designed to represent the localised  input/ouput of particles from the static bed into the fluidized region (see, in particular, \cite{PhKiBr11}).  The nature of this source term is purely phenomenological, and empirically based on the very particular situation under immediate consideration and must be recalibrated in every specific example; as such, the ability of this approach to capture, in general, a  decent representation of the key rational mechanism of entrainment/detrainment at the interface of the fluidized region and the static bed is very limited, and depends entirely, and critically, upon the specific choice of this term. This serious limitation is well addressed in the paper of \cite{PhKiBr11}. In their conclusions, they state that further developments in the theory for the brownout problem depend crucially on an understanding ``\ldots particularly of how sediment leaves the ground and enters the flow around the helicopter in the first place \ldots''. Moreover, these authors state \citep[p.126]{PhKiBr11} that ``\ldots the physics of sediment entrainment from the ground \ldots is no doubt the area which could most benefit from further research and, indeed, where a breakthrough in understanding could contribute most to improving our confidence in the ability of predictive methods to capture the detailed mechanisms at the origin of the brownout cloud \ldots''. The key message is that, although this approach does address local particle entrainment and detrainment from the static particle bed, a thorough understanding and implementation of the fundamental interactive mechanics governing these two principal processes and their coupling with the fluidized region, is essential for improving predictability in the problem. It is this serious defect which we address from fundamental first principles in this paper, which leads to a natural macroscopic boundary condition on void fraction to be applied at the interface of the fluidized region and the static bed, which closes the continuum scale problem two phase flow in the fluidized region.

In more general terms, the development of fluidized clouds in gas or liquid flows above otherwise static particle beds has many applications - for example clouds generated by the motion of desert vehicles, motion of sub-sea vehicles close to the ocean bed and large scale particle clouds raised by localised desert or sub-surface oceanic storms and disturbances. Indeed, the motion of desert dunes via bed load and suspended load sand transport, and the related physical processes, were originally considered in the classic text of \cite{RAB}, in which it was acknowledged that for a generally applicable predictive theory for both of these transport phenomena, a detailed understanding of the local interactive mechanisms in the layer adjacent to the interface between the fluid and the otherwise static particle bed would be essential. In the absence of such an understanding, \cite{RAB}, whilst acknowledging their shortcomings, resorted to the development of empirically fitted sediment transport relations. The physical aspects of wind-blown sand have more recently been addressed in the extensive review article of \cite{KPMK}.

In this paper we address these issues directly and generically. We consider the general situation of an incompressible fluid overlying an otherwise static particle bed, and allow for the general motion of the fluid to interact with the particle bed in a thin transition layer, in which the local force balance between the fluid flow and the incipiently fluidized particles is fully accounted for on the thickness scale of this thin transition region, which rationally accounts for both the entrainment and detrainment of particles from the static bed and into the fully fluidized region and vice versa. The resolution of the dynamics on the scale of this thin transition layer leads to an additional macroscopic boundary condition on local void fraction which must be applied at the interface of the fully fluidized region and the particle bed surface. This closes the macroscopic two phase flow problem in the fully fluidized region, with now the entrainment and detrainment mechanism rationally accounted for and the need for phenomenological parameters removed. This closed model can now be used to formulate the problem in the fluidized region for any particular circumstances under consideration. To illuminate this theoretical model, we apply it in detail to a simple version of the  helicopter cloud problem in the latter part of the paper.  Although direct experimental comparisons for the results from our model are not possible at present without a specifically designed experimental program, it is worthwhile  referring to the high speed photography in \citet{thesis}, through the images labelled as Figure 1.1, Figure 1.2, and, in particular Figure 2.6, which gives a good qualitative comparison with our theoretical examples presented in \S7 of this paper.

An outline of the paper is as follows.  In~\S\ref{sec:2} we give a detailed description of the physical model, and in particular our approach to rationally modelling the dynamics of particle entrainment and detrainment in an interfacial layer. This is followed in~\S\ref{sec:3} by the formulation of the physical model as a mathematical model, with a detailed treatment of the local dynamics in the interfacial layer, and the subsequent development of the corresponding macroscopic boundary conditions  which close  the full two-phase flow problem in the fully fluidized region. This is followed in ~\S\ref{sec:4} by bringing together these components  in the formulation of the full closed mathematical problem in the fluidized region. This mathematical problem is made dimensionless in~\S\ref{sec:5}, where the formulation is reduced to the determination of the voidage field and a fluid velocity potential.  The application to the helicopter cloud problem is developed in~\S\ref{sec:6}, where the effect of the helicopter close to the surface is modelled by a suitably chosen fluid dipole and fluid line-vortex. The boundary value problems which are associated with this application are formulated in a form suitable for numerical solution via finite-difference approximation; in~\S\ref{sec:7} we present numerical examples demonstrating how varying certain parameters may change qualitative features of the voidage field, with precise details of our numerical method and its accuracy given in Appendix~B.

\section{The physical model}
\label{sec:2}
\begin{figure}
  \centerline{\includegraphics[width=0.7\textwidth]{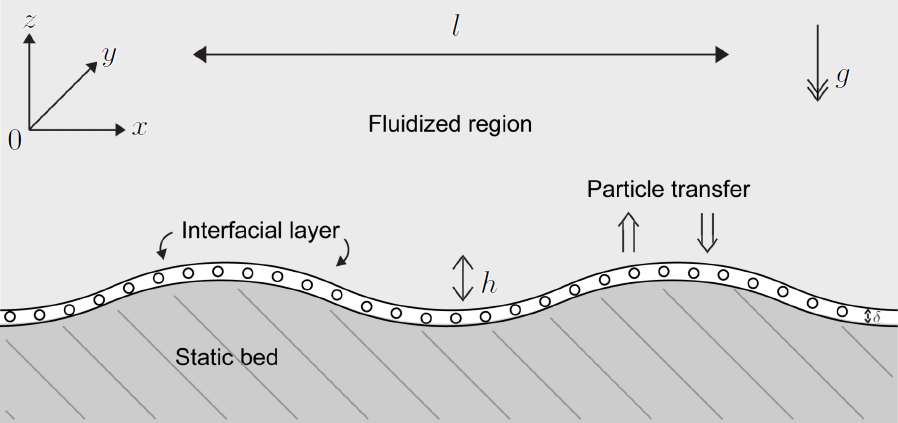}}
  \caption{Schematic diagram of the physical model.}
\label{fig:1}
\end{figure}

The physical model that we develop here is aimed at describing the particle and fluid flow fields generated above a ground surface composed of a packed bed of solid particles, when surface particles are raised into suspension via the flow of an incompressible fluid above the surface. The physical model is developed in three structured regions, namely:\\
\begin{enumerate}[label=(\arabic*)]
  \item{\, The {\em fluidized region} in which the particles lifted from the ground surface are in suspension in the fluid. The flow in this region is a fully developed two-phase flow.}\\
  \item{\, The {\em interfacial layer} in which ground surface particles are transferred from the static bed of particles into the fluidized region and/or are returned from the fluidized region into the static bed of particles.}\\
  \item{\, The {\em static particle bed region} where the particles are stored in the static particle bed.}\\
\end{enumerate}
 A schematic diagram of the above structure is shown in figure~\ref{fig:1}.
In the fluidized region, the two-phase flow varies on a length scale $l$ which is very much greater than the particle-spacing length scale $l_s$ and the particle radius $a$, so that
\begin{equation}
  a \ll l, \quad l_s \ll l.
  \label{eqn:1}
\end{equation}
Under condition~(\ref{eqn:1}), the particle and fluid fields in the fluidized region can be modelled as two inter-penetrating continua, and the flow is governed by conservation of mass and momentum in the two interacting phases. The interfacial layer is a thin layer, of approximately uniform thickness, separating the static particle bed from the fully developed fluidized region above, and accommodates particle mass transfer to and from the static bed and the fluidized region. The thickness of the interfacial layer is, typically, of the order of ten particle diameters.  Thus, with $\delta$ being the interfacial layer thickness,
\begin{equation}
  \delta \approx 20a,
  \label{eqn:2}
\end{equation}
so that, formally, $a\ll\delta$ and the interfacial layer is thin compared to the two-phase flow continuum length scale $l$, but sufficiently wide to contain enough particles for local average particle velocities to be used. Thus we formally require,
\begin{equation}
 a\ll \delta \ll l.
  \label{eqn:3}
\end{equation}
The fluid flow in the thin interfacial layer is taken as being tangential to the surface of the static particle bed when viewed relative to the normal motion of the surface of the static particle bed. The particle motion within the interfacial layer, based on~(\ref{eqn:3}), is modelled as a force balance between the drag and lift on the particles induced by the surface tangential fluid flow in the interfacial layer, and the gravitational and particle collisional forces on the particles, together with the fluid shearing force at the upper surface of the interfacial layer and the particle-particle frictional force at the lower surface of the interfacial layer. Mass transfer between the static particle bed and the fluidized region, via the interfacial layer, then determines the surface deformation of the static particle bed. With $h$ representing the transverse static particle bed surface deformation length scale (see figure~\ref{fig:1}), we consider the situation when
\begin{equation}
  \delta \ll h \ll l.
  \label{eqn:4}
\end{equation}
Connection between the interfacial layer and the bulk fluidized region is made by requiring continuity of the normal particle and fluid mass flow fields and the voidage field at a well defined interface. Of course, this interface is a model device to replace the rapid transition of normal length scale from the interfacial layer into the bulk fluidized region above. Provided such a transition is passive, its modelling as an interface is justified, and the theory bears this out. The static particle bed region is taken as a particle store, with the packed particles in static equilibrium (without strong cohesion). For the specific problem addressed in~\S\ref{sec:6}, namely the helicopter cloud problem, the continuum length scale $l$ will be taken as the geometric dimension of the helicopter blade length.  The hydrodynamic effect of the helicopter is modelled as a fluid dipole, aligned with the axis of blade rotation and located at the helicopter blade rotation point, to represent the fluid down-draft, together with a line-vortex through the axis of rotation of the blades, to represent the fluid swirl.

\section{The mathematical model}
\label{sec:3}
Based on the physical model described in~\S\ref{sec:2}, we derive the dynamical equations of motion in the fluidized region and the interfacial layer. We begin in the fluidized region. Throughout, reference will be made to a fixed Cartesian co-ordinate system with co-ordinates $x$, $y$ and $z$, and with $z$ pointing vertically upwards.  The usual associated unit vectors are $\bi$, $\bj$ and $\bk$, respectively.  The position vector is $\br = (x,y,z)$, and $\br_h = (x,y)$ corresponding to the position vector in the horizontal $xy$-plane. The location of the lower surface of the thin interfacial layer (which is also the upper surface of the static particle bed) is taken as being at
\begin{equation}
  z = \xi (\br_h,t),
  \label{eqn:5}
\end{equation}
for $\br_h\in\R^2$ and $t\geq 0$, with $t$ being time. It follows from~(\ref{eqn:5}) that the unit normal vector field at the thin interfacial layer, and pointing into the fluidized region, is given by
\[
  \hat{\bn}(\br_h,t) = \frac{(-\nabla_h\xi,1)}{(1+|\nabla_h\xi|^2)^{1/2}},
\]
for $\br_h\in \R^2$ and $t\geq 0$.  Here,
\begin{equation}
  \nabla_h(\cdot) = \frac{\partial}{\partial x}(\cdot)\bi + \frac{\partial}{\partial y}(\cdot)\bj
  \label{eqn:7}
\end{equation}
is the usual horizontal gradient operator.  We now consider the equations of motion in the fluidized region.

\subsection{The fluidized region}
\label{subsec:31}
In the fluidized region the fluid and particle phases are modelled as inter-penetrating continua, with the fluid being {\em incompressible} and {\em inviscid} on the two-phase continuum length scale, but {\em viscous} on the particle length scale.  Conservation of mass in the fluid and particle phases then requires, respectively,
\begin{eqnarray}
  E_t + \nabla \cdot (E\bu) & = & 0, \label{eqn:8a} \\
 -E_t + \nabla \cdot ((1-E)\bv) & = & 0, \label{eqn:8b}
\end{eqnarray}
with $\br$ in $D(t)$ and $t > 0$. Here $D(t)$ is the interior of the fluidized region occupied by the inter-penetrating continua at time $t\geq 0$, $E(\br,t)$ is the voidage field (representing the volume of fluid per unit volume in $D(t)$ and taking values between the packing voidage of the particles $E_s$ ($<1$) and unity), $\bu(\br,t)$ is the fluid velocity field and $\bv(\br,t)$ is the particle velocity field. Subscript $t$ represents partial differentiation with respect to time and $\nabla$ is the usual gradient operator,
\[
  \nabla(\cdot) = \frac{\partial}{\partial x}(\cdot)\bi + \frac{\partial}{\partial y}(\cdot)\bj + \frac{\partial}{\partial z}(\cdot)\bk.
\]
Conservation of momentum in the fluid and particle phases also requires, respectively,
\begin{eqnarray}
  \rho_f E(\bu_t + (\bu\cdot\nabla)\bu) & = & -\nabla p - \rho_f g E \bk - \bD, \label{eqn:10a} \\
  \rho_s (1-E)(\bv_t + (\bv\cdot\nabla)\bv) & = & -\nabla p_s - \rho_s g (1-E) \bk + \bD, \label{eqn:10b}
\end{eqnarray}
with $\br$ in $D(t)$ and $t>0$.  Here $p=p(\br,t)$ is the fluid pressure field, $p_s=p_s(E(\br,t))$ is the particle-particle collisional pressure field and $\bD$ is the total drag force per unit volume exerted by the fluid on the particles (which we anticipate, at the single particle scale \emph{in the fluidized flow field}, will significantly dominate the single particle lift force).  In addition, $\rho_f$ and $\rho_s$ are the constant fluid and particle material densities, respectively, and $g$ is the acceleration due to gravity.  It remains to relate the particle collision pressure to the voidage field, and determine the form of the drag force. In general the particle-particle collisional pressure is a decreasing function of voidage, which approaches zero as the voidage approaches unity. In the absence of detailed experimental evidence, we adopt, as a first approximation, the simple linear form
\begin{equation}
  p_s=p_s(E) = p_0(1-E),
  \label{eqn:11}
\end{equation}
with $p_0 > 0$ being a material constant. For the drag force per unit volume, we write
\[ \bD = \beta(E) n_p (6\pi a \mu (\bu-\bv)), \]
where $n_p$ is the number of particles per unit volume, so that $n_p = (1-E)/((4/3)\pi a^3)$, whilst the bracketed term is the Stokes' drag on a single particle, and $\beta=\beta(E)$ represents the effect of neighbouring particles on the Stokes' drag. In general, $\beta(E)\geq 1$ is a decreasing function of $E\in[E_s,1]$ and $\beta(1)=1$.  Thus we have
\begin{equation}
  \bD = \frac{9}{2} \frac{\mu}{a^2} (1-E)\beta(E)(\bu-\bv),
  \label{eqn:12}
\end{equation}
with $\mu$ being the kinematic viscosity of the fluid.  On substituting from~(\ref{eqn:11}) and~(\ref{eqn:12}) into~(\ref{eqn:10a}) and~(\ref{eqn:10b}), we have,
\begin{eqnarray}
  \rho_f E(\bu_t + (\bu\cdot\nabla)\bu) & = & -\nabla p - \rho_f g E \bk - (9\mu/2a^2) (1-E)\beta(E)(\bu-\bv), \label{eqn:13a} \\
  \rho_s (1\!-\!E)(\bv_t + (\bv\cdot\nabla)\bv) & = & p_0\nabla E \!-\! \rho_s g (1\!-\!E) \bk + (9\mu/2a^2) (1\!-\!E)\beta(E)(\bu-\bv), \label{eqn:13b}
\end{eqnarray}
with $\br$ in $D(t)$ and $t > 0$. In obtaining~(\ref{eqn:10a}) and~(\ref{eqn:10b}) (and consequently~(\ref{eqn:13a}) and~(\ref{eqn:13b})), for the micro-scale particle-fluid interaction, we have only included the drag term, which we expect to dominate over lift and buoyancy in the bulk fluidized region, where the fluid flow is primarily inviscid and irrotational on the two-phase continuum length scale (except in the interfacial layer, where ground effect vorticity will enhance lift; see~\S\ref{subsec:32}). Thus the dynamics in the fluidized region is represented by the four partial differential equations~(\ref{eqn:8a}) and~(\ref{eqn:8b}) with~(\ref{eqn:13a}) and~(\ref{eqn:13b}).  We now consider the dynamics in the interfacial layer.

\subsection{The interfacial layer}
\label{subsec:32}
\begin{figure}
  \centerline{\includegraphics[width=0.7\textwidth]{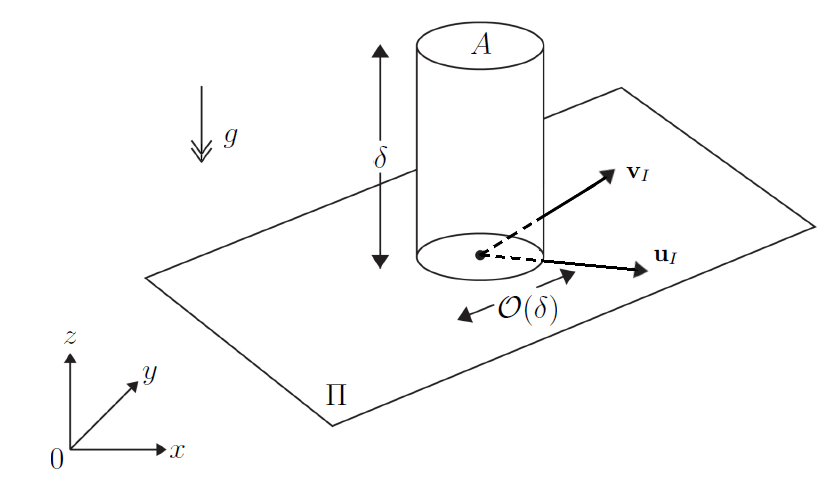}}
  \caption{Cylindrical element across the interfacial layer.  The element is centred at $\br=(\br_h,\xi(\br_h,t))$ which is represented by $\bullet$.  $\Pi$ is the interfacial layer tangent plane at $\br=(\br_h,\xi(\br_h,t))$.  The cross-sectional area of the cylindrical element is $A$.}
\label{fig:2}
\end{figure}

Within the thin interfacial layer we consider the dynamics on the thickness scale $\delta$ to be in local equilibrium relative to the slow (compared to fluid velocity scale) normal motion of the interfacial layer, with the fluid velocity $\bu_I$ and particle velocity $\bv_I$ being tangential to the interfacial layer relative to the normal motion of the interfacial layer, and without variation across the interfacial layer. Normal to the interfacial layer, the local lift force balances the gravity force on the particles, whilst tangential to the interface the drag and fluid shearing forces on the particles are balanced by the gravity force and the static particle bed friction force on the particles. We consider a cylindrical cross sectional element, of radius \textit{O}$(\delta)$, passing through the interfacial layer, and centred at $\br = (\br_h,\xi(\br_h,t))$, as shown in figure~\ref{fig:2}.

A force balance on the particles in the cylindrical element in figure~\ref{fig:2} gives, in equilibrium,
\begin{equation}
  \bF_g + \bF_L + \bF_D + \bF_s + \bF_f = 0,
  \label{eqn:14}
\end{equation}
where $\bF_g$ is the gravitational force on the particles, $\bF_L$ is the lift force on the particles and $\bF_D$ is the drag force on the particles, within the element. In addition, $\bF_s$ is the fluid shearing force exerted on the particles over the upper surface of the element, whilst $\bF_f$ is the frictional force exerted on the particles over the lower surface of the element.  As we are considering the local dynamics in the interfacial layer, on the length scale $\delta (\ll h \ll l)$, to be in equilibrium, then $\bu_I$ and $\bv_I$, together with the voidage in the interfacial layer $E_I\in[E_s,1]$, may be considered as fixed throughout the cylindrical element.  We then have
\begin{equation}
  \bF_g = -(1-E_I) \rho_s g \delta A ((\bk\cdot\hat{\bn})\hat{\bn} + (\bk - (\bk\cdot\hat{\bn})\hat{\bn})),
  \label{eqn:15}
\end{equation}
noting that the vector $\bk - (\bk\cdot\hat{\bn})\hat{\bn}$ lies in the tangent plane $\Pi$ to the interfacial layer.  Next, we have
\begin{equation}
  \bF_L = n_p\delta A\Phi(E_I)L_p\hat{\bn},
  \label{eqn:16}
\end{equation}
where $L_p$ is the lift force on a single particle in the cylindrical element and $\Phi=\Phi(E_I)$ accounts for the effect of neighbouring particles.  For a single particle in the interfacial layer, moving through the fluid as a rolling grain, we have
\begin{equation}
  L_p = 4\pi a^2 \rho_f | \bu_I - \bv_I|^2,
  \label{eqn:17}
\end{equation}
which is the Magnus force induced on an individual rolling particle grain as it moves along the interfacial layer. We anticipate that in this incipiently fluidized interfacial layer this Magnus force dominates over the Saffman lift force due to the local shear in the fluid flow. The function $\Phi(E_I)$ is anticipated to be, in general, monotone increasing in $E_I$, with $\Phi(E_s) <1$ and $\Phi(1)=1$.  Without detailed experimental evidence, we adopt the simple linear form approximation
\begin{equation}
  \Phi(E_I) = E_I.
  \label{eqn:18}
\end{equation}
On substituting from~(\ref{eqn:17}) and~(\ref{eqn:18}) into~(\ref{eqn:16}), with $n_p = 3(1-E_I)/(4\pi a^3)$, we obtain
\begin{equation}
  \bF_L = \left(\frac{3\rho_f}{a}\right)\delta A E_I(1-E_I)| \bu_I - \bv_I|^2\hat{\bn}.
  \label{eqn:19}
\end{equation}
Similarly, adopting~(\ref{eqn:12}), with $E=E_I$, $\bu=\bu_I$ and $\bv=\bv_I$, we obtain
\begin{equation}
  \bF_D = \frac{9}{2} \frac{\mu}{a^2} \delta A (1-E_I) \beta(E_I) (\bu_I - \bv_I),
  \label{eqn:20}
\end{equation}
with $\beta(\cdot)$ as introduced in~\S\ref{subsec:31}.  As discussed earlier, we recall that the vector $(\bu_I - \bv_I)$ is taken to lie in the tangent plane $\Pi$.  For the upper surface shearing force, we may thus write
\begin{equation}
  \bF_s = A (1-E_I) \tau | \bu_I - \bv_I| (\bu_I - \bv_I),
  \label{eqn:21}
\end{equation}
where $\tau>0$ is the turbulent fluid shear coefficient.  Finally, the particle-particle friction force at the lower surface is taken as
\begin{equation}
  \bF_f = A (1-E_I) F \bv_I,
  \label{eqn:22}
\end{equation}
where $F > 0$ is the dynamic friction coefficient. We now substitute from~(\ref{eqn:15}), (\ref{eqn:19}), (\ref{eqn:20}), (\ref{eqn:21}) and~(\ref{eqn:22}) into~(\ref{eqn:14}) and separate components normal and tangential to the tangent plane~$\Pi$.  We obtain the scalar equation
\begin{equation}
  \Lambda E_I |\bu_I - \bv_I|^2 - (\bk\cdot\hat{\bn}) = 0,
  \label{eqn:23}
\end{equation}
and the tangent plane vector equation
\begin{equation}
  - \rho_s g \delta (\bk - (\bk\cdot\hat{\bn})\hat{\bn}) + \frac{9}{2} \frac{\delta \mu}{a^2} \beta(E_I) (\bu_I - \bv_I)  + \tau |\bu_I - \bv_I|(\bu_I - \bv_I) - F \bv_I = \bzero_T,
  \label{eqn:24}
\end{equation}
where $\bzero_T$ is the zero vector in the tangent plane $\Pi$, and
\[
  \Lambda = \frac{3\rho_f}{a \rho_s g}.
\]
Since the interfacial layer is thin ($\delta \ll h \ll l$), following~(\ref{eqn:2}), we conclude that the dominant terms on the left hand side of~(\ref{eqn:24}) are those representing upper surface shear and lower surface friction (surface forces dominating body forces).  With $u_I^s$ and $v_I^s$ being typical scales for $\bu_I$ and $\bv_I$, then a balance between upper surface shear $\bF_s$ and lower surface friction $\bF_f$ gives
\begin{equation}
  v_I^s = O\left(\frac{\tau}{F}(u_I^s)^2\right),
  \label{eqn:26}
\end{equation}
and
\[
  \delta \ll \min \left( \frac{\tau(u_I^s)^2}{\rho_s g}, \frac{\tau a^2 u_I^s}{\mu} \right).
\]
With this approximation, the dominant form of equation~(\ref{eqn:24}) is
\begin{equation}
  \tau |\bu_I - \bv_I|(\bu_I - \bv_I) - F \bv_I = \bzero_T.
  \label{eqn:28}
\end{equation}
Also, we anticipate that
\[
  v_I^s \ll u_I^s,
\]
which follows from (\ref{eqn:26}), when $\tau u_I^s \ll F$.  Thus
\[ |\bv_I| \ll |\bu_I| \]
which allows us to approximate both of equations~(\ref{eqn:23}) and~(\ref{eqn:28}), which become
\begin{eqnarray}
  &\Lambda E_I |\bu_I|^2 - (\bk\cdot\hat{\bn}) = 0,& \label{eqn:31} \\
  &\tau |\bu_I|\bu_I - F \bv_I = \bzero_T.& \label{eqn:32}
\end{eqnarray}
A further simplification of~(\ref{eqn:31}) can be made in recalling from~(\ref{eqn:4}) that $h\ll l$, and so the interfacial layer slope, which is \textit{O}$(h/l)$, is small.  In particular, then
\begin{equation}
  \hat{\bn} = \bk + \mbox{\textit{O}}(h/l),
  \label{eqn:star}
\end{equation}
and so we may take
\[ \hat{\bn} \sim \bk, \]
after which~(\ref{eqn:31}) becomes
\begin{equation}
  \Lambda E_I |\bu_I|^2 - 1 = 0.
  \label{eqn:33}
\end{equation}
Since $E_s \leq E_I \leq 1$, with $E_s$ being the packing voidage for the particles, it follows from~(\ref{eqn:33}) that the interfacial layer {\em collapses} for $|\bu_I|^2 < 1/\Lambda$ (with $E_I = 1$), whilst it becomes saturated (with $E_I=E_s$) for $|\bu_I|^2 > 1/(\Lambda E_s)$.  Between these limiting values of $|\bu_I|^2$ we have, from~(\ref{eqn:32}) and~(\ref{eqn:33}) that in the interfacial layer
\begin{eqnarray}
  &\bv_I = (\tau/F)|\bu_I|\bu_I,& \label{eqn:34} \\
  &E_I = \displaystyle{\frac{1}{\Lambda|\bu_I|^2}}.& \label{eqn:35}
\end{eqnarray}
In (\ref{eqn:35}) the local voidage in the interfacial layer $E_I$ is related to the local tangential fluid velocity in the interfacial layer, $\bu_I$, whilst~(\ref{eqn:34}) relates the local tangential particle velocity in the interfacial layer $\bv_I$ to the local tangential fluid velocity in the interfacial layer $\bu_I$, and the form is similar to a mobile bed-load sediment transport function (see, for example, \cite{RAB} for the classical discussion of bed-load sediment transport formula in a wind-blown sand environment).  It should also be noted that throughout the above determination of the overall force balance on particles in the small continuum (containing many particles) element across the interfacial layer, we have taken the primary single particle representative forces for drag, friction, shear and lift to be as simple as possible, whilst retaining their principal features, and this is sufficient for our present purpose. However it is recognised that, if required, these basic forms could be refined, and thus improved, by introducing higher order contributions, using, for example, the recent results of \cite{SMITH}, \cite{Palmer_Smith_2020}, \cite{Palmer_Smith_2021} and \cite{SMITH1}, amongst others, arising from detailed studies of the motion of small free bodies located in the thin viscous boundary layer next to a wall, in high and moderate Reynolds number flows. We next consider the motion of the interfacial layer.

\subsection{Interfacial layer motion}
\label{subsec:33}
\begin{figure}
  \centerline{\includegraphics[width=0.7\textwidth]{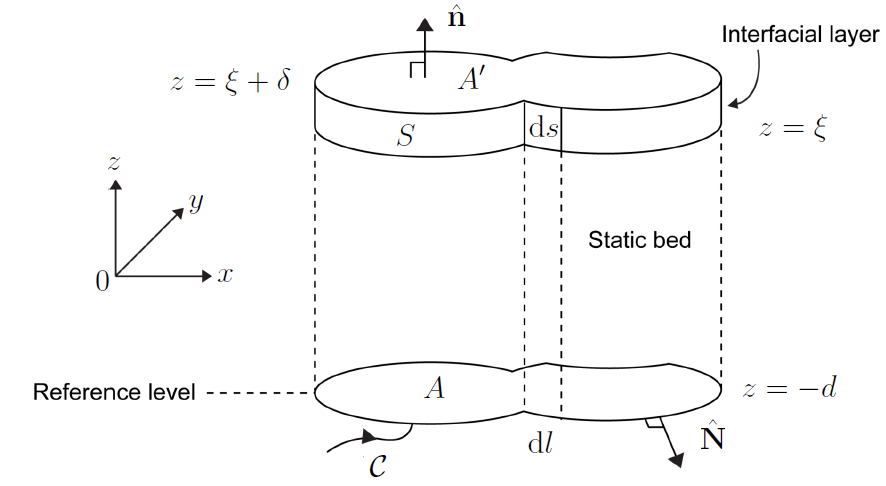}}
  \caption{A qualitative schematic diagram representing a small element as a vertical slice through the static bed and the interfacial layer, with an illustration of the notation used in analysing this element.}
\label{fig:3}
\end{figure}

We consider a cylindrical volume from $z=-d$ in the static bed region ($z=-d$ forming a reference level in the static particle bed region) and extending vertically upwards to the upper surface of the interfacial layer at $z=\xi + \delta$. The situation and nomenclature is illustrated in figure~\ref{fig:3}.

Now the rate of change with respect to time $t$ of the total particle mass in the cylindrical element must be equal to the total mass flux of particles into or out of the cylindrical element. The total particle mass in the cylindrical element is $M$, where
\begin{eqnarray}
   M & = & \int\!\!\int_{A} \, \int_{z=-d}^{z=\xi} \rho_s (1-E_s) \, \rd z \, \rd A + \int\!\!\int_{A} \, \int_{z=\xi}^{z=\xi+\delta} \rho_s (1-E_I) \, \rd z \, \rd A, \nonumber \\
  & = & \rho_s(1-E_s) \int\!\!\int_{A} (\xi+d) \, \rd A + \rho_s \delta \int\!\!\int_{A} (1-E_I) \, \rd A, \label{eqn:37}
\end{eqnarray}
recalling that $E_s$ is the packing voidage of the particles in the static particle bed and $E_I$ is independent of $z$ in the interfacial layer. The total mass flux of particles into the cylindrical element is $Q$, where
\begin{eqnarray*}
   Q & = & - \int\!\!\int_{S} \rho_s (1-E_I) \bv_I \cdot \hbN \, \rd s - \int\!\!\int_{A'} \rho_s (1-E_I) \bv(\br_h,\xi+\delta,t)\cdot\hbn \, \rd A', \nonumber \\
  & = & - \int_{\cal C} \rho_s \delta (1-E_I) \bv_I \cdot \hbN \, \rd \ell - \int\!\!\int_{A'} \rho_s (1-E_I) \bv(\br_h,\xi+\delta,t)\cdot\hbn \, \rd A', \nonumber \\
  & = & - \rho_s\delta\int\!\!\int_{A} \nabla_h\cdot((1-E_I)\bv_I) \, \rd A - \rho_s \int\!\!\int_{A'} (1-E_I) \bv(\br_h,\xi+\delta,t)\cdot\hbn \, \rd A',
\end{eqnarray*}
up to \textit{O}$(h/l)$, via Green's theorem in the plane, with $\nabla_h(\cdot)$ as defined in~(\ref{eqn:7}) and making use of~(\ref{eqn:star}).  Finally, we may re-write the surface integral over $A'$ as a surface integral over $A$ to obtain, on using~(\ref{eqn:star}),
\begin{equation}
  Q =  \int\!\!\int_{A} \left( - \rho_s\delta \nabla_h\cdot((1-E_I)\bv_I) - \rho_s (1-E_I) \bv(\br_h,\xi+\delta,t)\cdot\hbn \right) \, \rd A.
  \label{eqn:39}
\end{equation}
It now follows from the earlier conservation statement, via~(\ref{eqn:37}) and~(\ref{eqn:39}), that
\begin{equation}
  (1-E_s)\xi_t - \delta {E_I}_t + \delta\nabla_h \cdot((1-E_I)\bv_I) = - (1-E_I) \bv(\br_h,\xi+\delta,t)\cdot\hbn, \quad \br_h\in\R^2, \quad t>0.
  \label{eqn:40}
\end{equation}
A further simplification can be made here, based upon~(\ref{eqn:4}).  The ratio of the second two terms on the left side of~(\ref{eqn:40}) to the first term on the left side of~(\ref{eqn:40}) is of \textit{O}$(\delta/h)$, and so equation~(\ref{eqn:40}) may be approximated by
\begin{equation}
  (1-E_s)\xi_t = - (1-E_I) \bv(\br_h,\xi,t)\cdot\hbn, \quad \br_h\in\R^2, \quad t>0,
  \label{eqn:41}
\end{equation}
which finally describes the dynamics of the interfacial layer motion.

\subsection{Boundary conditions between the fluidized region and the interfacial layer}
\label{subsec:34}
Firstly, the normal fluid velocity in the fluidized region at $z=\xi+\delta$ must agree with the normal velocity of the interfacial layer.  This requires
\begin{equation}
  \xi_t = \bu(\br_h,\xi,t)\cdot\hbn, \quad \br_h\in\R^2, \quad t>0,
  \label{eqn:42}
\end{equation}
where $\xi+\delta$ is approximated as $\xi$ via~(\ref{eqn:4}).  Secondly, the tangential fluid velocity in the fluidized region at $z=\xi+\delta$ must agree with
the tangential fluid velocity in the interfacial layer.  This requires, after using~(\ref{eqn:4}), that
\begin{equation}
  \bu_I(\br_h,t) = \bu(\br_h,\xi,t) - (\bu(\br_h,\xi,t)\cdot\hbn)\hbn \equiv \bu_h(\br_h,t), \quad \br_h\in\R^2, \quad t>0.
  \label{eqn:43}
\end{equation}
Finally the voidage field in the fluidized region at $z=\xi+\delta$ must agree with the voidage field in the interfacial layer which, on using~(\ref{eqn:4}), requires
\begin{equation}
  E_I(\br_h,t) = E(\br_h,\xi,t), \quad \br_h\in\R^2, \quad t>0.
  \label{eqn:44}
\end{equation}
The three equations (\ref{eqn:42})--(\ref{eqn:44}) provide boundary conditions relating conditions in the fluidized region to conditions in the interfacial layer.

\section{Full problem in the fluidized region}
\label{sec:4}
Our main objective is to describe the dynamics in the fluidized region, together with the motion of the interfacial layer. To this end, we can, via elimination, obtain a decoupled problem in the fluidized region. Via~(\ref{eqn:8a}) and~(\ref{eqn:8b}) with~(\ref{eqn:13a}) and~(\ref{eqn:13b}), we have
\begin{eqnarray}
  &E_t + \nabla \cdot (E\bu)  =  0,& \label{eqn:45} \\
  &-E_t + \nabla \cdot ((1-E)\bv) = 0,& \label{eqn:46} \\
  &\rho_f E(\bu_t + (\bu\cdot\nabla)\bu) = -\nabla p - \rho_f g E \bk - (9\mu/2a^2) (1-E)\beta(E)(\bu-\bv),& \label{eqn:47} \\
  &\rho_s (1-E)(\bv_t + (\bv\cdot\nabla)\bv) = p_0\nabla E - \rho_s g (1-E) \bk + (9\mu/2a^2) (1-E)\beta(E)(\bu-\bv), \label{eqn:48}
\end{eqnarray}
for $\br$ in $D(t)$, $t > 0$.  Also, we must have
\begin{equation}
  ((1-E_s)\bu + (1-E) \bv)\cdot\hbn = 0, \quad \mbox{on }z=\xi, \mbox{ with }\br_h\in\R^2, \quad t>0,
  \label{eqn:49}
\end{equation}
via~(\ref{eqn:41}), (\ref{eqn:42}) and~(\ref{eqn:44}), whilst~(\ref{eqn:44}) together with~(\ref{eqn:35}) and~(\ref{eqn:43}) requires
\begin{equation}
  E = \left\{ \begin{array}{ll}
                1, & \mbox{for }|\bu_h|^2<1/\Lambda, \\
                \displaystyle{\frac{1}{\Lambda|\bu_h|^2}}, & \mbox{for }1/\Lambda\leq|\bu_h|^2\leq 1/(\Lambda E_s), \\
                E_s, & \mbox{for }|\bu_h|^2 > 1/(\Lambda E_s), \end{array} \right.
                \quad \mbox{on }z=\xi, \mbox{ with }\br_h\in\R^2, \quad t>0,
  \label{eqn:50}
\end{equation}
where
\begin{equation}
  \bu_h(\br_h,t) = \bu(\br_h,\xi,t) - (\bu(\br_h,\xi,t)\cdot\hbn)\hbn, \quad \br_h\in\R^2, \quad t>0,
  \label{eqn:51}
\end{equation}
is the tangential component of the fluid velocity in the fluidized region at $z=\xi$.  Finally, we have, via~(\ref{eqn:42}), that
\begin{equation}
  \xi_t = \bu\cdot\hbn, \quad \mbox{on }z=\xi, \mbox{ with }\br_h\in\R^2, \quad t>0.
  \label{eqn:52}
\end{equation}
Thus, the decoupled problem in the fluidized region is composed of the partial differential equations (\ref{eqn:45})--(\ref{eqn:48}) together with the boundary conditions (\ref{eqn:49})--(\ref{eqn:52}) on $z=\xi$. In particular we observe that the detailed consideration of the dynamics in the interfacial layer has led to  the introduction of the key boundary condition (\ref{eqn:50}) on the voidage at the interface. It is now convenient to write (\ref{eqn:45})--(\ref{eqn:52}) in dimensionless form.

\section{The dimensionless problem}
\label{sec:5}
\begin{figure}
  \centerline{\includegraphics[width=0.7\textwidth]{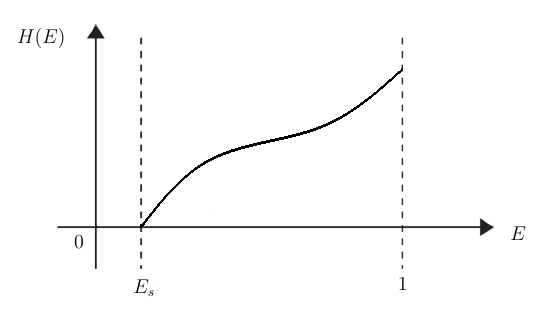}}
  \caption{A typical graph of the function $H(E)$ against $E$.}
\label{fig:4}
\end{figure}

Let $u_s$ be a typical velocity scale in the fluidized region (for both the fluid and particle phases), with the length scales $l$ and $h$ as introduced in~\S\ref{sec:2}. The time scale in the fluidized region is then $(l/u_s)$, whilst a balance between pressure and drag in the momentum equation~(\ref{eqn:47}) leads to a pressure scale
\begin{equation}
  p_s = \frac{\mu u_s l}{a^2}.
  \label{eqn:53}
\end{equation}
Based upon these scales, we introduce the dimensionless variables
\begin{equation}
  x=lx', \, y=ly', \, z=lz', \, \xi=l\xi', \, t=\frac{l}{u_s}t', \, \bu=u_s\bu', \, \bv=u_s\bv', \, p=p_s p'.
  \label{eqn:54}
\end{equation}
On substitution from~(\ref{eqn:54}) and~(\ref{eqn:53}) into (\ref{eqn:45})--(\ref{eqn:52}), we obtain the dimensionless problem in the fluidized region as, after dropping primes for convenience,
\begin{eqnarray}
  &E_t + \nabla \cdot (E\bu)  =  0,& \label{eqn:55} \\
  &-E_t + \nabla \cdot ((1-E)\bv) = 0,& \label{eqn:56} \\
  &\bar{R}_fE(\bu_t + (\bu\cdot\nabla)\bu) = -\nabla p - G E \bk - (1-E)\bar{\beta}(E)(\bu-\bv),& \label{eqn:57} \\
  &\rho \bar{R}_f (1-E)(\bv_t + (\bv\cdot\nabla)\bv) = \alpha\nabla E - \rho G (1-E) \bk + (1-E)\bar{\beta}(E)(\bu-\bv), \label{eqn:58}
\end{eqnarray}
for $\br$ in $D(t)$, $t > 0$, together with
\begin{eqnarray}
  &((1-E_s)\bu + (1-E) \bv)\cdot\hbn  = 0,& \label{eqn:59} \\
  &E = \displaystyle{\left\{ \begin{array}{ll}
                1, & \mbox{for }|\bu_h|^2<1/\gamma, \\
                \displaystyle{\frac{1}{\gamma|\bu_h|^2}}, & \mbox{for }1/\gamma\leq|\bu_h|^2\leq 1/(\gamma E_s), \\
                E_s, & \mbox{for }|\bu_h|^2 > 1/(\gamma E_s), \end{array} \right.}& \label{eqn:60} \\
 &\xi_t = \bu \cdot \hbn,& \label{eqn:61}
\end{eqnarray}
on $z=\xi$, with $\br_h\in\R^2$, $t>0$.  Here we have introduced the dimensionless parameters $\rho=\rho_s/\rho_f$, giving the density ratio of the particulate material to the fluid material, and
\[
  \bar{R}_f=\epsilon^2 R_f
\]
where $\epsilon = a/l$ is the ratio of the microscopic scale particle radius and the macroscopic continuum length scale, and in general can be considered very small, whilst the fluid phase Reynold's number, $R_f=\rho_f u_s l/\mu$, based on the macroscopic length scale l, is expected to be large, justifying our neglect of fluid phase macroscopic viscous terms in (\ref{eqn:57}). Correspondingly $\bar{R}_f=\epsilon^2 R_f$ measures the ratio of the inertia terms to the drag terms in the momentum equations~(\ref{eqn:57}) and~(\ref{eqn:58}). In what follows we will restrict attention to the situation when the macroscopic Reynolds number is large, but the ratio of the microscopic to macroscopic length scales is sufficiently small so that
\begin{equation}
    1 \ll R_f \ll \epsilon^{-2},
\end{equation}
and hence $\epsilon^2 \ll \bar{R}_f\ll 1$. As a consequence of this, in the fluid phase macroscopic momentum balance, inertia will play a secondary role. This is not uncommon in two-phase gas/solid particle flow, as seen in models of gas fluidized beds (see, for example, \cite{NM}). In addition $G = \rho_f g a^2 / \mu u_s$ measures the ratio of the gravity terms to the drag terms in the momentum equations~(\ref{eqn:57}) and~(\ref{eqn:58}), whilst $\alpha=p_0 a^2 / l\mu u_s$ measures the ratio of the particle-particle pressure term to the drag term in the momentum equation~(\ref{eqn:58}).  Finally, $\gamma=3u_s^2 \rho_f / a g \rho_s$ measures the ratio of the lift force to the gravity force acting on the particles in the interfacial layer.  For convenience, we have also written
\[
  \bar{\beta}(E) = \left(\frac{9}{2}\right)\beta(E).
\]
The partial differential equations (\ref{eqn:55})--(\ref{eqn:58}), together with the boundary conditions (\ref{eqn:59})--(\ref{eqn:61}) govern the dynamics in the fluidized region, and the dynamics of the location of the interfacial layer.  For the problems we wish to examine (in particular for the helicopter cloud problem, where the length scale $l$ is based upon the helicopter rotary circle radius, and the velocity scale $u_s$ is based upon the downdraft generated by the rotor, with the fluid being air and the particles being sand grains), we estimate that
\[
  \rho \gg 1, \quad \bar{R}_f\ll \frac{\alpha}{\rho} \ll 1, \quad G \ll \frac{\alpha}{\rho} \ll 1.
\]
Based upon these estimates of the dimensionless parameters for the cloud problem we may approximate equations~(\ref{eqn:57}) and~(\ref{eqn:58}) by
\begin{eqnarray}
  & \bzero = - \nabla p - (1-E)\bar{\beta}(E)(\bu-\bv), & \label{eqn:65} \\
  & \bzero = \alpha\nabla E + (1-E)\bar{\beta}(E)(\bu-\bv), & \label{eqn:66}
\end{eqnarray}
for $\br$ in $D(t)$, $t > 0$, which henceforth replace equations~(\ref{eqn:57}) and~(\ref{eqn:58}).  In both phases, the fundamental balance in~(\ref{eqn:65}) and~(\ref{eqn:66}) is between drag and pressure gradient.  We can now make some direct progress with equations~(\ref{eqn:65}) and~(\ref{eqn:66}).  An addition of~(\ref{eqn:65}) and~(\ref{eqn:66}) gives
\[
  \nabla(p-\alpha E) = \bzero, \quad \br\in D(t), \quad t > 0,
\]
which gives, on integration,
\begin{equation}
  p(\br,t) = \alpha E(\br,t) + F(t), \quad \br\in \bar{D}(t), \quad t \geq 0,
  \label{eqn:67}
\end{equation}
where $\bar{D}(t)$ is the closure of $D(t)$, which replaces~(\ref{eqn:65}).  Here $F(t)$ is an arbitrary function of $t\geq 0$. We note from~(\ref{eqn:67}) that fluid isobars and isovoids coincide in $\bar{D}(t)$. A rearrangement of~(\ref{eqn:66}) then leads to
\begin{equation}
  \bv(\br,t) = \bu(\br,t) + \frac{\alpha\nabla(H(E(\br,t)))}{(1-E(\br,t))}, \quad \br\in \bar{D}(t), \quad t \geq 0,
  \label{eqn:68}
\end{equation}
where for later convenience we have introduced
\[
  H(E) = \int_{E_s}^E \frac{\rd\theta}{\bar{\beta}(\theta)}, \quad E_s \leq E \leq 1,
\]
as sketched in figure~\ref{fig:4}.

We next add equations~(\ref{eqn:55}) and~(\ref{eqn:56}) to obtain
\begin{equation}
  \nabla \cdot (E\bu + (1-E)\bv) = 0, \quad \br\in D(t), \quad t >0,
  \label{eqn:70}
\end{equation}
which replaces equation~(\ref{eqn:56}).  We also observe from~(\ref{eqn:68}) that
\begin{equation}
  \nabla \times \bv = \nabla \times \bu, \quad \br\in D(t), \quad t >0.
  \label{eqn:71}
\end{equation}
Attention is now restricted to irrotational flow in the fluid phase, so that $\nabla \times \bu = \bzero$ in $\bar{D}(t)$ for all $t\geq 0$.  It then follows from~(\ref{eqn:71}) that $\nabla \times \bv = \bzero$ in $\bar{D}(t)$ for all $t\geq 0$, and so the flow in the particle phase is also irrotational.  Under these conditions there exist scalar potentials $\phi$ and $\psi$ such that
\begin{equation}
  \bu = \nabla \phi, \quad \bv  = \nabla \psi,
  \label{eqn:72}
\end{equation}
for all $\br$ in $\bar{D}(t)$, $t\geq 0$.  It follows from~(\ref{eqn:68}) that
\begin{equation}
  \nabla \psi = \nabla\phi + \frac{\alpha}{(1-E)}\nabla(H(E)), \quad \br\in \bar{D}(t), \quad t\geq 0,
  \label{eqn:73}
\end{equation}
after which substituting into the two remaining equations~(\ref{eqn:70}) and~(\ref{eqn:55}) gives
\begin{eqnarray}
  &\nabla^2(\phi + \alpha H(E)) = 0,& \label{eqn:74} \\
  &E_t + \nabla \phi \cdot \nabla E - \alpha E \nabla^2(H(E)) = 0,& \label{eqn:75}
\end{eqnarray}
for $\br$ in $D(t)$, $t> 0$, which are two coupled nonlinear partial differential equations determining $\phi$ and $E$, after which $\psi$ and $p$ are obtained from~(\ref{eqn:73}) and~(\ref{eqn:67}).  Finally, we substitute from~(\ref{eqn:72}) and~(\ref{eqn:73}) into the boundary conditions (\ref{eqn:59})--(\ref{eqn:61}), which become
\begin{eqnarray}
   &E = \displaystyle{\left\{ \begin{array}{ll}
                1, & \mbox{for }|\nabla_h \phi|^2<1/\gamma, \\
                \displaystyle{\frac{1}{\gamma|\nabla_h \phi|^2}}, & \mbox{for }1/\gamma\leq|\nabla_h \phi|^2\leq 1/(\gamma E_s), \\
                E_s, & \mbox{for }|\nabla_h \phi|^2 > 1/(\gamma E_s), \end{array} \right.}& \label{eqn:76} \\
  &(2-E_s-E)\nabla\phi\cdot\hbn+ \alpha\nabla H(E)\cdot\hbn = 0,& \label{eqn:77} \\
  &\xi_t  =  \nabla\phi \cdot \hbn,& \label{eqn:78}
\end{eqnarray}
on $z=\xi$, with $\br_h$ in $\R^2$, $t>0$.  We observe from~(\ref{eqn:74}) that $(\phi+\alpha H(E))$ is harmonic in $D(t)$ at all $t > 0$, whilst $E$ satisfies a convection-diffusion equation (\ref{eqn:75}) in $D(t)$ for all $t>0$, which is nonlinear, with diffusion coefficient
\[
  \alpha = \frac{p_0 a^2}{l \mu u_s}.
\]
The boundary conditions (\ref{eqn:76})--(\ref{eqn:78}) are coupled and nonlinear.

With a view to the helicopter cloud problem, in general we expect that the wind velocity scale will be sufficiently large so that
\[ u_s \gg \frac{p_0 a^2}{l\mu}. \]
Therefore the parameter $\alpha$ may be regarded as small.  It then follows that we may write
\begin{equation}
  \phi = \bar{\phi} + \mbox{\textit{O}}(\alpha),
  \label{eqn:new_star}
\end{equation}
with $\bar{\phi}=\mbox{\textit{O}}(1)$ as $\alpha\rightarrow 0$, whilst, from~(\ref{eqn:77}) and~(\ref{eqn:78}) we may write
\begin{equation}
  \xi = \alpha\bar{\xi} + \mbox{\textit{O}}(\alpha^2),
  \label{eqn:dagger}
\end{equation}
with $\bar{\xi}=\mbox{\textit{O}}(1)$ as $\alpha\rightarrow 0$, from which it follows that $\hbn = \bk + \mbox{\textit{O}}(\alpha)$ as $\alpha\rightarrow 0$.  On substituting from~(\ref{eqn:new_star}) and~(\ref{eqn:dagger}) into~(\ref{eqn:74}) and~(\ref{eqn:77}), the leading order problem for $\bar{\phi}$ decouples, and is given by
\begin{equation}
  \nabla^2 \bar{\phi} = 0, \quad \br\in D_0, \quad t>0,
  \label{eqn:ddagger}
\end{equation}
subject to
\begin{equation}
  \nabla \bar{\phi} \cdot \bk = 0, \quad z=0, \quad \br_h\in \R^2, \quad t>0.
  \label{eqn:tdagger}
\end{equation}
Here $D_0$ is the fixed domain
\[
  D_0 = \R^2 \times (0,\infty).
\]
With $E=\mbox{\textit{O}}(1)$ as $\alpha\rightarrow0$, the remaining problem for the voidage $E$ is given, from~(\ref{eqn:75}) and~(\ref{eqn:76}), as
\begin{equation}
  E_t + \nabla \bar{\phi} \cdot \nabla E - \alpha E \nabla^2(H(E)) = 0, \quad \br\in D_0, \quad t>0,
  \label{eqn:v}
\end{equation}
subject to
\begin{equation}
  E = \left\{ \begin{array}{ll}
                1, & \mbox{for }|\nabla_h \bar{\phi}|^2<1/\gamma, \\
                \displaystyle{\frac{1}{\gamma|\nabla_h \bar{\phi}|^2}}, & \mbox{for }1/\gamma\leq|\nabla_h \bar{\phi}|^2\leq 1/(\gamma E_s), \\
                E_s, & \mbox{for }|\nabla_h \bar{\phi}|^2 > 1/(\gamma E_s), \end{array} \right.
                \quad z=0, \quad \br_h\in \R^2, \quad t>0.
  \label{eqn:vv}
\end{equation}
It should be noted here that to maintain the spatial uniformity of the approximation in $E$ when $0 < \alpha \ll 1$, the dominant terms in {\em both} convection {\em and} diffusion have been retained in~(\ref{eqn:v}).  Finally, on substitution from~(\ref{eqn:dagger}), (\ref{eqn:star}) and~(\ref{eqn:77}) into~(\ref{eqn:78}), we obtain, at leading order,
\[
  \bar{\xi}_t =-\frac{1}{(2-E_s-E)}\nabla H(E)\cdot\bk, \quad z=0, \quad \br_h\in \R^2, \quad t>0.
\]
We are now in a position to formulate the helicopter cloud problem in detail.

\section{The helicopter cloud problem}
\label{sec:6}
\begin{figure}
  \centerline{\includegraphics[width=0.5\textwidth]{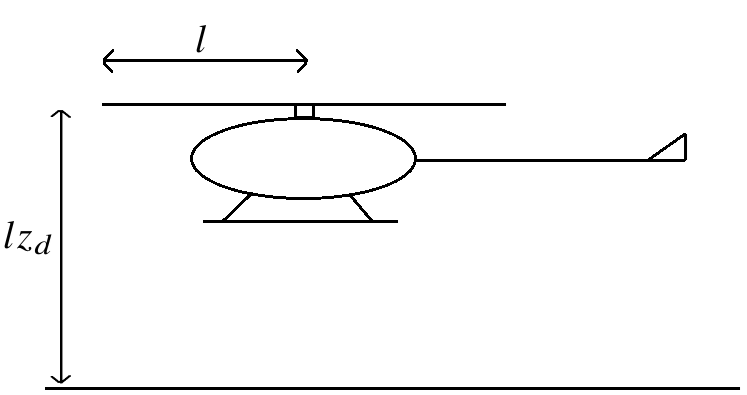}}
  \caption{The parameter $z_d$ measures the ratio of the hovering height of the helicopter rotor to the rotor blade radius - though precise dimensions will vary, we might expect $z_d\approx0.5$ to correspond to the helicopter resting on the sand bed.}
\label{fig:copter}
\end{figure}

In this section, we introduce an elementary model for the helicopter cloud problem and examine this model in detail  using the generic framework established in the preceding sections.  Specifically we model a helicopter hovering steadily, with rotor blades rotating in a horizontal plane, in otherwise still air above a sand bed, which has undisturbed level at $z=0$.  The effect of the helicopter, with rotor blade length $l$, hovering steadily with the rotor at a vertical distance $lz_d$ above the undisturbed particle bed level at $z=0$, as shown in figure~\ref{fig:copter}, is modelled as follows:\\
\begin{enumerate}[label=(\roman*)]
  \item{\, A half-space fluid dipole is placed at location
    \[
      \br = (0,0,lz_d),
    \]
    with $z_d>0$ (and dimensionless).  The axis of the dipole aligns with the unit vector ($-\bk$), and the moment of the dipole is $M_s>0$.  This represents the down draught effect of the helicopter rotor motion.  The associated fluid velocity scale is then
    \[
      u_s = \frac{M_s}{l^3}.
    \]
    The dipole is placed at the centre point of the axis of a cylindrical shell aligned with the $z$-axis, which has axial length and radius $\Delta l$, with $\Delta \ll 1$.  The cylindrical shell is permeable to fluid and particles, with the normal fluid and particle velocities taken as equal on the surface of the cylindrical shell (which represents drag domination close to the dipole, where the fluid velocity becomes unbounded approaching the dipole) and equal to that normal velocity field associated with the dipole at the centre of the shell.  In dimensionless variables the dipole is located at $\br=(0,0,z_d)$, and the cylindrical shell length and radius is $\Delta$.}\\
    \item{\, A fluid line-vortex is placed with its core along the positive $z$-axis.  The strength of the line vortex is taken as $\Gamma_s>0$.  This represents the swirl generated by the helicopter rotor motion.}\\
\end{enumerate}

We might expect this simplistic model of the flow field to at least represent the overall qualitative  features of that flow field generated by a helicopter  in hovering mode, at least away from the immediate neighbourhood of the helicopter body. The key objective is to provide a relevant case to illustrate an application of the generic theory developed in the preceding sections. Representing the interior of the cylindrical shell containing the dipole as $C$, with boundary $\partial C$, the fluidized region now occupies the domain $D_c = D_0 \backslash \bar{C}$.  The modelling structures~(i) and~(ii) require us to introduce the boundary conditions, in dimensionless form,
\begin{eqnarray}
  & \nabla\bar{\phi} \cdot \hat{\bN} = \nabla \phi_d \cdot \hat{\bN}, \quad \br \in \partial C,& \label{eqn:h3} \\
  & \nabla\bar{\phi} = \frac{\Omega}{2\pi R} \hat{\boldsymbol{\theta}} + \mbox{\textit{O}}(1), \quad \mbox{as } R \rightarrow 0 \mbox{ uniformly in }\bar{D}_c.& \label{eqn:h4}
\end{eqnarray}
Here $(R,\theta,z)$ are the usual cylindrical polar coordinates in relation to the Cartesian coordinates $(x,y,z)$ which were introduced earlier, with unit vectors $\hat{\bR}$, $\hat{\boldsymbol{\theta}}$ and $\bk$.  In~(\ref{eqn:h3}), $\phi_d$ represents the harmonic half-space dipole, given by
\begin{equation}
  \phi_d = \sum_{i=1}^2 \frac{(-1)^{i-1} (z-z_i)}{4\pi(R^2 + (z-z_i)^2)^{3/2}}, \quad \mbox{in }\bar{D}_c,
  \label{eqn:85}
\end{equation}
with
\[ z_1 = z_d, \quad z_2 = -z_d, \]
whilst the dimensionless parameter $\Omega$ in~(\ref{eqn:h4}) is given by
\[
  \Omega = \frac{\Gamma_s l^2}{M_s}.
\]
Here $\Omega$ measures the ratio of the fluid velocity scale induced by the line-vortex to that induced by the dipole.  In addition to~(\ref{eqn:h3}) and~(\ref{eqn:h4}), it follows from~(i) and~(\ref{eqn:72}) with (\ref{eqn:73}) that we must have
\begin{equation}
  \nabla E \cdot \hat{\bN} = 0, \quad \br\in\partial C.
  \label{eqn:h6}
\end{equation}
Throughout the above $\hat{\bN}$ is the unit outward normal on the cylindrical shell.  It remains to consider the far field boundary conditions at large distances from the cylindrical shell surface $\partial C$.  We require
\begin{eqnarray}
  & E \rightarrow 1 \mbox{ as } (R^2 + z^2) \rightarrow \infty \mbox{ uniformly in }\bar{D}_c,& \label{eqn:h7} \\
  & \nabla\bar{\phi} \rightarrow 0 \mbox{ as } (R^2 + z^2) \rightarrow \infty \mbox{ uniformly in }\bar{D}_c,& \label{eqn:h8} \\
  & \bar{\xi} \rightarrow 0 \mbox{ as } R\rightarrow \infty \mbox{ uniformly in } 0 \leq \theta < 2\pi.& \nonumber
\end{eqnarray}

To conclude, we note that the dynamics of the helicopter rotor will generally relate the dipole strength $M_s$ and the line-vortex strength $\Gamma_s$ in the model.  The simplest form, which we adopt here, is a linear relation
\[
  \Gamma_s = k M_s,
\]
with $k>0$ a constant of proportionality, depending upon the particular helicopter rotor under consideration.  The dimensionless parameter $\Omega$ is then given by
\[
  \Omega = k l^2,
\]
independent of the dipole and line-vortex strengths.  In its final form, the steady helicopter cloud model has reduced to the following problem for the fluid velocity potential $\bar{\phi}$, namely, from~(\ref{eqn:ddagger}), (\ref{eqn:tdagger}), (\ref{eqn:h3}), (\ref{eqn:h4}) and~(\ref{eqn:h8}),
\begin{eqnarray*}
  & \nabla^2 \bar{\phi} = 0, \quad \br\in D_c, & \\
  & \nabla \bar{\phi} \cdot \bk = 0, \quad z=0, \quad \br_h\in \R^2, & \\
  & \nabla\bar{\phi} \cdot \hat{\bN} = \nabla \phi_d \cdot \hat{\bN}, \quad \br \in \partial C,& \\
  & \nabla\bar{\phi} = \frac{\Omega}{2\pi R} \hat{\boldsymbol{\theta}} + \mbox{\textit{O}}(1), \quad \mbox{as } R \rightarrow 0 \mbox{ uniformly in }\bar{D}_c.& \\
  & \nabla\bar{\phi} \rightarrow 0 \mbox{ as } |\br| \rightarrow \infty \mbox{ uniformly in }\bar{D}_c.&
\end{eqnarray*}
The unique solution to this problem is readily established to be $\bar{\phi}: \bar{D}_c \rightarrow \R$, given by
\begin{equation}
  \bar{\phi}(R,\theta,z) = \phi_d(R,z) + \frac{\Omega}{2\pi}\theta,
  \label{eqn:h12}
\end{equation}
for $\br(R,\theta,z)\in\bar{D}_c$.  The steady problem for $E: \bar{D}_c\rightarrow\R$ is then, via~(\ref{eqn:v}), (\ref{eqn:vv}), (\ref{eqn:h6}) and~(\ref{eqn:h7}), given by
\begin{equation}
  \nabla \bar{\phi} \cdot \nabla E - \alpha E \nabla^2(H(E)) = 0, \quad \br\in D_c,
  \label{eqn:h13}
\end{equation}
with
\begin{equation}
  E = \left\{ \begin{array}{ll}
                1, & \mbox{for }|\nabla_h \bar{\phi}|^2<1/\gamma, \\
                \displaystyle{\frac{1}{\gamma|\nabla_h \bar{\phi}|^2}}, & \mbox{for }1/\gamma\leq|\nabla_h \bar{\phi}|^2\leq 1/(\gamma E_s), \\
                E_s, & \mbox{for }|\nabla_h \bar{\phi}|^2 > 1/(\gamma E_s), \end{array} \right.
                \quad z=0, \quad \br_h\in \R^2, \quad t>0,
  \label{eqn:h14}
\end{equation}
and
\begin{eqnarray}
  & \nabla E \cdot \hat{\bN} = 0, \quad \br\in\partial C, & \label{eqn:h15} \\
  & E \rightarrow 1 \mbox{ as } |\br| \rightarrow \infty \mbox{ uniformly in }\bar{D}_c. & \label{eqn:h16}
\end{eqnarray}
We now observe that $\bar{\phi}:\bar{D}_c\rightarrow\R$, as given in~(\ref{eqn:h12}) and~(\ref{eqn:85}), is such that
\begin{equation}
  \nabla \bar{\phi}(R,z) = a(R,z) \hat{\bR} + \frac{\Omega}{2\pi R} \hat{\boldsymbol{\theta}} + b(R,z) \bk,
  \label{eqn:h17}
\end{equation}
for $\br\in\bar{D}_c$, with
\begin{eqnarray}
  a(R,z) & = & \displaystyle{\frac{3R}{4\pi} \left[\frac{(z+z_d)}{(R^2 + (z + z_d)^2)^{5/2}} - \frac{(z-z_d)}{(R^2 + (z - z_d)^2)^{5/2}} \right]}, \label{eqn:h18} \\
  b(R,z) & = & \displaystyle{\frac{3}{4\pi} \left[\frac{(z+z_d)^2}{(R^2 + (z + z_d)^2)^{5/2}} - \frac{(z-z_d)^2}{(R^2 + (z - z_d)^2)^{5/2}} \right]} \nonumber \\
         &   & \quad + \displaystyle{\frac{1}{4\pi} \left[\frac{1}{(R^2 + (z - z_d)^2)^{3/2}} - \frac{1}{(R^2 + (z + z_d)^2)^{3/2}} \right]}. \label{eqn:h19}
\end{eqnarray}
Thus, via (\ref{eqn:h17})--(\ref{eqn:h19}), we have
\begin{equation}
  |\nabla_h \bar{\phi}(R,0)|^2 = \frac{9z_d^2 R^2}{4\pi^2(R^2+z_d^2)^5} + \frac{\Omega^2}{4\pi^2 R^2},
  \label{eqn:h20}
\end{equation}
for $R> 0$.  Using~(\ref{eqn:h20}), the problem (\ref{eqn:h13})--(\ref{eqn:h16}) may be re-written as
\begin{eqnarray}
  & \nabla \bar{\phi} \cdot \nabla E - \alpha E \nabla^2(H(E)) = 0, \quad \br\in D_c, & \nonumber \\
  & E = g (|\br_h|), \mbox{ on } z=0, \br_h\in\R^2, & \nonumber \\
  & \nabla E \cdot \hat{\bN} = 0, \mbox{ on } \br\in\partial C, & \nonumber \\
  & E \rightarrow 1 \mbox{ as } |\br| \rightarrow \infty \mbox{ uniformly in }\bar{D}_c. & \label{eqn:h24}
\end{eqnarray}
which we henceforth refer to as [BVP].  The function $g:[0,\infty)\rightarrow\R$ is continuous and piecewise smooth and is given by, via~(\ref{eqn:h14}) and~(\ref{eqn:h20}),
\begin{equation}
  g(X) = \left\{ \begin{array}{ll}
                E_s, & \mbox{for }|\nabla_h \bar{\phi}(X,0)|^2 > (\gamma E_s)^{-1}, \\
                \displaystyle{\gamma^{-1}\left( \frac{9z_d^2 X^2}{4\pi^2(X^2+z_d^2)^5} + \frac{\Omega^2}{4\pi^2 X^2} \right)^{-1}}, & \mbox{for }\gamma^{-1}\leq|\nabla_h \bar{\phi}(X,0)|^2\leq (\gamma E_s)^{-1}, \\
                1, & \mbox{for }|\nabla_h \bar{\phi}(X,0)|^2<\gamma^{-1}. \\
                \end{array} \right.
   \label{eqn:h25}
\end{equation}
We note that it is straightforward to refine the far field boundary condition~(\ref{eqn:h24}) in [BVP] to
\begin{equation}
  E(\br) = 1 + \mbox{\textit{O}}(|\br|^{-2}) \mbox{ as } |\br| \rightarrow \infty \mbox{ uniformly in }\bar{D}_c.
  \label{eqn:h27}
\end{equation}
On observing that
\[
  E_s \leq g(X) \leq 1 \mbox{ for all } X\in[0,\infty),
\]
an application of the strong elliptic maximum principle (see, for example, \citet[Chapter~9]{GiTr:98}) establishes that any solution $E:\bar{D}_c\rightarrow\R$ to [BVP] must satisfy the inequality
\begin{equation}
  E_s < E(\br) < 1 \mbox{ for all } \br\in D_c.
  \label{eqn:h29}
\end{equation}
It then follows from~(\ref{eqn:h29}) and~(\ref{eqn:h27}) that [BVP] has a classical unique solution (see, for example, \citet[Chapter~9]{GiTr:98}).  An immediate consequence of uniqueness for [BVP], together with the rotational symmetry of [BVP] about the $z$-axis, is that the solution to [BVP] is axisymmetric in the cylindrical polar coordinates $(R,\theta,z)$; that is, with $E:\bar{D}_c\rightarrow\R$ being the solution to [BVP], then $E=E(R,z)$.

\begin{figure}
  \centerline{\includegraphics[width=0.5\textwidth]{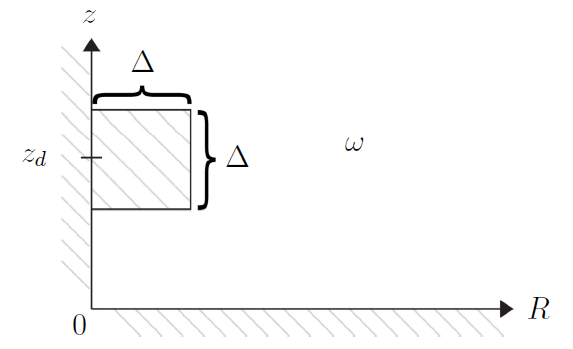}}
  \caption{The domain $\omega$ in the $(R,z)$ plane.}
\label{fig:6}
\end{figure}

A final simplification can be made to [BVP].  In general $\bar{\beta}:[E_s,1]\rightarrow\R$ is monotone decreasing, with $\bar{\beta}(1)=9/2$. In the absence of further information, we will take, as a first approximation, the simple linear form
\[
  \bar{\beta}(E) = \bar{\beta}_0 - (\bar{\beta}_0 - 9/2) E,
\]
for $E\in[E_s,1]$, with $\bar{\beta}_0 > 9/2$ a material constant.  It then follows, after an integration, that
\begin{equation}
  H(E) = (\bar{\beta}_0 - 9/2)^{-1} \log(\bar{\beta}_0(\bar{\beta}_0 - (\bar{\beta}_0 - 9/2) (E-E_s))^{-1}),
  \label{eqn:h31}
\end{equation}
for $E\in[E_s,1]$.  We anticipate that, in general, the variation in $\bar{\beta}(E)$ over $E\in[E_s,1]$ will be small, so we write
\[
  \bar{\beta}_0 = \frac{9}{2} + \delta \beta_0,
\]
with $0<\delta\beta_0\ll 1$, after which~(\ref{eqn:h31}) may be approximated as
\begin{equation}
  H(E) = \frac{2}{9} (E-E_s) + \mbox{\textit{O}}(\delta\beta_0),
  \label{eqn:h33}
\end{equation}
as $\delta\beta_0\rightarrow 0$ uniformly for $E\in[E_s,1]$.  The complete form of [BVP] when written in terms of the cylindrical polar coordinates $(R,\theta,z)$ is presented in Appendix A, and its domain $\omega$ is illustrated in figure~\ref{fig:6}.

\begin{figure}
  \centerline{\includegraphics[width=1.0\textwidth]{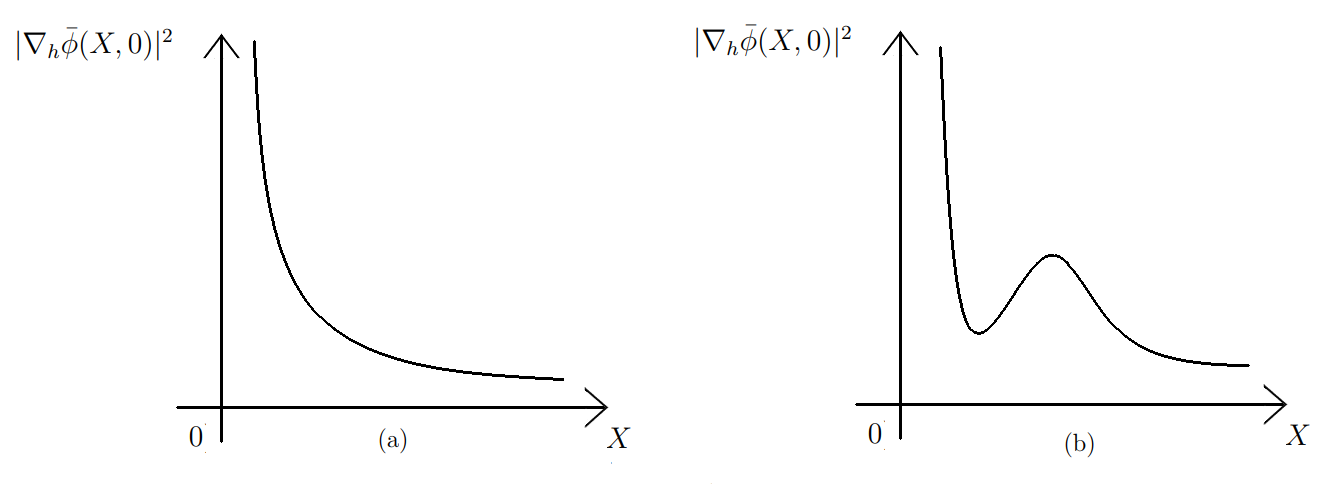}}
  \caption{A qualitative sketch of $|\nabla_h \bar{\phi}(X,0)|^2$ against $X$ for: (a) $\Omega > \Omega_c(z_d)$; and (b) $0< \Omega < \Omega_c(z_d)$.}
\label{fig:7}
\end{figure}

Before proceeding further with [BVP], we must consider the detailed form of $g:[0,\infty)\to\R$, as defined in~(\ref{eqn:h25}).  First we examine $|\nabla_h \bar{\phi}(X,0)|^2$ for $X\in(0,\infty)$, the form of which depends upon the two parameters $\Omega$ and $z_d$. We begin by observing that
\begin{equation}
  |\nabla_h \bar{\phi}(X,0)|^2 \sim \frac{\Omega^2}{4\pi^2 X^2},
  \label{eqn:h42}
\end{equation}
as $X\to 0^+$ and as $X\to\infty$.  Recalling~(\ref{eqn:h20}), we have
\begin{equation}
  \frac{\rd}{\rd X}\left|\nabla_h \bar{\phi}(X,0)\right|^2 = \frac{9 X^4 z_d^2(z_d^2- 4X^2)-\Omega^2(X^2+z_d^2)^6}{2\pi^2 X^3 (X^2+z_d^2)^6},
  \label{eqn:ddx}
\end{equation}
and hence $\left|\nabla_h \bar{\phi}(X,0)\right|^2$ is strictly monotone decreasing for $X\geq z_d/2$.  For fixed $z_d>0$, we see from~(\ref{eqn:h42}) and (\ref{eqn:ddx}) that there is a critical value $\Omega=\Omega_c(z_d)$ such that $\left|\nabla_h \bar{\phi}(X,0)\right|^2$ will also be strictly monotone decreasing for $X\in[0,z_d/2]$, unless $\Omega<\Omega_c(z_d)$, where
\[ \Omega_c(z_d) = \max_{X\in[0,z_d/2]} f(X), \]
with
\begin{equation}
  f(X):=\frac{3 X^2 z_d(z_d^2- 4X^2)^{1/2}}{(X^2+z_d^2)^3}, \quad X\in[0,z_d/2].
  \label{eqn:f}
\end{equation}
We see immediately that $f(X)>0$ for $X\in(0,z_d/2)$, with $f(0)=f(z_d/2)=0$, and it is straightforward to calculate
\[ f'(X) = \frac{36 X z_d}{(X^2+z_d^2)^4(z_d^2-4X^2)^{1/2}}\left(\left(\frac{2z_d^2}{3}-X^2\right)^2-\frac{5z_d^4}{18}\right), \]
hence $f'(X)=0$ if and only if $X=0$ (a local minimum) or
\[ X= X_c(z_d) := \sqrt{\frac{4-\sqrt{10}}{6}}z_d \approx 0.373658z_d, \]
a local maximum, with $f'(X)>0$ for $X\in(0,X_c(z_d))$, and $f'(X)<0$ for $X\in(X_c(z_d),z_d/2)$.  At $X=X_c(z_d)$, substitution into~(\ref{eqn:f}) and a little algebraic manipulation reveals that
\begin{equation}
  \Omega_c(z_d) = f(X_c(z_d)) = \frac{2}{5z_d^2}\sqrt{\frac{-505+188\sqrt{10}}{405}} \approx \frac{0.188046}{z_d^2}.
  \label{eqn:Omega_c}
\end{equation}
We summarise the cases $\Omega\geq\Omega_c(z_d)$ and $\Omega<\Omega_c(z_d)$ as follows:\\
\begin{enumerate}[label=(\alph*)]
  \item{\, $\displaystyle{\Omega \geq \Omega_c(z_d)}$:  In this case $|\nabla_h \bar{\phi}(X,0)|^2$ is strictly monotone decreasing in $X>0$, and for $\Omega>\Omega_c(z_d)$ has the qualitative form shown in figure~\ref{fig:7}(a).  If $\Omega=\Omega_c(z_d)$ then there is a point of inflection at $X=X_c$.}\\
  \item{\, $\displaystyle{0< \Omega < \Omega_c(z_d)}$:  In this case $|\nabla_h \bar{\phi}(X,0)|^2$ has a local minimum and a local maximum in $X>0$, and has the qualitative form shown in figure~\ref{fig:7}(b).}\\
\end{enumerate}

We can now construct $g:[0,\infty)\to\R$, via~(\ref{eqn:h25}).  We have,
\[
  g(X) = \left\{ \begin{array}{ll}
                E_s, & X\in I_1, \\
                \displaystyle{\gamma^{-1}\left( \frac{9z_d^2 X^2}{4\pi^2(X^2+z_d^2)^5} + \frac{\Omega^2}{4\pi^2 X^2} \right)^{-1}}, & X\in I_2, \\
                1, & X\in I_3. \\
                \end{array} \right.
\]
In case (a),
\[
  I_1 = [0,R_1^-], \quad I_2 = (R_1^-,R_1^+), \quad I_3 = [R_1^+,\infty).
\]
However, in case (b), depending upon the choice of $z_d$ and $\Omega$, there are four possibilities, as follows:\\
\begin{enumerate}[label=(\roman*)]
  \item{\, $\displaystyle{I_1 = [0,R_1^-], \quad I_2 = (R_1^-,R_1^+), \quad I_3 = [R_1^+,\infty)}$;}
  \item{\, $\displaystyle{I_1 = [0,R_1^-], \quad I_2 = (R_1^-,R_1^+)\cup(R_2^+,R_3^+), \quad I_3 = [R_1^+,R_2^+]\cup[R_3^+,\infty)}$;}
  \item{\, $\displaystyle{I_1 = [0,R_1^-]\cup[R_2^-,R_3^-], \quad I_2 = (R_1^-,R_2^-)\cup(R_3^-,R_1^+), \quad I_3 = [R_1^+,\infty)}$;}
  \item{\, $\displaystyle{I_1 = [0,R_1^-]\cup[R_2^-,R_3^-], \,\, I_2 = (R_1^-,R_1^+)\cup(R_2^+,R_2^-)\cup(R_3^-,R_3^+), \,\,
        I_3 = [R_1^+,R_2^+]\cup[R_3^+,\infty)}$.}\\
\end{enumerate}
Here $R_i^-$, $i=1,2,3$, are the consecutive positive zeros of the algebraic equation
\begin{equation}
  |\nabla_h \bar{\phi}(X,0)|^2 = (\gamma E_s)^{-1},
  \label{eqn:h45}
\end{equation}
whilst $R_i^+$, $i=1,2,3$, are the consecutive positive zeros of the corresponding algebraic equation
\begin{equation}
  |\nabla_h \bar{\phi}(X,0)|^2 = \gamma^{-1}.
  \label{eqn:h46}
\end{equation}
We note that in case (b), with $z_d$, $\Omega$ and $E_s$ fixed, small values of $\gamma$ correspond to case~(i), whilst increasing $\gamma$ moves through cases~(ii)--(iv).

In order to understand more precisely for which choices of $\gamma$, $E_s$, $z_d$ and $\Omega$ we will see each of cases (i)--(iv), we now investigate a little further how the local maximum and minimum of $|\nabla_h \bar{\phi}(X,0)|^2$ in case~(b) depend on $z_d$ and $\Omega$.  Defining $x_{max}$ and $x_{min}$, respectively, as the values of $X$ at which the local maximum and minimum of $|\nabla_h \bar{\phi}(X,0)|^2$ are achieved, we see from~(\ref{eqn:ddx}), with the change of variables
\[ \bar{x}=\left(\frac{X}{z_d}\right)^2, \quad \bar{\Omega}=\Omega^2 z_d^4, \]
that $\bar{x}_{max}:=(x_{max}/z_d)^2$ and $\bar{x}_{min}:=(x_{min}/z_d)^2$ are the solutions $\bar{x}\in(0,1/4)$ of
\begin{equation}
  9\bar{x}^2(1-4\bar{x})=\bar{\Omega}(1+\bar{x}^6),
  \label{eqn:t}
\end{equation}
where if
\[ \bar{\Omega}> \bar{\Omega}_c:=\Omega_c^2 z_d^4 =\frac{4}{25}\left(\frac{-505+188\sqrt{10}}{405}\right) \approx 0.0353613, \]
then~(\ref{eqn:t}) has no solutions, if $\bar{\Omega}=\bar{\Omega}_c$ then~(\ref{eqn:t}) has one solution at $\bar{x}_*=(4-\sqrt{10})/6\approx 0.139620$, and if $0<\bar{\Omega}<\bar{\Omega}_c$ then~(\ref{eqn:t}) has precisely two solutions, $\bar{x}_{max}$ and $\bar{x}_{min}$.  As $\bar{\Omega}\to 0$, we see that $\bar{x}_{min}\to 0$ and $\bar{x}_{max}\to 1/4$.  A little algebraic manipulation then reveals that
\begin{equation}
  |\nabla_h \bar{\phi}(X,0)|^2 = \frac{9\bar{x}(2-3\bar{x})}{4\pi^2(1+\bar{x})^6z_d^6}, \quad \mbox{for }X=x_{max},x_{min},
  \label{eqn:Xminmax}
\end{equation}
where $\bar{x}$ solves~(\ref{eqn:t}).  From this, we see that
\[ |\nabla_h \bar{\phi}(x_{min},0)|^2 \to 0, \quad |\nabla_h \bar{\phi}(x_{max},0)|^2 \to \frac{576}{3125\pi^2z_d^6}\approx \frac{0.0186755}{z_d^6}, \quad \mbox{as }\Omega\to 0, \]
and
\[ |\nabla_h \bar{\phi}(x_{min},0)|^2 = |\nabla_h \bar{\phi}(x_{max},0)|^2 = \frac{955+424\sqrt{10}}{10125\pi^2z_d^6}\approx \frac{0.0229742}{z_d^6}, \quad \mbox{for }\Omega=\Omega_c(z_d), \]
and hence the range of possible values of $|\nabla_h \bar{\phi}(x_{min},0)|^2$ and $|\nabla_h \bar{\phi}(x_{max},0)|^2$ is rather narrow.

We now consider the parameters in [BVP].  There are six dimensionless parameters, namely, $E_s$, $\alpha$, $\gamma$, $\Omega$, $z_d$ and $\Delta$. We can give order of magnitude estimates of these parameters for the helicopter cloud problem using typical helicopter parameters as recorded, for example, in the papers by \citet{WaWhKeMcGiDo08,GLG,PoMiPaPeVe20,TaGaBaLiZhHu21,TGZWCW,TaYoHeYuWa22,LSXW,LXJHL} together with the standard properties of air and sand, to approximate $l,~M_s,~\gamma_s$. The parameter $\Delta$ measures the ratio of the helicopter rotor core radius to the rotor blade radius, and typically $\Delta\approx 10^{-2}$.  The parameter $z_d$ measures the ratio of the hovering height of the helicopter rotor to the rotor blade radius (recall figure~\ref{fig:copter}), and typically $z_d\approx10^{-1}$--$10^1$.  For the fluid the ratio of the hovering helicopter induced swirl velocity to the induced downwash velocity is measured by $\Omega$, and using typical values extracted from the above papers, we have $\Omega\approx10^{-1}$--$10^0$.  The parameter $E_s$ is the particle packing voidage, and for sand in air this is typically of magnitude $10^{-2}$.  The ratio of the lift force per unit volume to the gravity force per unit volume on particles in the interfacial layer is given by $\gamma$, and again using the typical values extracted from the above papers, we have $\gamma\approx10^{2}$--$10^3$.  Finally, $\alpha$ measures the ratio of particle collisional pressure to drag in the particle flow phase and  we tentatively estimate, for sand fluidized in air, that $\alpha\approx10^{-1}$.  With these estimates of all dimensionless parameters appearing in [BVP], in the next section we consider the numerical solution to [BVP].

\section{Numerical examples}
\label{sec:7}
The accurate and efficient numerical solution of [BVP] requires a degree of nontrivial consideration, and as such, we present the technical details and assessment in Appendix B, where we also establish convergence of our numerical scheme (see example~\ref{eg1}).  Here, we now consider how the solution behaviour depends on our parameter choices.  In examples~\ref{eg2}--\ref{eg4}, we fix
\[  E_s = 0.01,\quad \alpha = 0.1,\quad \Delta = 0.04,\]
and consider the nine possible combinations of
\[ \gamma = 1000, 500, 100, \quad \Omega = 1, 0.5, 0.1 \]
(with $\gamma=500$, $\Omega=0.5$, corresponding to example~\ref{eg1}).  As described in Appendix~B, the accuracy of our numerical scheme depends on a number of discretisation parameters.
From our assessment there of the scheme, we here choose the parameters $N=8$, $L_R=32N$, $L_z=16N$ (compared to $L_R=12N$ and $L_z=6N$ in example~\ref{eg1}), and $\epsilon=1/(10N^2)$, in order to ensure that the computational domain is large enough that all relevant solution behaviour will be captured across this wide range of values of $\gamma$ and $\Omega$.  These parameter choices lead to the total number of degrees of freedom in our numerical solution being $\mbox{DOF}=2317056$ for $z_d = 0.5$, $\mbox{DOF}=2481536$ for $z_d = 0.9$ and $\mbox{DOF}=2646016$ for $z_d = 1.3$, for each of examples~\ref{eg2}--\ref{eg4}.
In each of the figures below, we truncate the computational domains where appropriate for presentational purposes, noting that the solution satisfies $E\approx1$ everywhere outside the plotted range - note that all plots are over the same range within each figure, but that the plotted range varies between examples~\ref{eg2}--\ref{eg4} ($R\in[0,6]$, $z\in[0,5]$ for example~\ref{eg2};  $R\in[0,5]$, $z\in[0,4]$ for example~\ref{eg3}; $R\in[0,3]$, $z\in[0,2]$ for example~\ref{eg4}), though the computational domain for any given value of $z_d$ is identical for each example.

\begin{figure}
     \centering
     \includegraphics[width=0.48\linewidth]{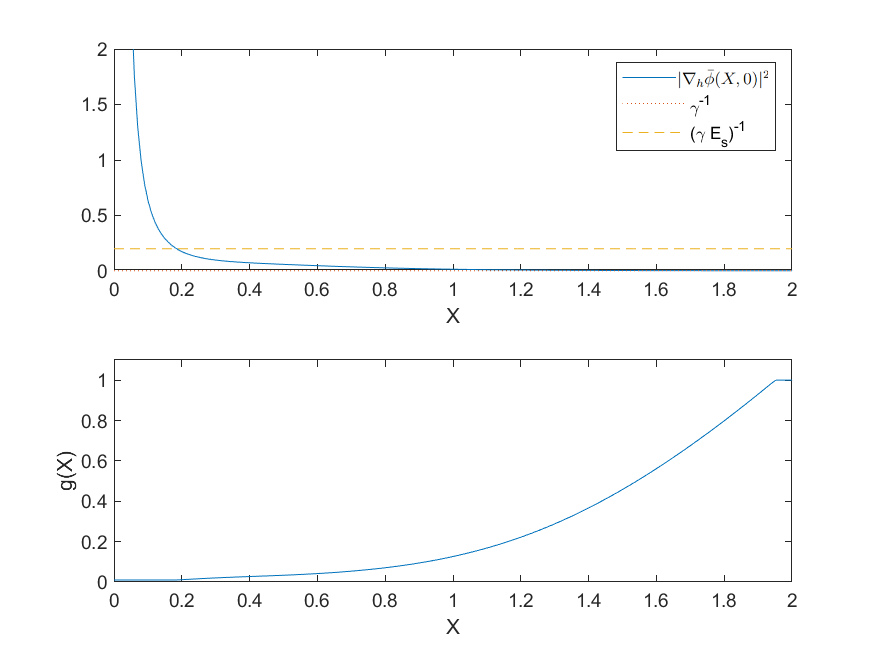}
	 \includegraphics[width=0.48\linewidth]{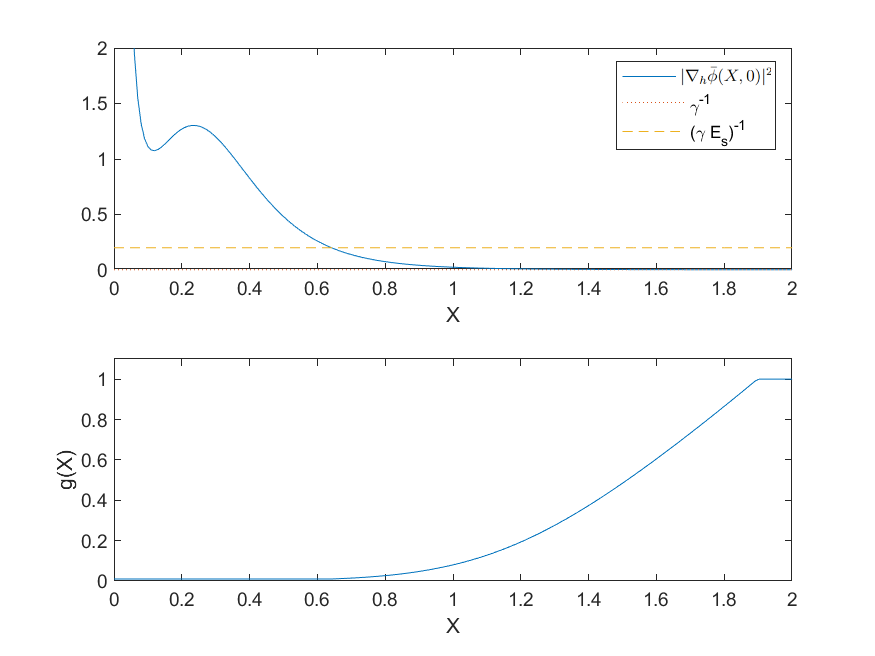}
  \caption{Boundary data $g(X)$ and $|\nabla_h\bar{\phi}(X,0)|^2$, $\gamma^{-1}=0.002$ and $(\gamma E_s)^{-1}=0.2$ for $\gamma=500$, $\Omega=0.5$, $z_d=0.9$ (case (a), left), and $z_d=0.5$ (case (b)(i), right).  Note that the dotted line $\gamma^{-1}=0.002$ almost overlays the $X$-axis.}
  \label{fig:a_bi}
\end{figure}

For each combination of $\gamma$ and $\Omega$ we choose $z_d=1.3$, $z_d=0.9$ and $z_d=0.5$, representing the helicopter landing on the bed of sand (recall figure~\ref{fig:copter}).  Recalling~(\ref{eqn:Omega_c}), the critical value $\Omega_c$ of $\Omega$ that determines whether the boundary data is in case~(a) or~(b), as shown in figure~\ref{fig:7}, depends only on $z_d$, and for the three choices of $z_d$ considered here we have:
\[ \Omega_c(0.5)\approx 0.7522, \quad \Omega_c(0.9)\approx 0.2322, \quad \Omega_c(1.3)\approx 0.1113. \]
For all parameter choices in examples~\ref{eg2} and~\ref{eg3}, and all in examples~\ref{eg4} except one (detailed below) we are either in case~(a) or case~(b)(i).  This is illustrated for the case $\gamma=500$, $\Omega=0.5$, and $z_d=0.9$ or $z_d=0.5$ (as in example~\ref{eg1}) in figure~\ref{fig:a_bi}:  for $z_d=0.9$, $\Omega=0.5>\Omega_c(0.9)$, and hence case~(a) holds;  for $z_d=0.5$, $\Omega=0.5<\Omega_c(0.5)$, and hence case~(b) holds, however in this case both the local maximum and the local minimum take values greater than $(\gamma E_s)^{-1}$ and $\gamma^{-1}$, hence case~(b)(i) holds.  The boundary data $g$ (shown in the lower part of figure~\ref{fig:a_bi}) is qualitatively comparable in each of these cases.

Finally, we note that our expectation, from the physical interpretation of the parameters, is that there will be more sand in the air for large $\gamma$, since large $\gamma$ means that lift dominates gravity in the lifting layer (recalling that $\gamma$ is the ratio of the lift force per unit volume to the gravity force per unit volume on particles in the interfacial layer), so that larger $\gamma$ corresponds to more sand being entrained into the fluidized region.  We recall also that $\Omega$ measures the ratio of the swirl velocity to the downdraft velocity, and recalling the definition of $g$ (\ref{eqn:h25}) we expect that this will also lead to more sand being entrained into the air for large $\Omega$.
\begin{example}[Large $\gamma$]
  \label{eg2}
  \textnormal{We first consider $\gamma=1000$.  In figure~\ref{fig:eg2} we plot the voidage field $E$ for $\Omega=1$ (left column), $\Omega=0.5$ (middle column) and $\Omega=0.1$ (right column), each for $z_d=1.3$ (top row), $z_d=0.9$ (middle row) and $z_d=0.5$ (bottom row).}
   \begin{figure}
     \centering
     \includegraphics[width=0.32\linewidth]{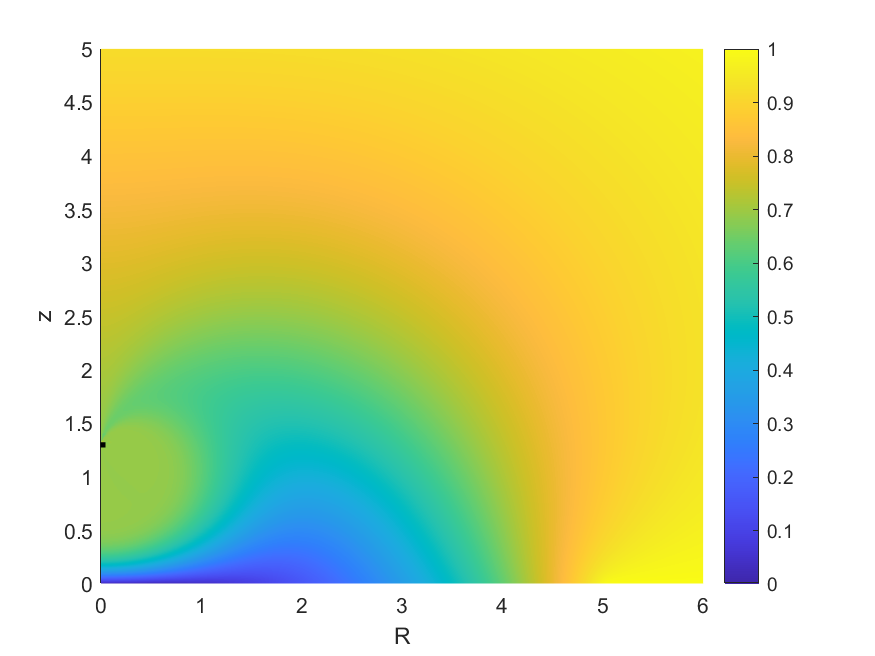}
	 \includegraphics[width=0.32\linewidth]{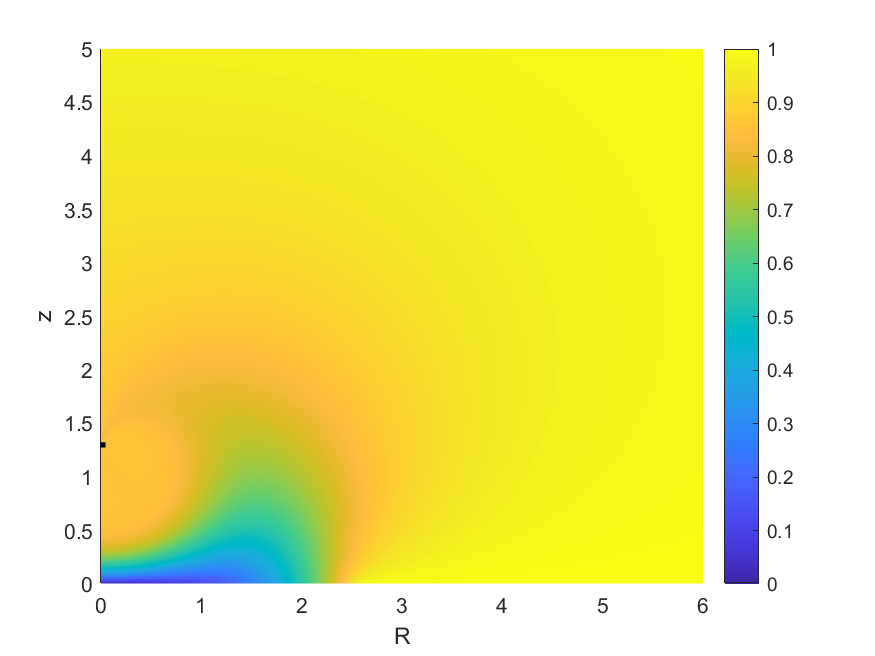}
	 \includegraphics[width=0.32\linewidth]{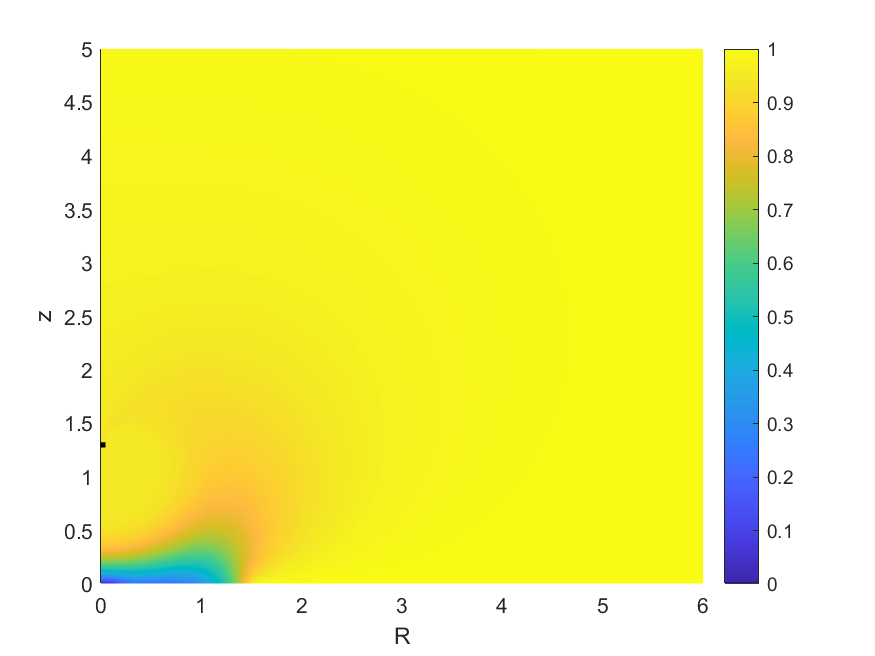}
	 \includegraphics[width=0.32\linewidth]{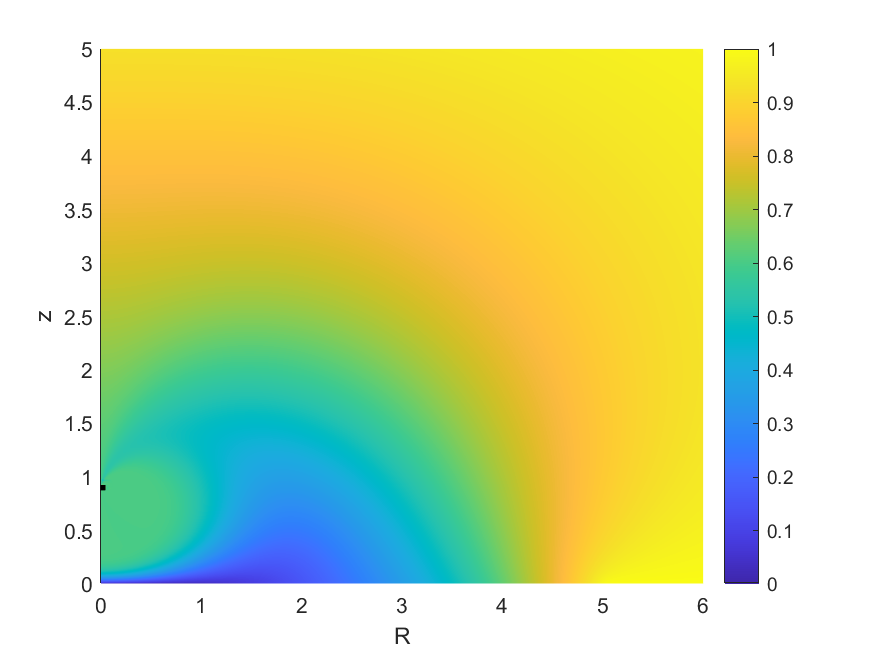}
	 \includegraphics[width=0.32\linewidth]{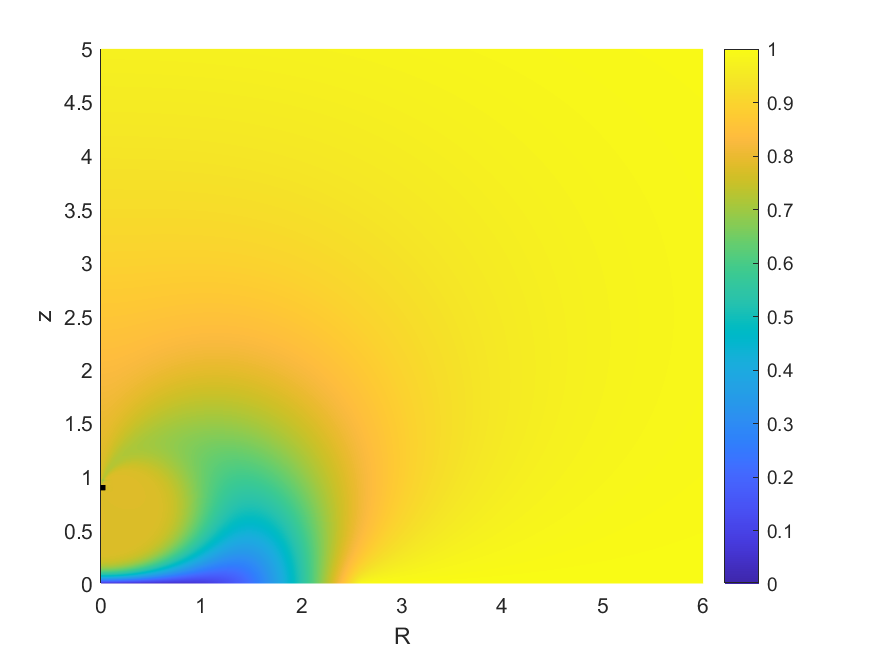}
	 \includegraphics[width=0.32\linewidth]{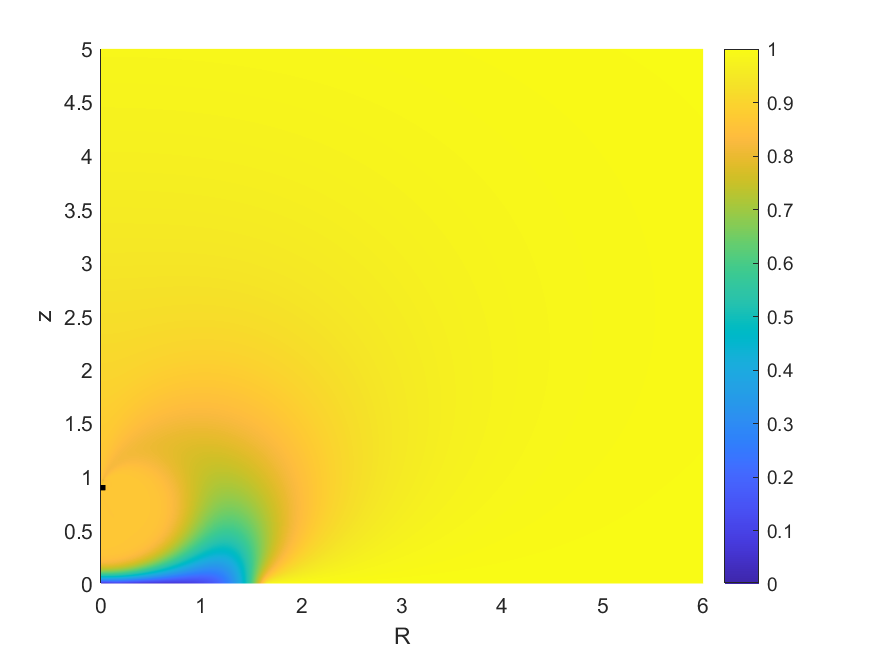}
	 \includegraphics[width=0.32\linewidth]{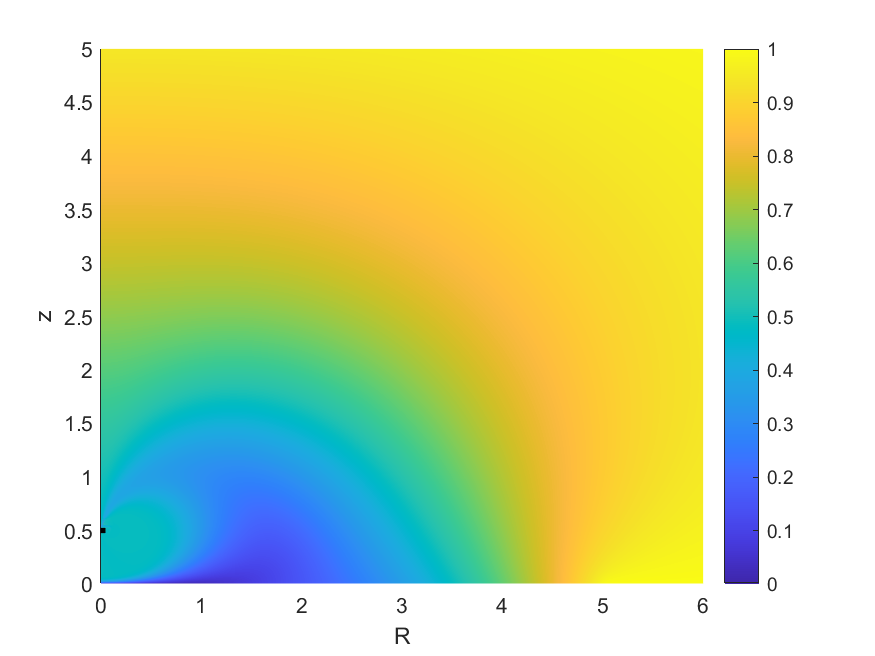}
	 \includegraphics[width=0.32\linewidth]{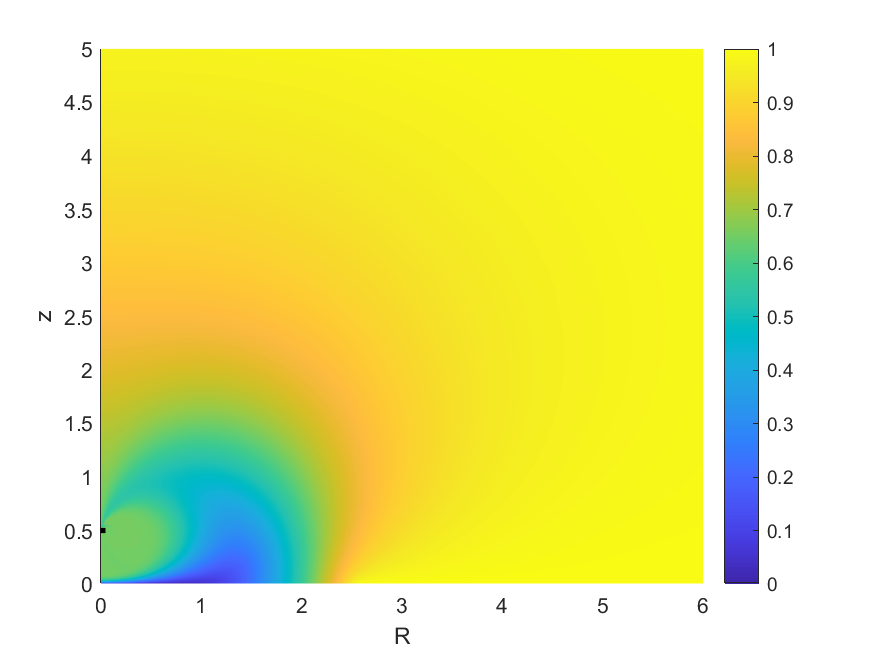}
	 \includegraphics[width=0.32\linewidth]{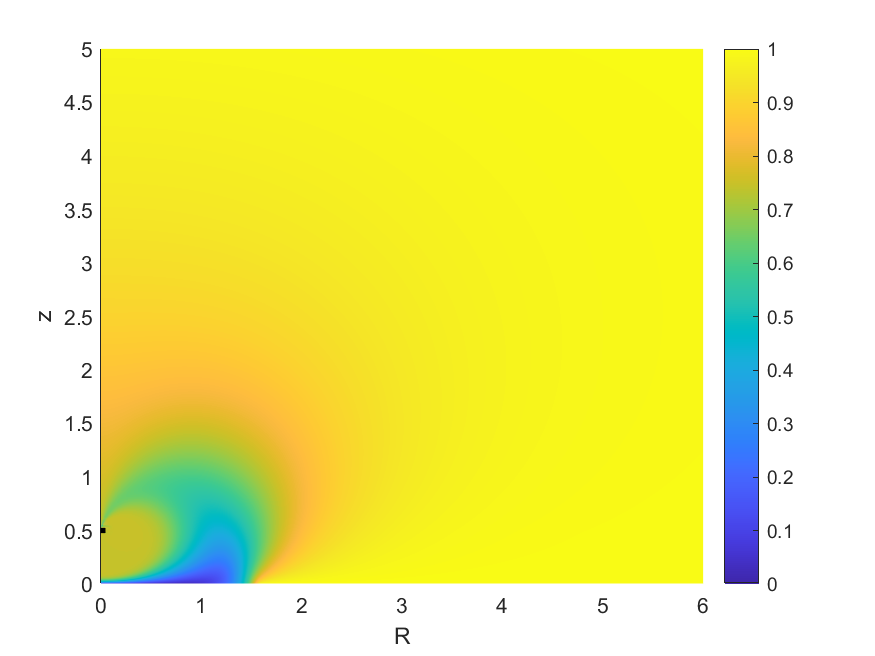}
     \caption{Voidage field $E$, example~\ref{eg2}, plotted for $\gamma=1000$ and for $\Omega=1$ (left column), $\Omega=0.5$ (middle column) and $\Omega=0.1$ (right column), each for $z_d=1.3$ (top row), $z_d=0.9$ (middle row) and $z_d=0.5$ (bottom row).  In each plot, $R\in[0,6]$, $z\in[0,5]$.}
   \label{fig:eg2}
   \end{figure}
   \textnormal{We see more sand in the air (lower values of $E$) corresponding to larger values of $\Omega$, and to lower values of $z_d$.  For $\Omega=1$, and for $\Omega=0.5$ and $z_d=0.9, 1.3$, the boundary data fits case~(a) and is qualitatively comparable to that seen on the left of figure~\ref{fig:a_bi}.  For $\Omega=0.1$, and for $\Omega=0.5$ and $z_d=0.5$, the boundary data fits case~(b)(i) and is qualitatively comparable to that seen on the right of figure~\ref{fig:a_bi}.  In either case, this tallies with the single plume that we see rising up around the helicopter as $z_d$ decreases.}
\end{example}

\begin{example}[Medium $\gamma$]
  \label{eg3}
  \textnormal{Next we consider $\gamma=500$.  In figure~\ref{fig:eg3} we plot the voidage field $E$ for $\Omega=1$ (left column), $\Omega=0.5$ (middle column) and $\Omega=0.1$ (right column), each for $z_d=1.3$ (top row), $z_d=0.9$ (middle row) and $z_d=0.5$ (bottom row).}
   \begin{figure}
     \centering
     \includegraphics[width=0.32\linewidth]{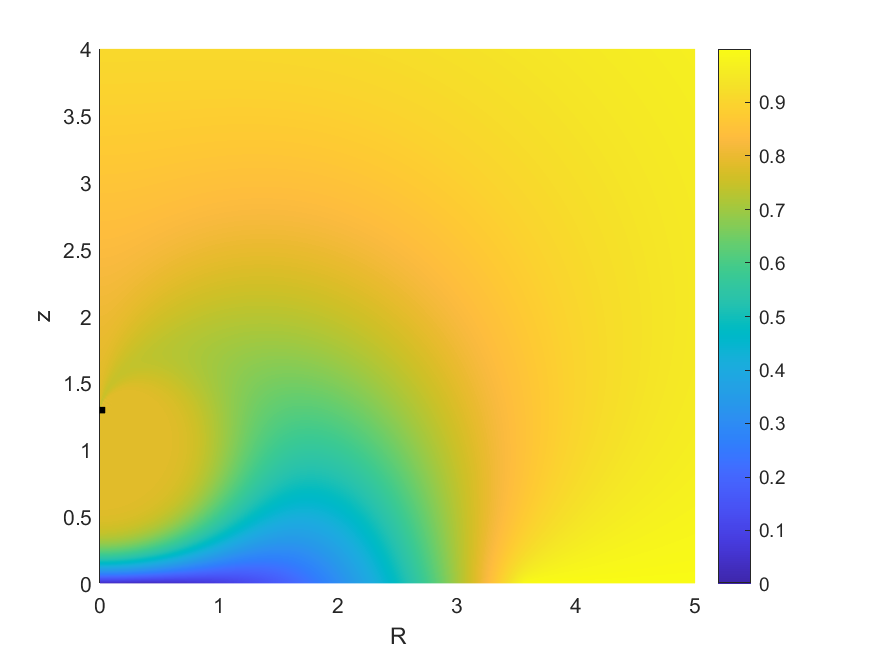}
	 \includegraphics[width=0.32\linewidth]{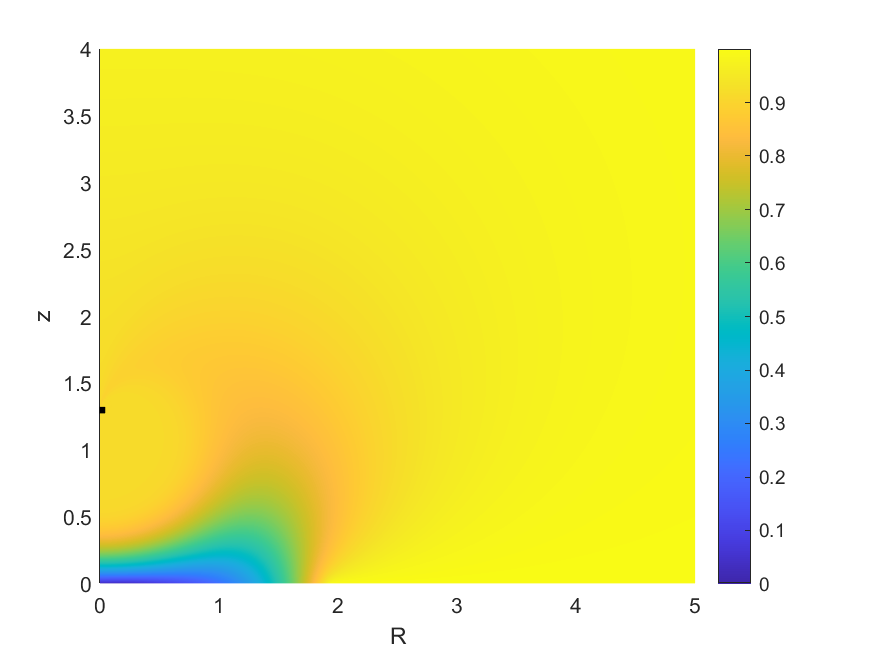}
	 \includegraphics[width=0.32\linewidth]{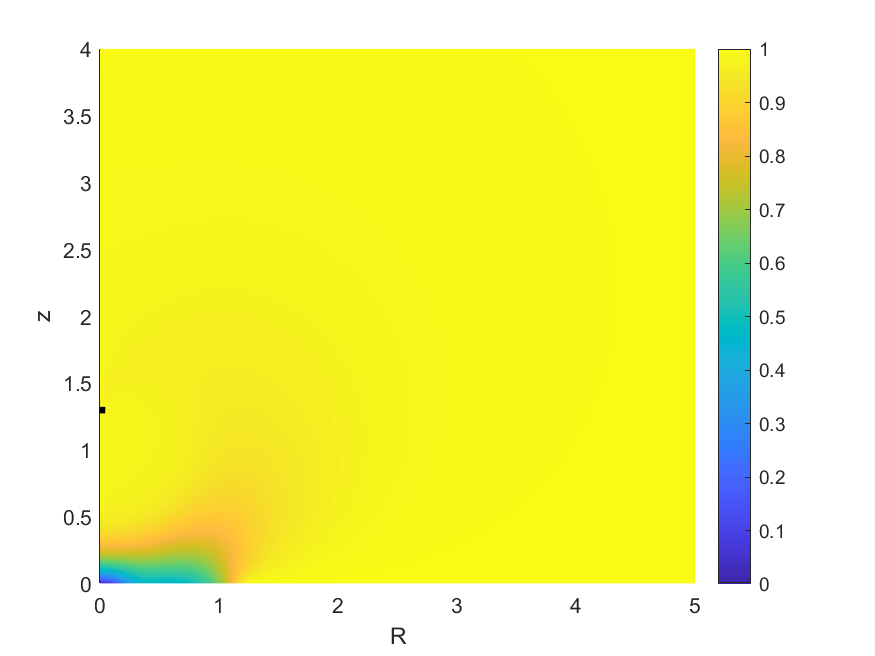}
	 \includegraphics[width=0.32\linewidth]{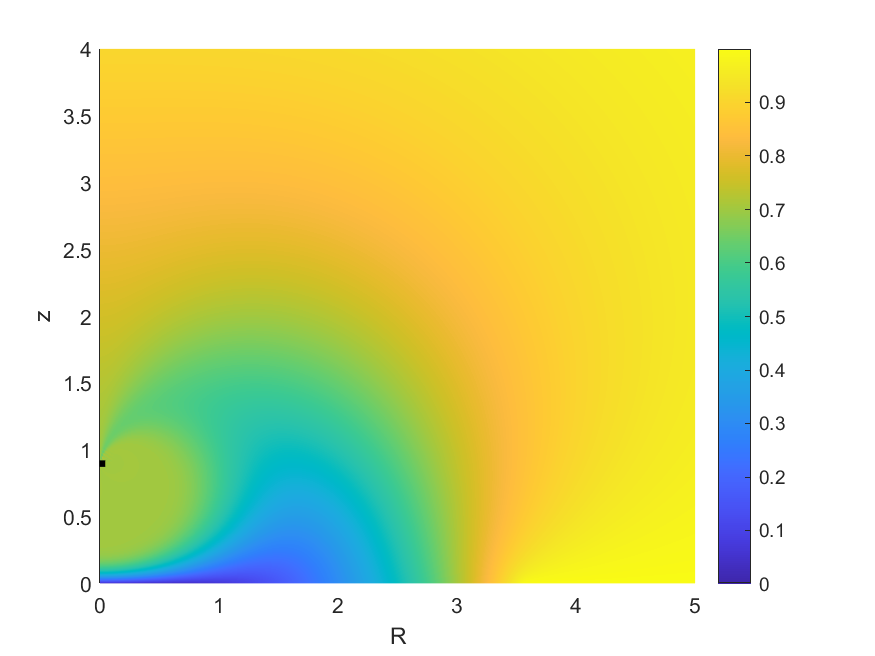}
	 \includegraphics[width=0.32\linewidth]{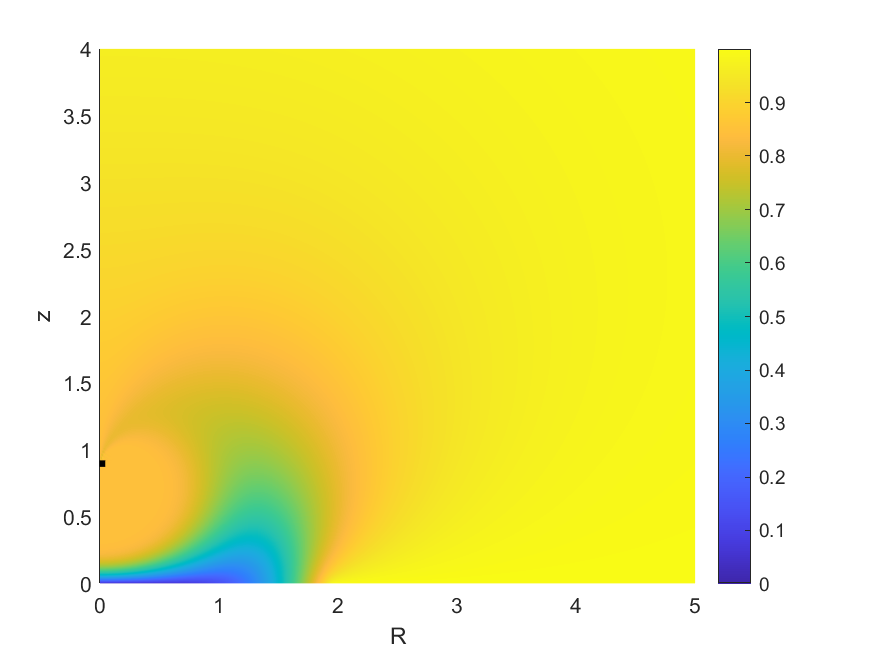}
	 \includegraphics[width=0.32\linewidth]{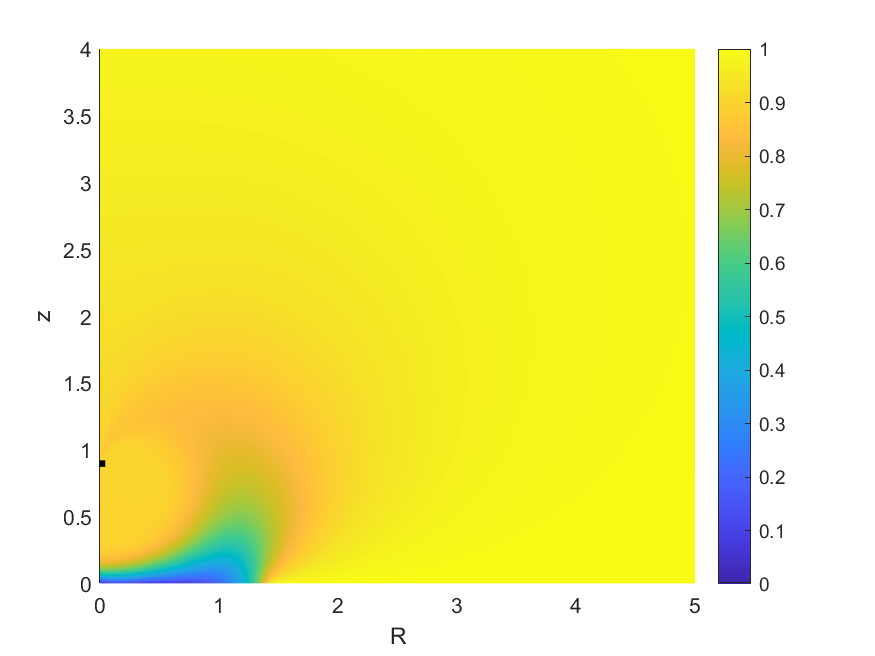}
	 \includegraphics[width=0.32\linewidth]{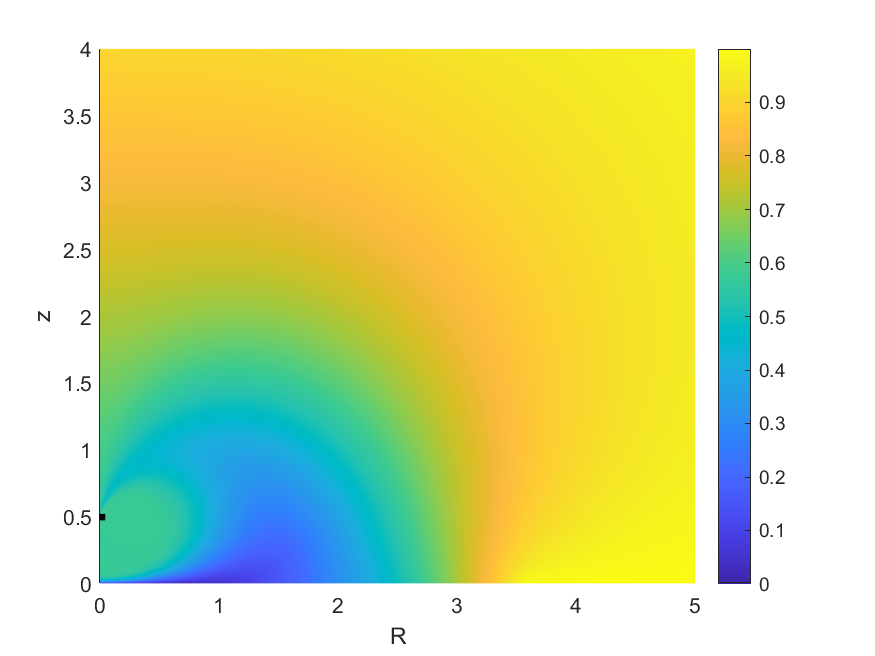}
	 \includegraphics[width=0.32\linewidth]{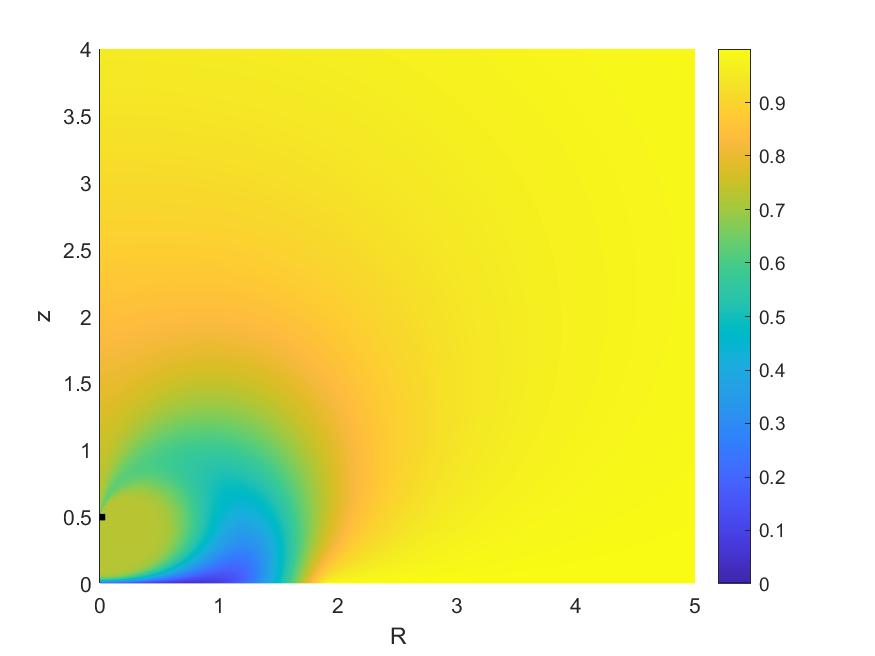}
	 \includegraphics[width=0.32\linewidth]{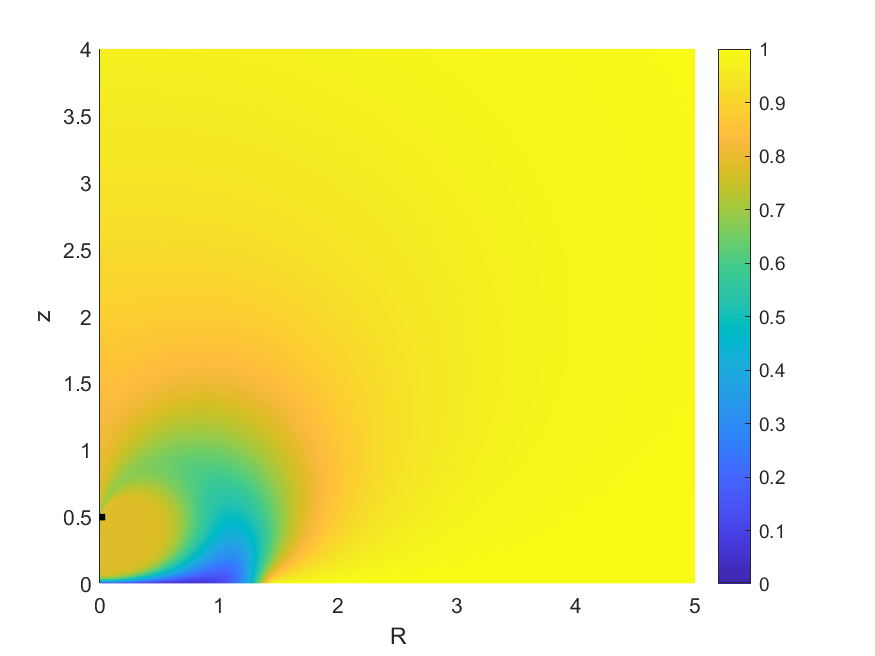}
     \caption{Voidage field $E$, example~\ref{eg3}, plotted for $\gamma=500$ and for $\Omega=1$ (left column), $\Omega=0.5$ (middle column) and $\Omega=0.1$ (right column), each for $z_d=1.3$ (top row), $z_d=0.9$ (middle row) and $z_d=0.5$ (bottom row).  In each plot, $R\in[0,5]$, $z\in[0,4]$.}
   \label{fig:eg3}
   \end{figure}
   \textnormal{Again, we see more sand in the air (lower values of $E$) corresponding to larger values of $\Omega$, and to lower values of $z_d$.  Comparing to figure~\ref{fig:eg2}, we see less sand in the air for the lower value of $\gamma$.  Note that the plotted range is smaller in figure~\ref{fig:eg3} than for figure~\ref{fig:eg2}, though the computational domain is identical.  Exactly as for example~\ref{eg2}, for $\Omega=1$, and for $\Omega=0.5$ and $z_d=0.9, 1.3$, the boundary data fits case~(a), whilst for $\Omega=0.1$, and for $\Omega=0.5$ and $z_d=0.5$, the boundary data fits case~(b)(i), with similar qualitative solution behaviour.}
\end{example}

\begin{example}[Small $\gamma$]
  \label{eg4}
  \textnormal{Next we consider $\gamma=100$.  In figure~\ref{fig:eg4} we plot the voidage field $E$ for $\Omega=1$ (left column), $\Omega=0.5$ (middle column) and $\Omega=0.1$ (right column), each for $z_d=1.3$ (top row), $z_d=0.9$ (middle row) and $z_d=0.5$ (bottom row).}
   \begin{figure}
     \centering
     \includegraphics[width=0.32\linewidth]{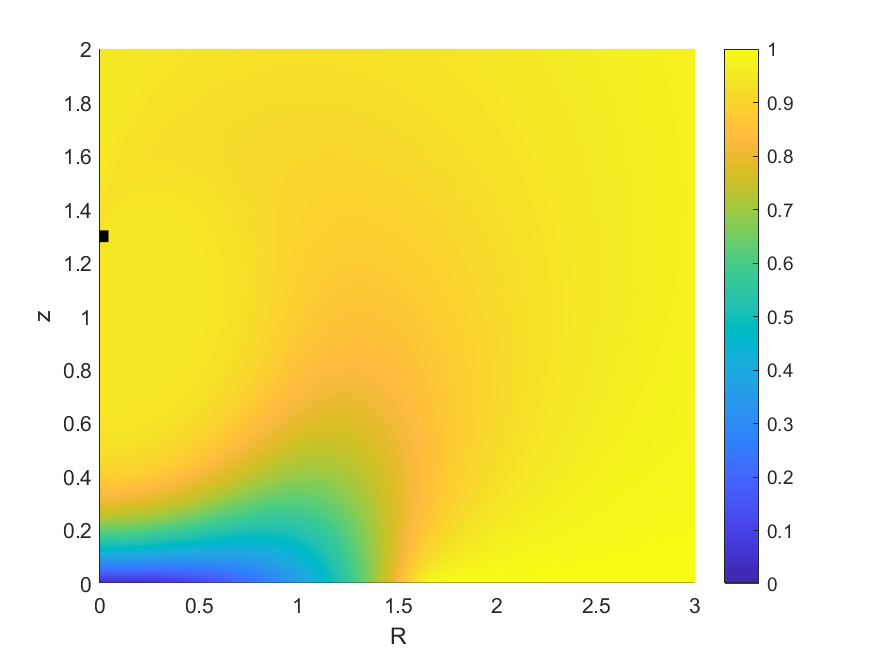}
	 \includegraphics[width=0.32\linewidth]{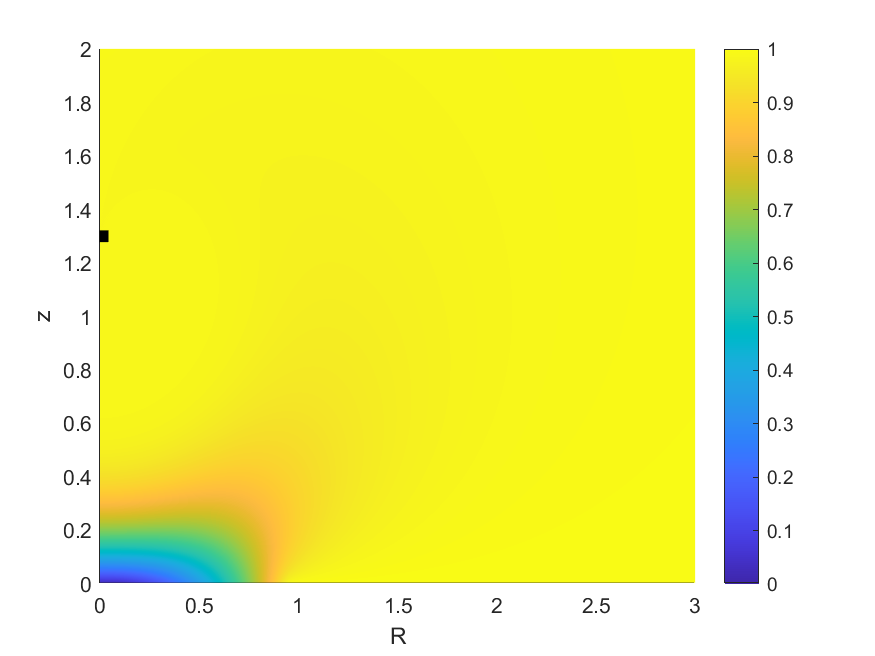}
	 \includegraphics[width=0.32\linewidth]{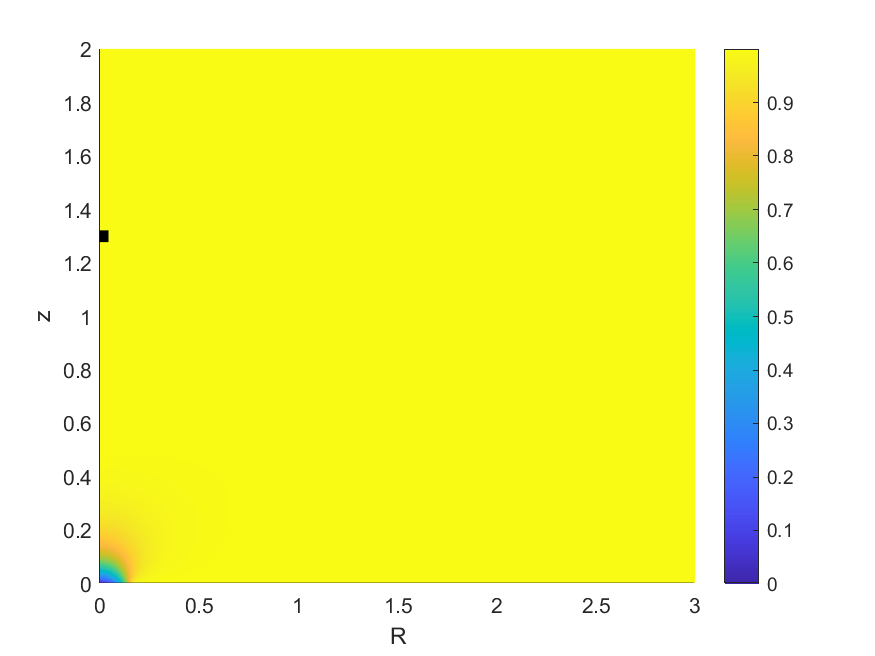}
	 \includegraphics[width=0.32\linewidth]{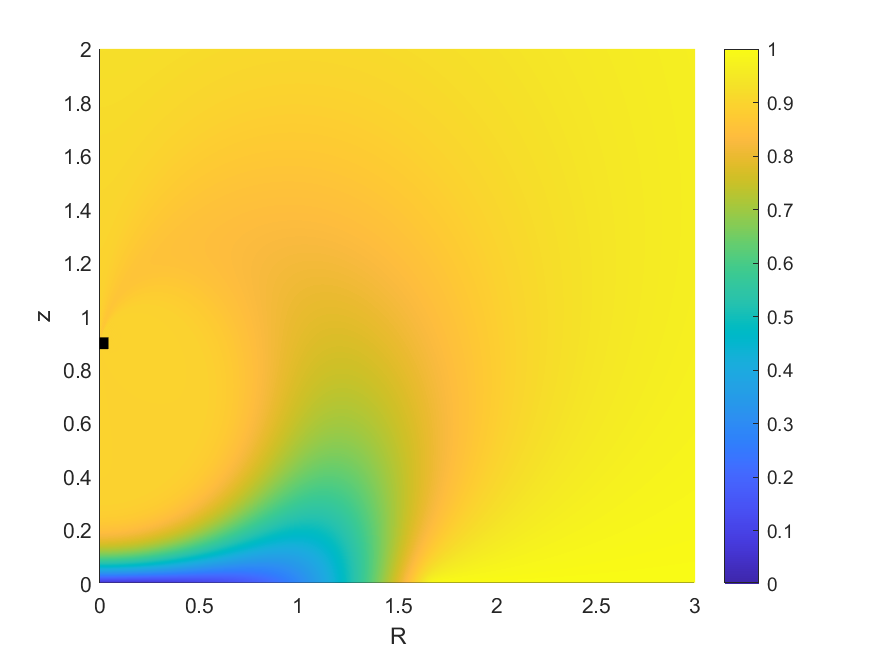}
	 \includegraphics[width=0.32\linewidth]{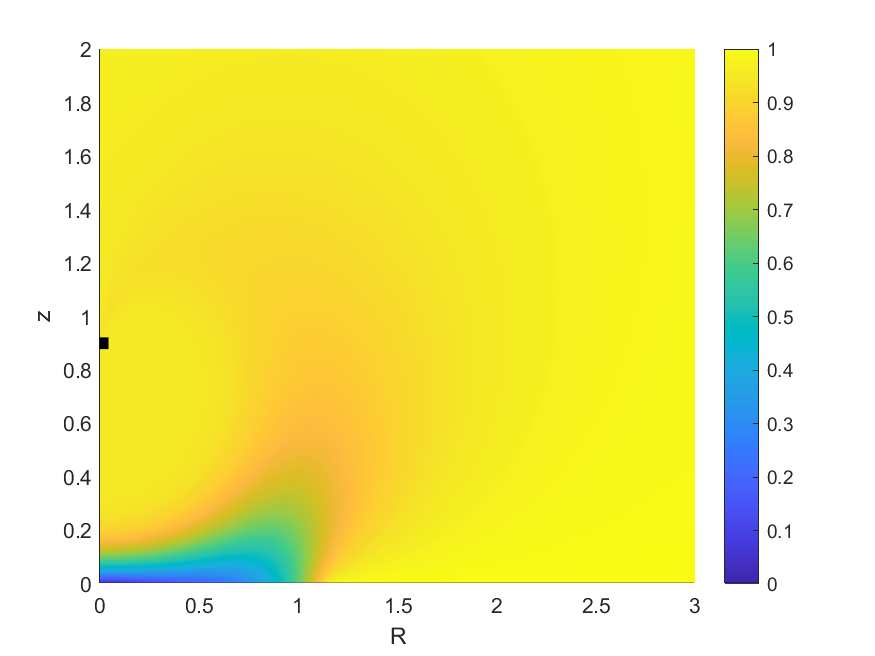}
	 \includegraphics[width=0.32\linewidth]{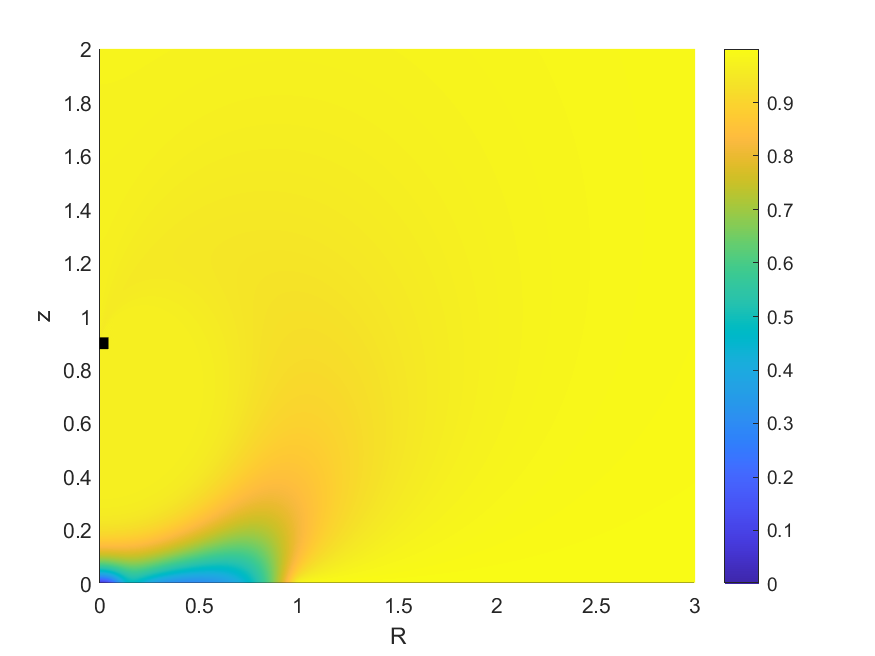}
	 \includegraphics[width=0.32\linewidth]{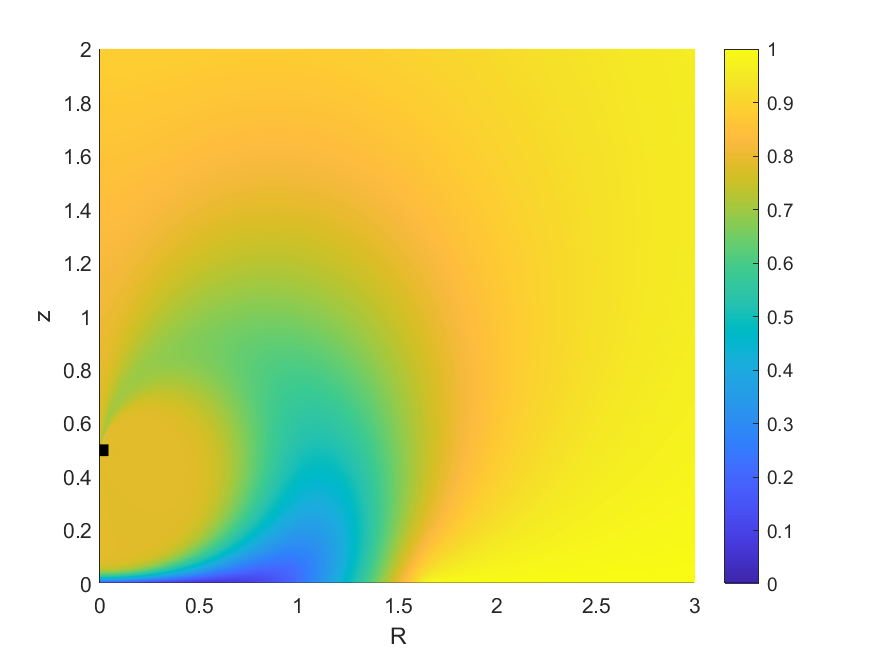}
	 \includegraphics[width=0.32\linewidth]{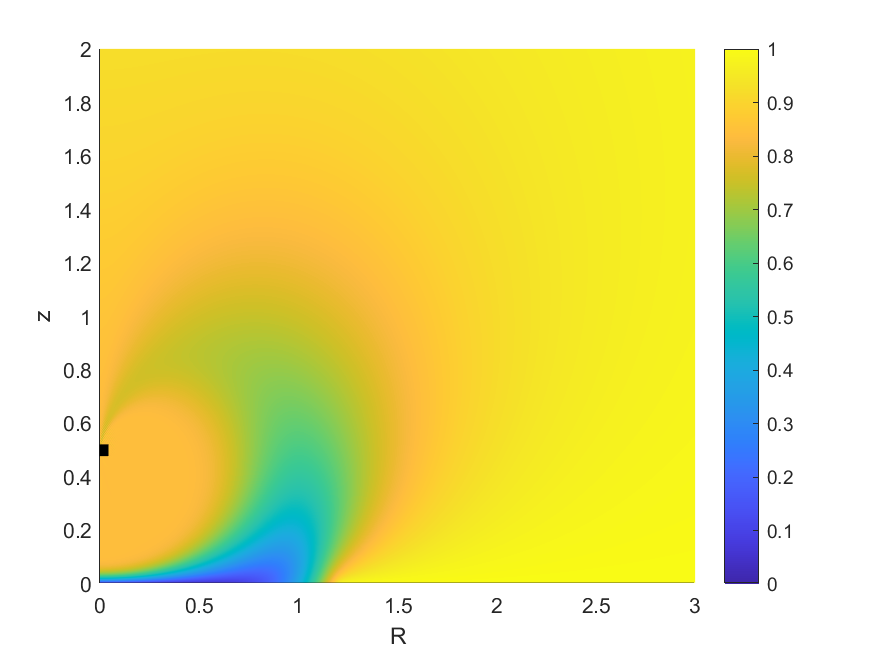}
	 \includegraphics[width=0.32\linewidth]{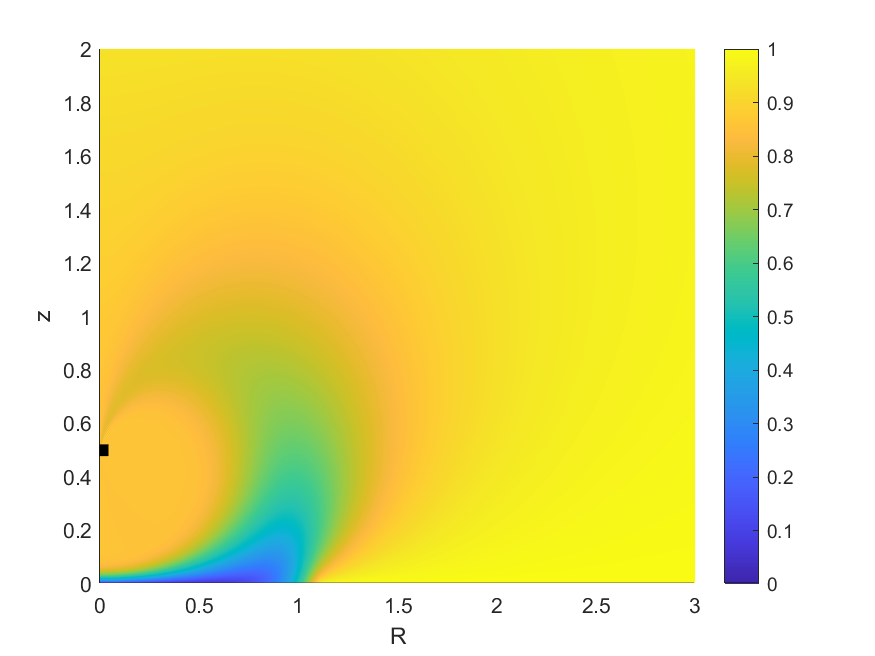}
     \caption{Voidage field $E$, example~\ref{eg4}, plotted for $\gamma=100$ and for $\Omega=1$ (left column), $\Omega=0.5$ (middle column) and $\Omega=0.1$ (right column), each for $z_d=1.3$ (top row), $z_d=0.9$ (middle row) and $z_d=0.5$ (bottom row).  In each plot, $R\in[0,3]$, $z\in[0,2]$.}
   \label{fig:eg4}
   \end{figure}
   \textnormal{Again, we see more sand in the air (lower values of $E$) corresponding to larger values of $\Omega$, and to lower values of $z_d$.  Comparing to figures~\ref{fig:eg2} and~\ref{fig:eg3}, we see less sand in the air for the lower value of $\gamma$.  Note that the plotted range is smaller in figure~\ref{fig:eg4} than for both figure~\ref{fig:eg3} and figure~\ref{fig:eg2}, though the computational domain is identical in each case.  As for examples~\ref{eg2} and~\ref{eg3}, for $\Omega=1$, and for $\Omega=0.5$ and $z_d=0.9, 1.3$, the boundary data fits case~(a), whilst for $\Omega=0.5$ and $z_d=0.5$, and for $\Omega=0.1$ and $z_d=0.9,1.3$ the boundary data fits case~(b)(i), with similar qualitative solution behaviour.  However, for $\Omega=0.1$ and $z_d=0.5$ we are now in case~(b)(iii), as illustrated in figure~\ref{fig:eg4bdy}.  In this case, we see that the local maximum is greater than $(\gamma E_s)^{-1}$, whilst the local minimum is less than $(\gamma E_s)^{-1}$.  As a result, the boundary data $g$ has a small hump near $X\approx0.05$, though it is hard to see any discernible effect from this on the qualitative solution behaviour (bottom right plot of figure~\ref{fig:eg4}).}
   \begin{figure}
     \centering
     \includegraphics[width=0.9\linewidth]{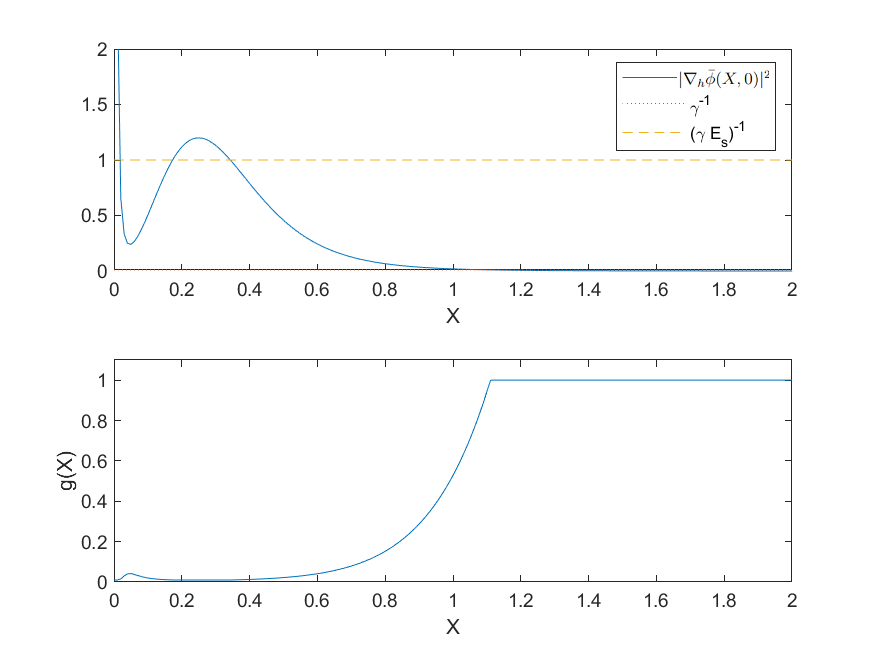}
  \caption{Boundary data $g(X)$ and $|\nabla_h\bar{\phi}(X,0)|^2$, $\gamma^{-1}=0.01$ and $(\gamma E_s)^{-1}=1$ for $\Omega=0.1$, $z_d=0.5$, example~\ref{eg4}.  Note that the dotted line $\gamma^{-1}=0.01$ almost overlays the $X$-axis.}
  \label{fig:eg4bdy}
\end{figure}
\end{example}

\begin{example}[Two-plume example]
  \label{eg5}
  \textnormal{In our final example, we choose
  \[  E_s = 0.01,\quad \Delta = 0.04, \quad \gamma=70, \quad \Omega=0.1, \]
  with first $\alpha = 0.1$ (as above), and then $\alpha=0.02$.
  Our choice of $\gamma=70$ is smaller than in examples~\ref{eg2}--\ref{eg4}, hence $\gamma^{-1}$ is larger, enabling us to demonstrate case~(b)(ii) (recalling figure~\ref{fig:7}), which provides a new and qualitatively different phenomena in the solution behaviour compared to that seen in examples~\ref{eg2}--\ref{eg4}.}

  \textnormal{Here, we will start with $z_d=1.1$, and gradually reduce it, with the transition between cases (b)(i) and (b)(ii) being rather sensitive to the choice of $z_d$, for $z_d\in[0.9,1.1]$.  This is illustrated in figure~\ref{fig:eg5b}, where we plot $|\nabla_h\bar{\phi}(X,0)|^2$, the value $\gamma^{-1}$ and the boundary data $g$ for a range of values of $z_d$.}
  \begin{figure}
     \centering
     \includegraphics[width=0.45\linewidth]{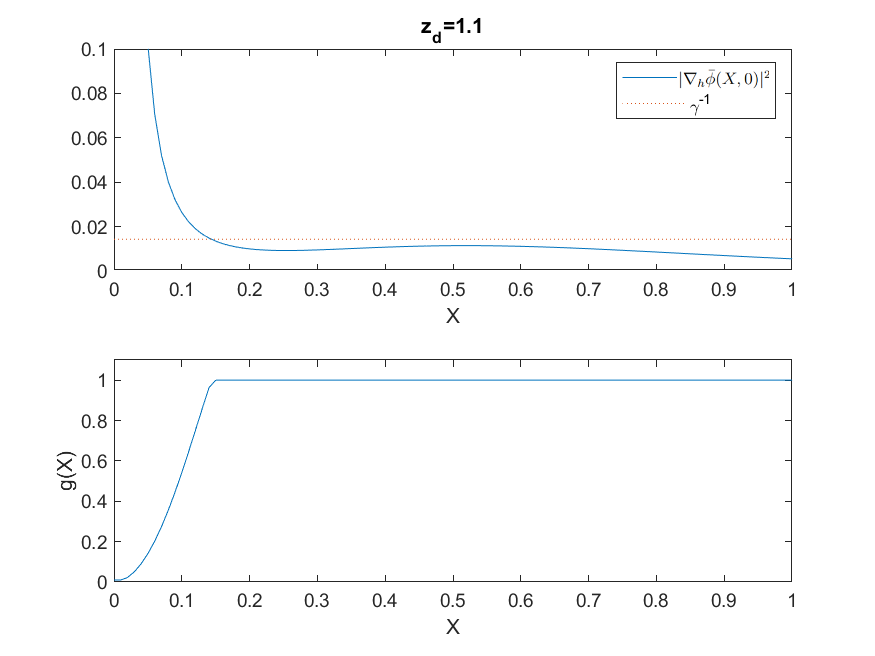}
     \includegraphics[width=0.45\linewidth]{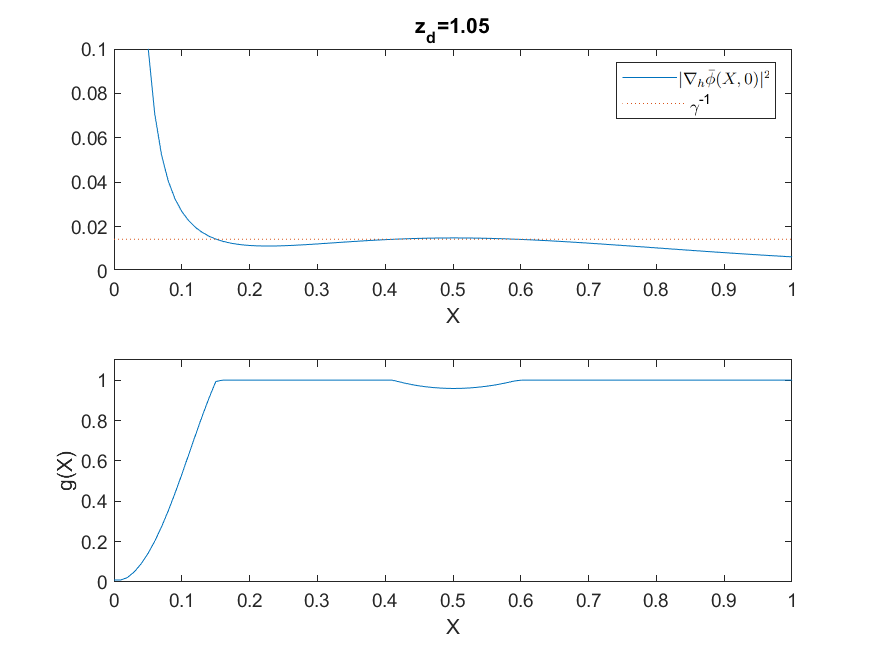}
     \includegraphics[width=0.45\linewidth]{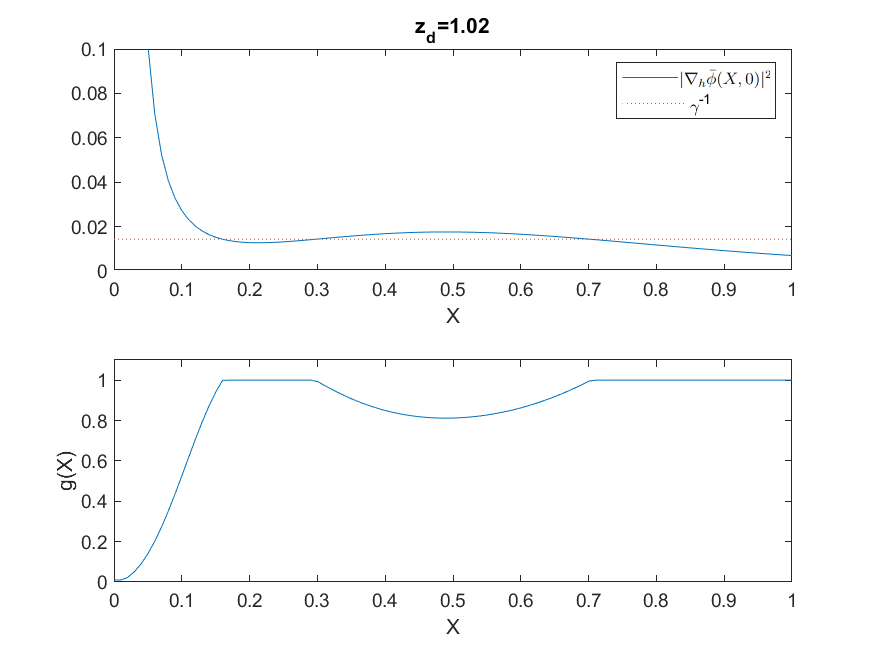}
     \includegraphics[width=0.45\linewidth]{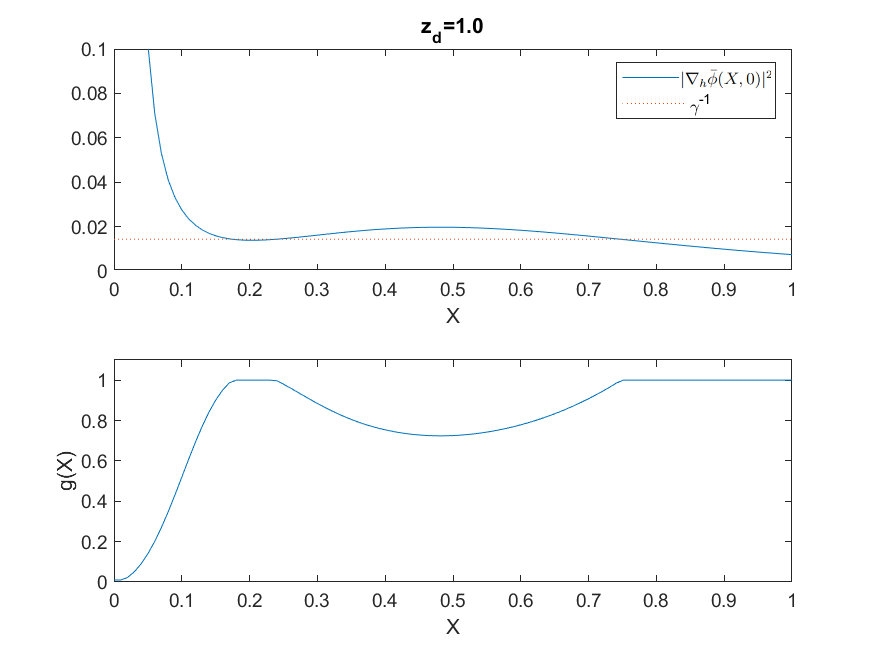}
     \includegraphics[width=0.45\linewidth]{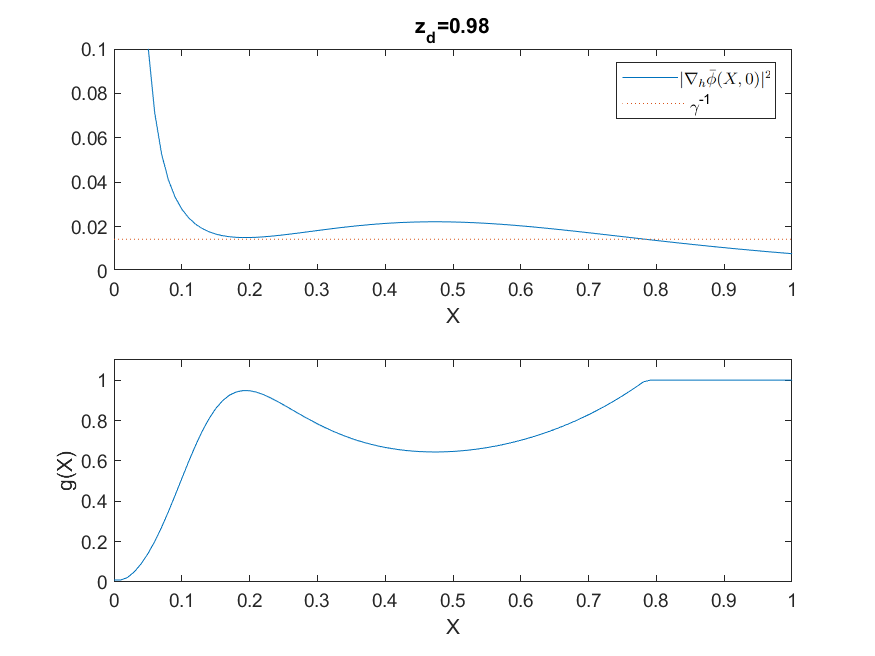}
     \includegraphics[width=0.45\linewidth]{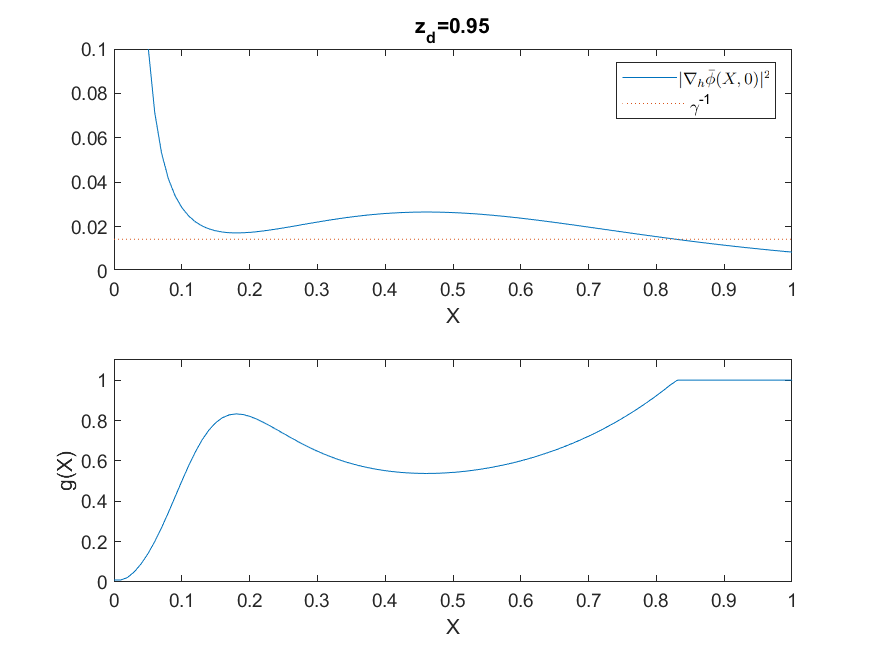}
     \includegraphics[width=0.45\linewidth]{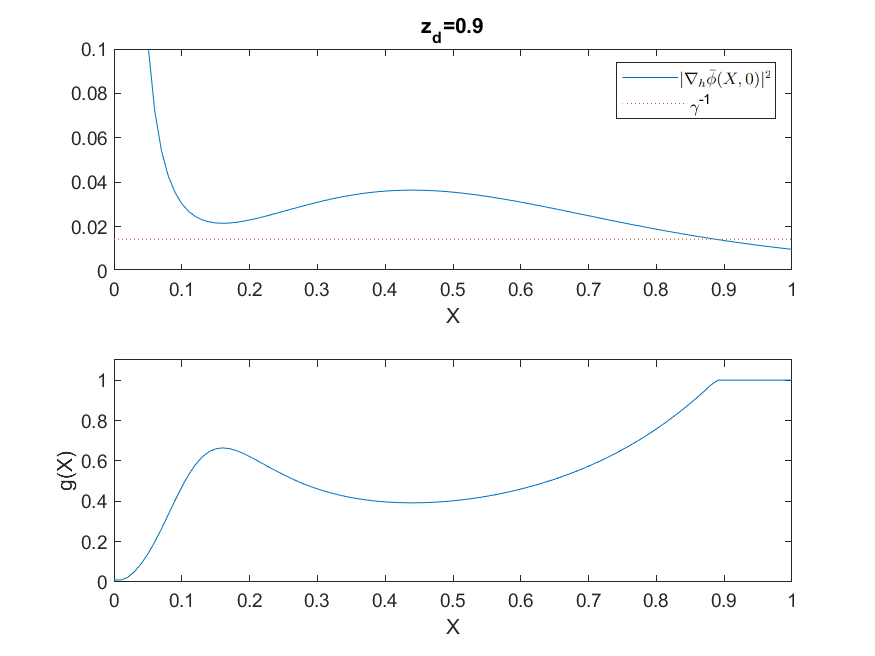}
     \includegraphics[width=0.45\linewidth]{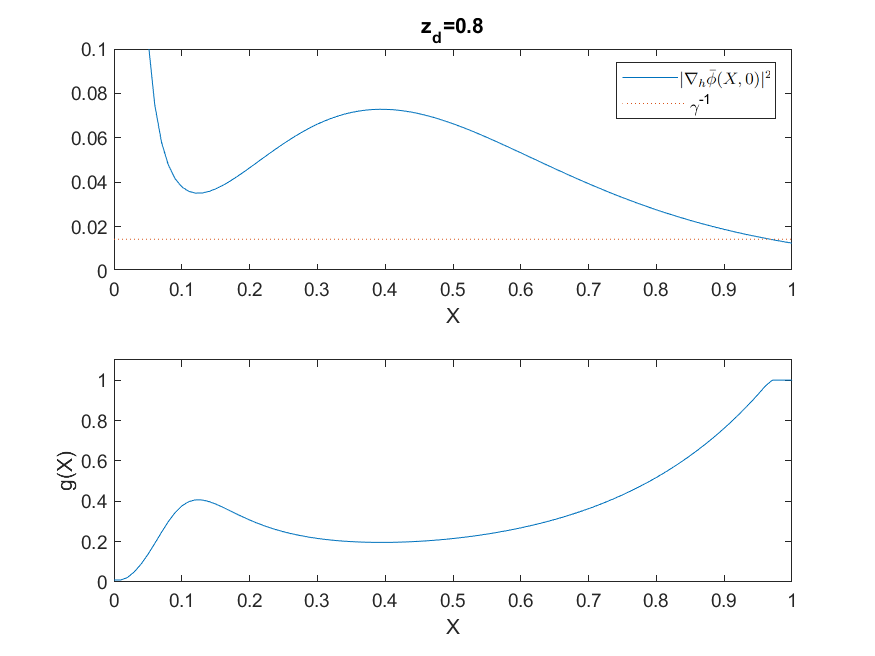}
     \caption{Boundary data $g(X)$, and $|\nabla_h\bar{\phi}(X,0)|^2$, $\gamma^{-1}=1/70$, example~\ref{eg5}, plotted for $z_d=1.1, 1.05, 1.02, 1.0, 0.98, 0.95, 0.9, 0.8$.  For $z_d=1.0,1.02,1.05$ we are in case (b)(ii); for all other values of $z_d$ we are in case (b)(i).  The line $(\gamma E_s)^{-1}=0.7$ is not shown on the plot.}
   \label{fig:eg5b}
   \end{figure}
   \textnormal{We see that for $z_d=1.1$, the local maximum is below $\gamma^{-1}$ (case (b)(i)).  As $z_d$ decreases, the local maximum becomes greater than $\gamma^{-1}$, causing a dip in the boundary data, where $g<1$ between two regions where $g=1$ (case (b)(ii)).  As $z_d$ decreases further, to $z_d=0.98$ and below, the local minimum rises above $\gamma^{-1}$, and we return to case (b)(i). The presence of two separated regions where $g=1$ drives the formation of a qualitatively new phenomena, compared to examples \ref{eg2}--\ref{eg4}, namely a second ``plume''.}

   \textnormal{Specifically, for $z_d=1.1$, we see a single concentration of sand below the helicopter.  As the helicopter descends to $z_d=1.05$, a second concentration of sand forms;  as the helicopter descends further, this second concentration grows and starts to form a second ``plume'' (compared to the single plume that we saw in examples~\ref{eg2}--\ref{eg4}), until at around $z_d=1.0$ the two concentrations of sand merge, forming a single plume again.  This behaviour is illustrated, for $\alpha=0.1$ in figure~\ref{fig:eg5a}.  All experiments in this example were carried out with $N=8$, and with $L_R=16N$, $L_z=8N$ (lower values than for examples~\ref{eg2}--\ref{eg4}, with the key phenomena more localised near the origin), and again $\epsilon=1/(10N^2)$.  As for the examples above, we again truncate the computational domains for presentational purposes, noting that the solution satisfies $E\approx1$ outside the plotted range.}
  \begin{figure}
     \centering
     \includegraphics[width=0.32\linewidth]{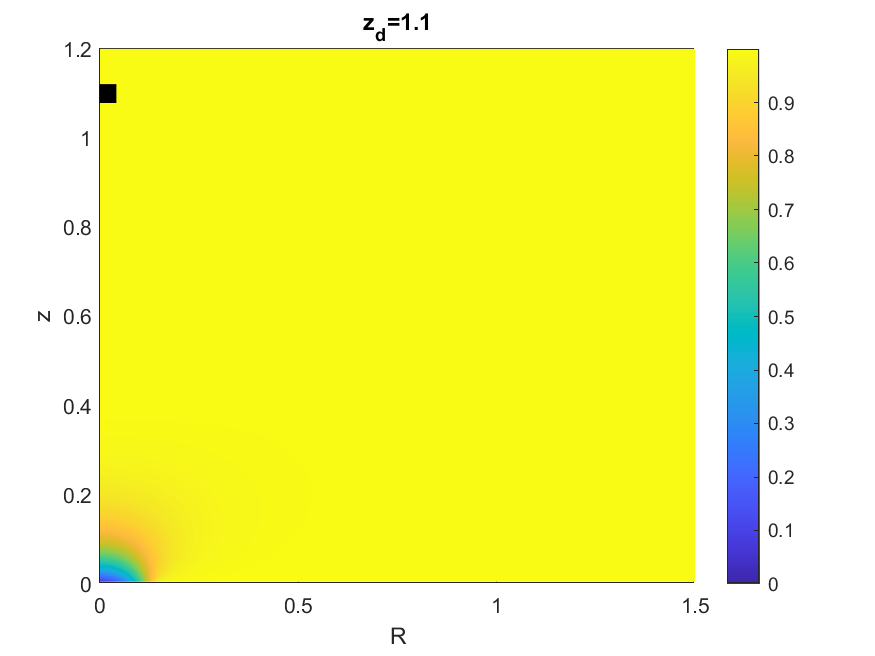}
	 \includegraphics[width=0.32\linewidth]{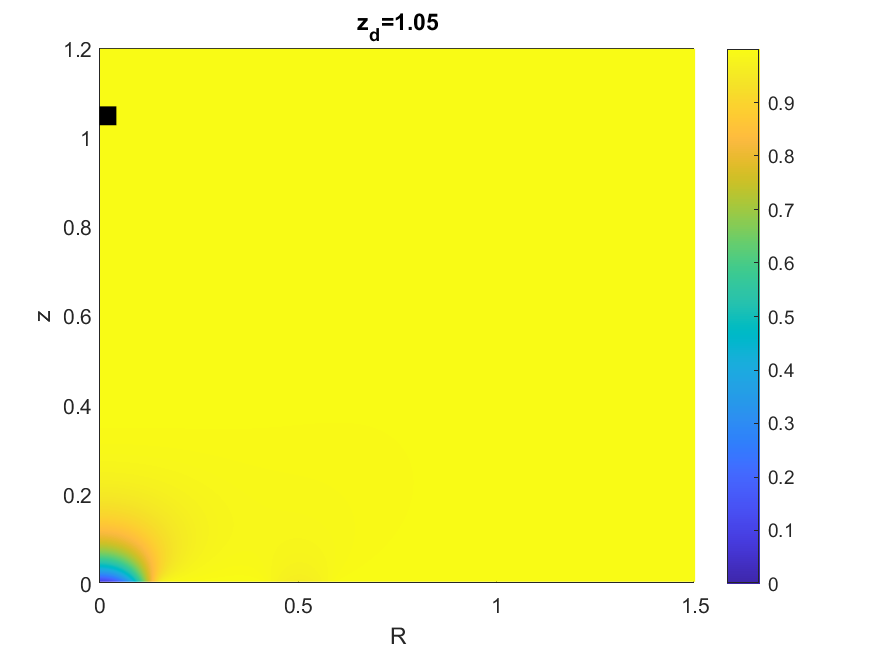}
	 \includegraphics[width=0.32\linewidth]{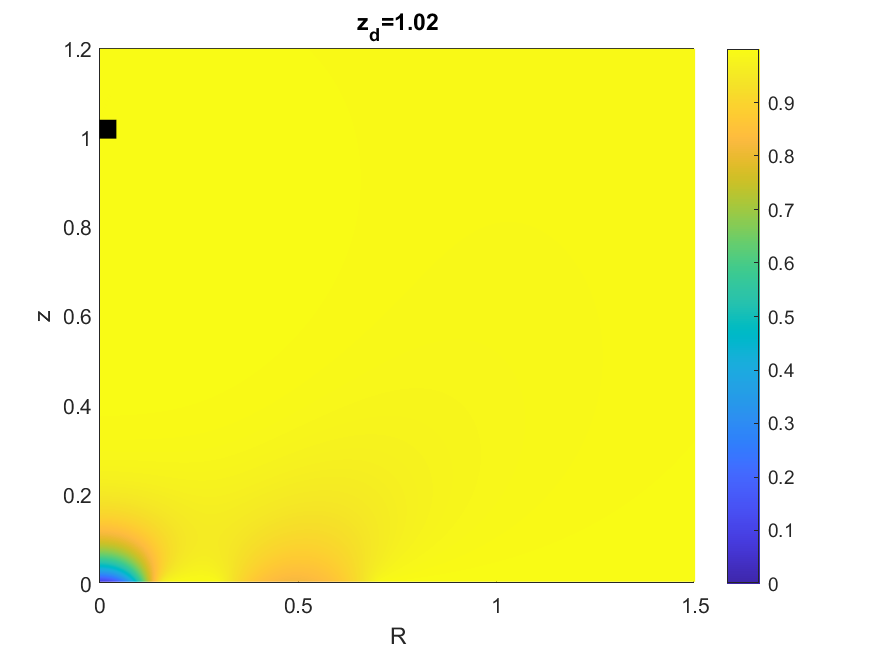}
	 \includegraphics[width=0.32\linewidth]{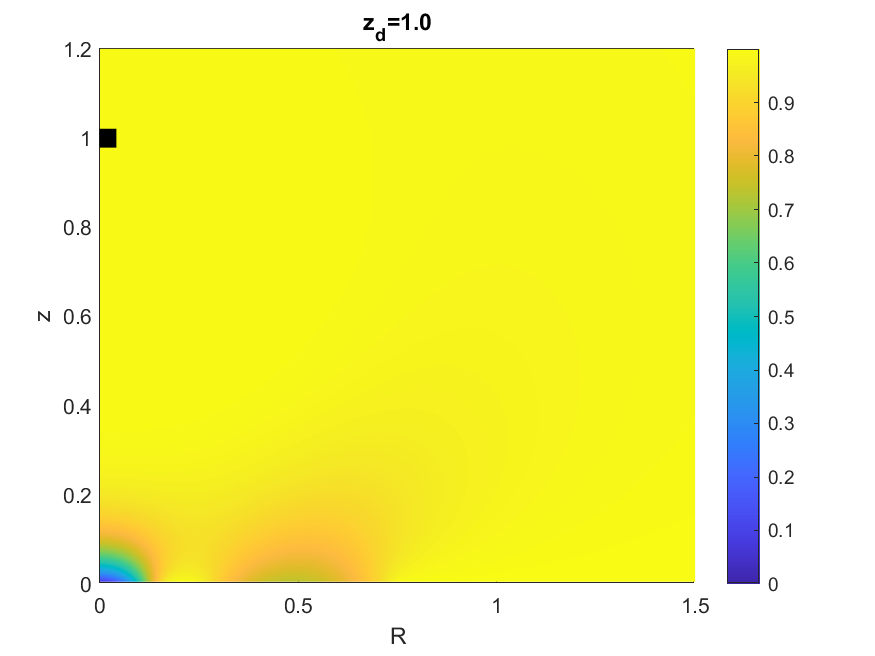}
	 \includegraphics[width=0.32\linewidth]{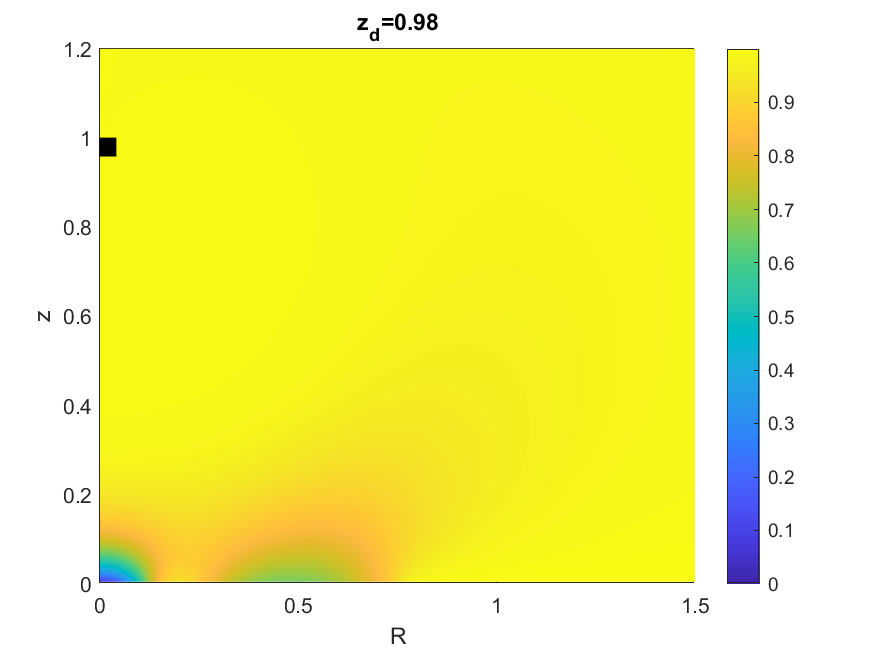}
	 \includegraphics[width=0.32\linewidth]{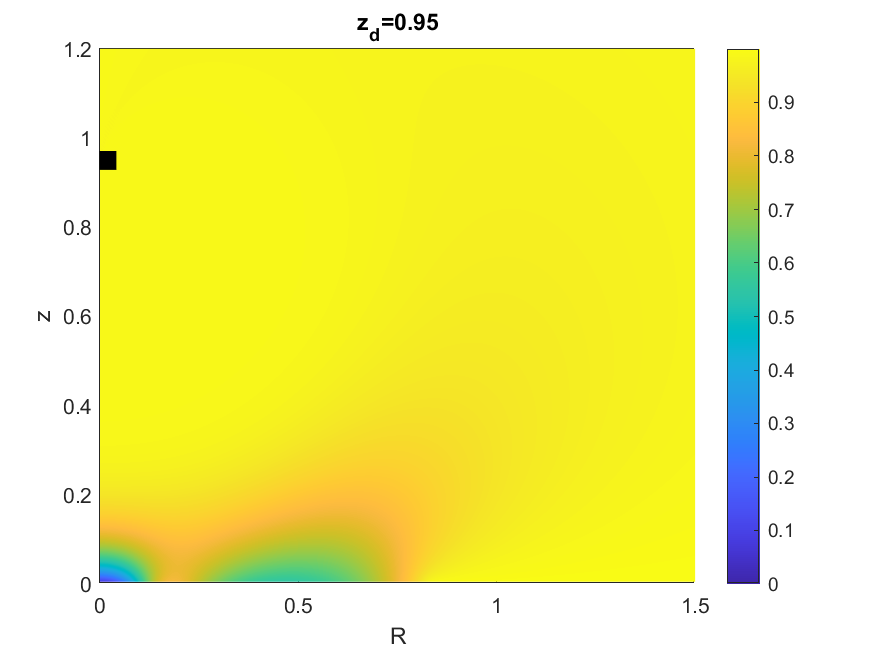}
	 \includegraphics[width=0.32\linewidth]{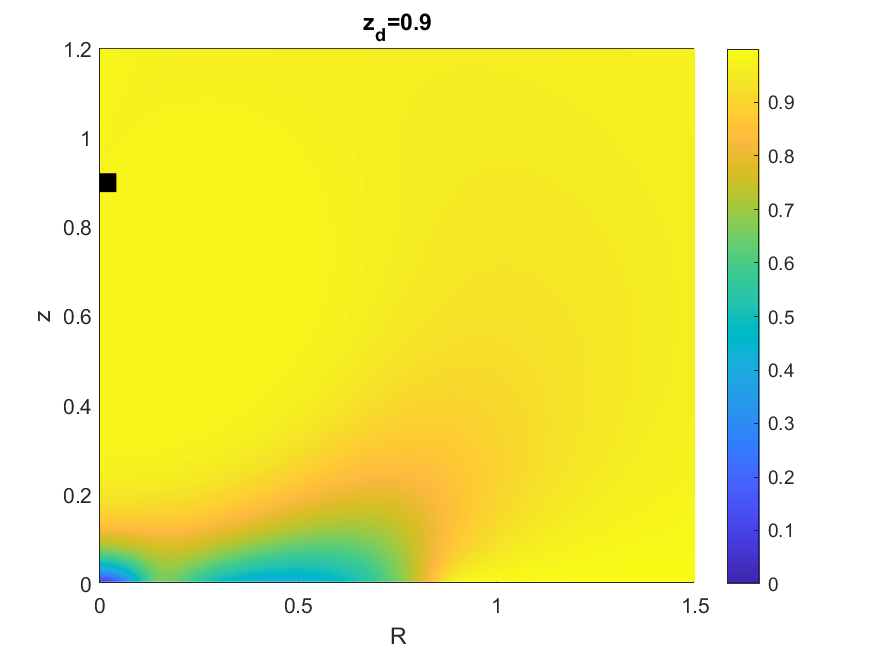}
	 \includegraphics[width=0.32\linewidth]{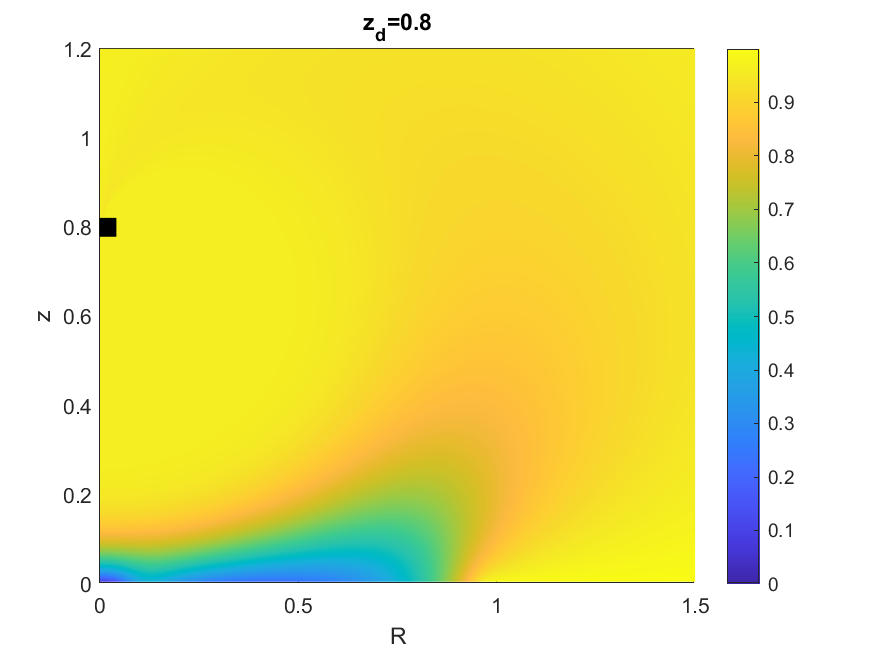}
	 \includegraphics[width=0.32\linewidth]{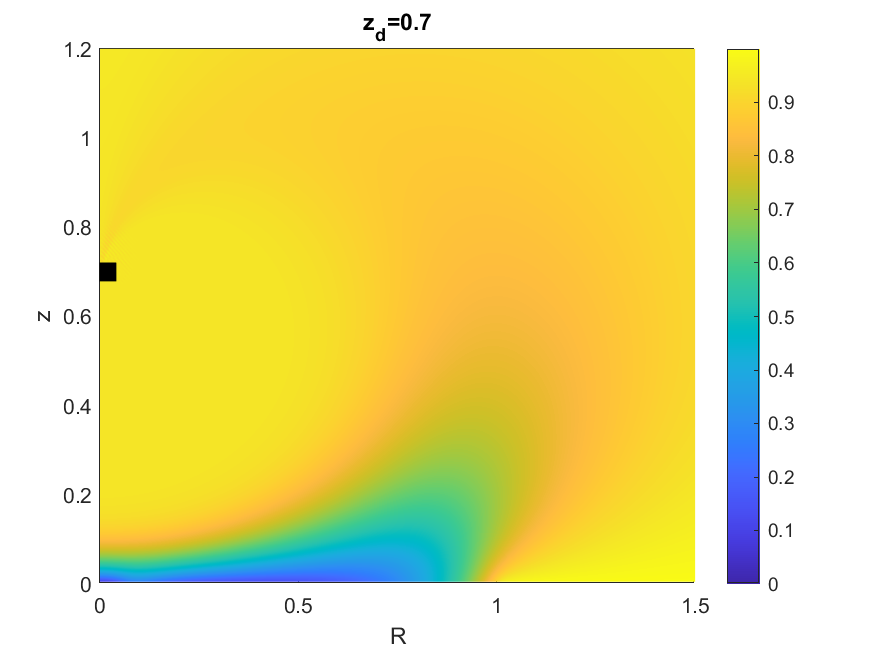}
	 \includegraphics[width=0.32\linewidth]{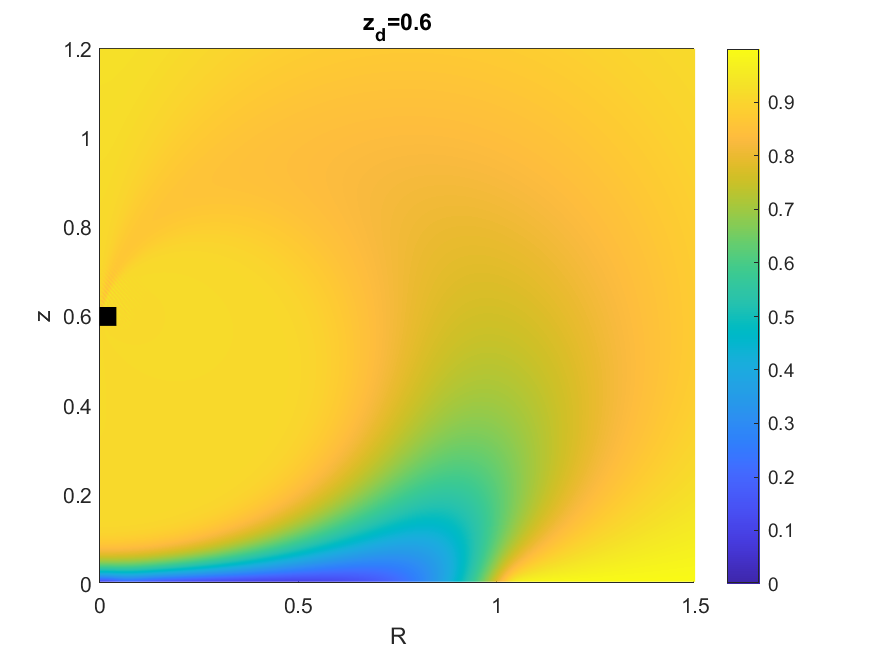}
	 \includegraphics[width=0.32\linewidth]{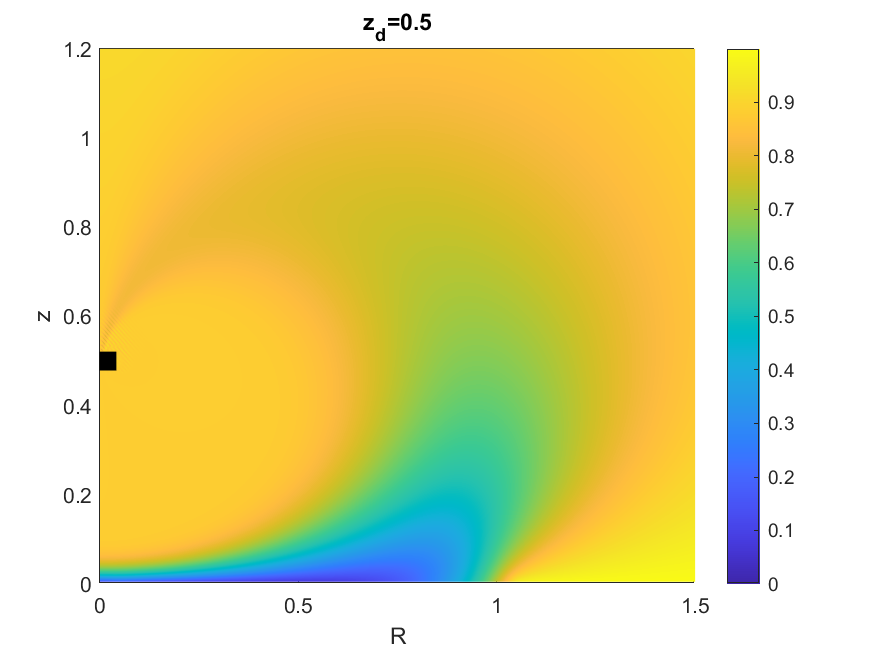}
	 \includegraphics[width=0.32\linewidth]{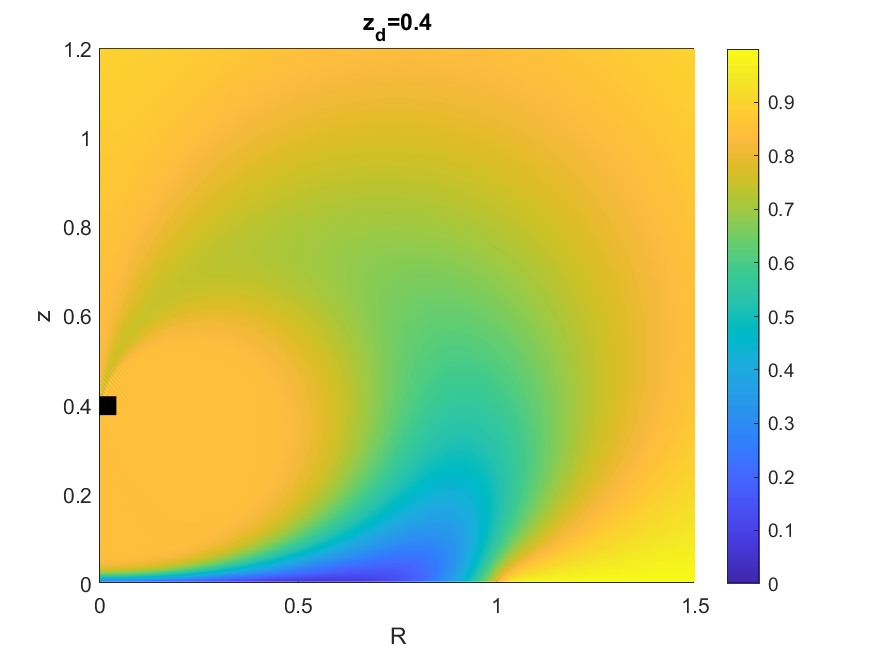}
     \caption{Voidage field $E$, example~\ref{eg5}, $\alpha=0.1$, $z_d\in[0.4,1.1]$.}
   \label{fig:eg5a}
   \end{figure}

    \textnormal{Finally, we repeat the same experiment with $\alpha=0.02$ instead of $\alpha=0.1$.  This smaller value of $\alpha$, compared to all other examples above, represents the case where convection is dominating more than diffusion, and we see in figure~\ref{fig:eg5c} that this leads to a ``wispier'' (thinner) plume, compared to the earlier examples.  Again, we see the formation of a second "plume", for the same reasons as explained above for the case $\alpha=0.1$.}
    \begin{figure}
     \centering
     \includegraphics[width=0.32\linewidth]{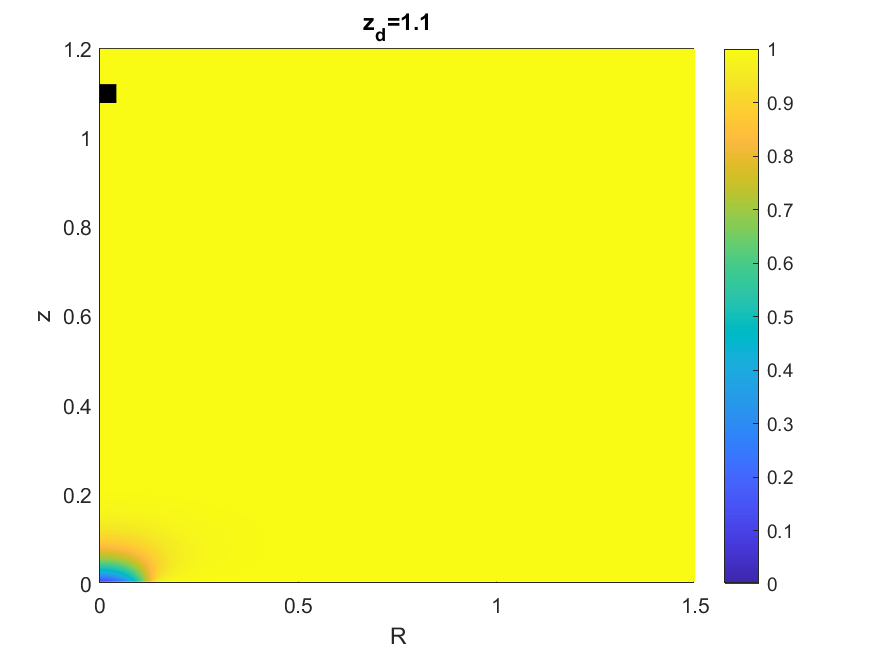}
	 \includegraphics[width=0.32\linewidth]{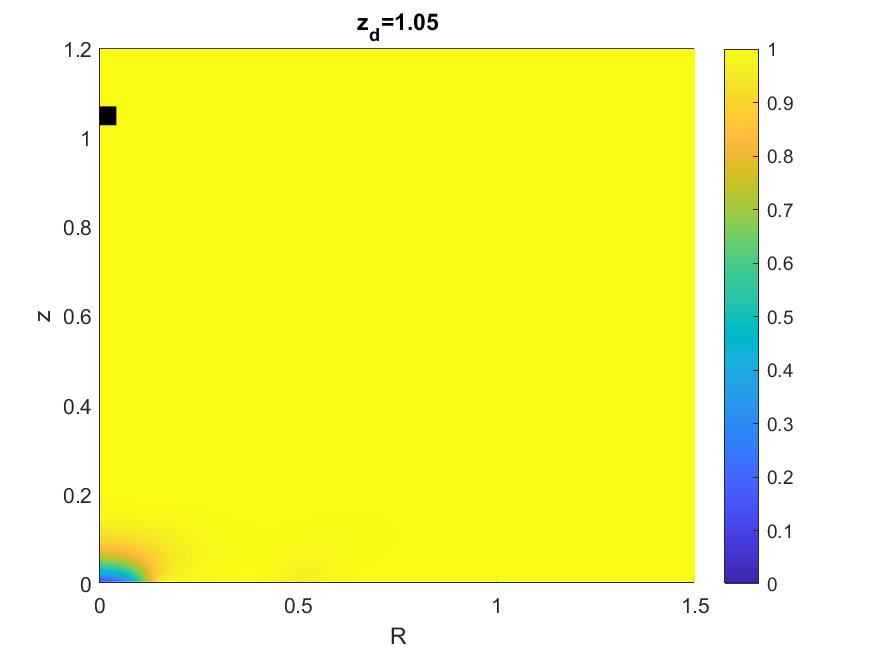}
	 \includegraphics[width=0.32\linewidth]{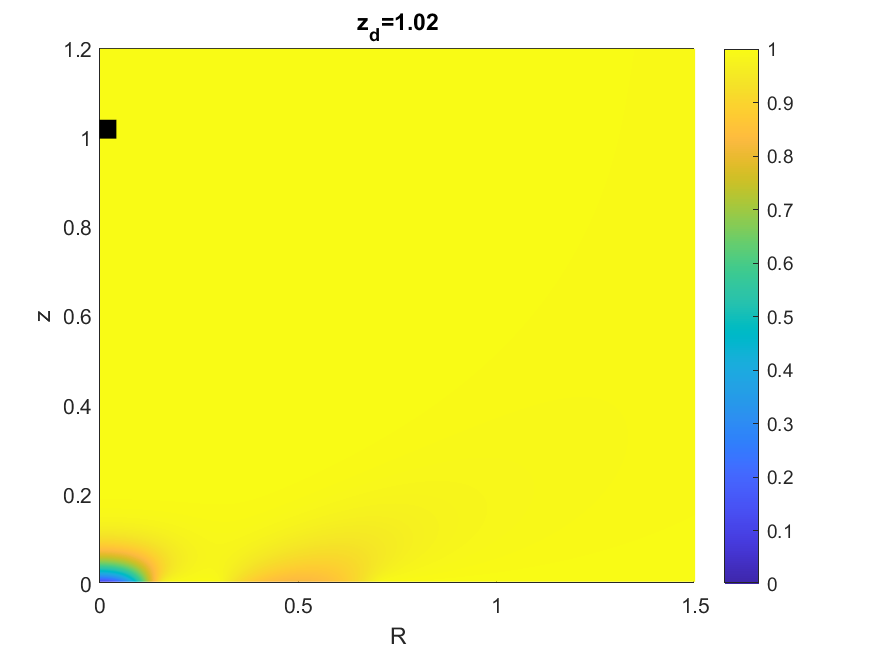}
	 \includegraphics[width=0.32\linewidth]{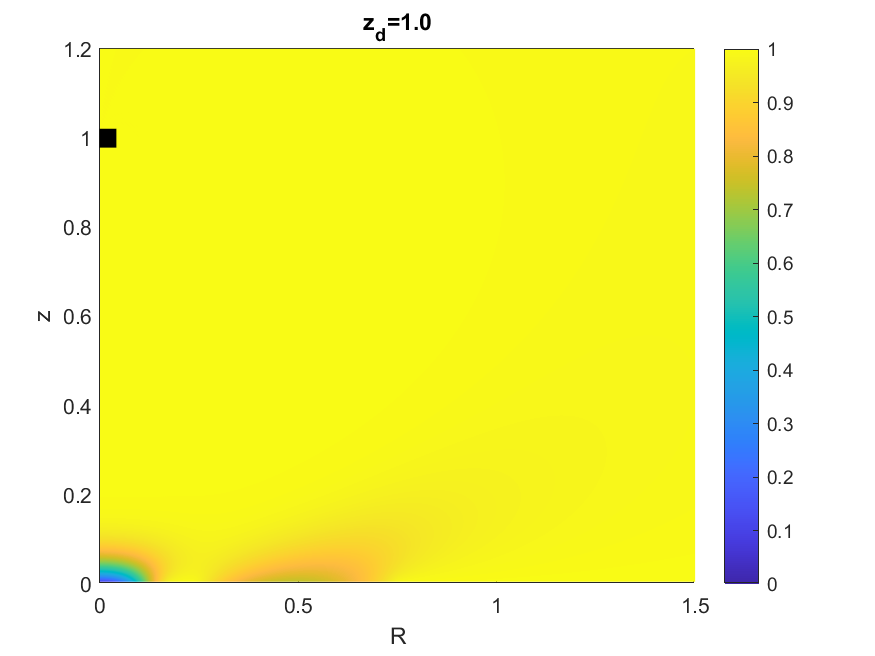}
	 \includegraphics[width=0.32\linewidth]{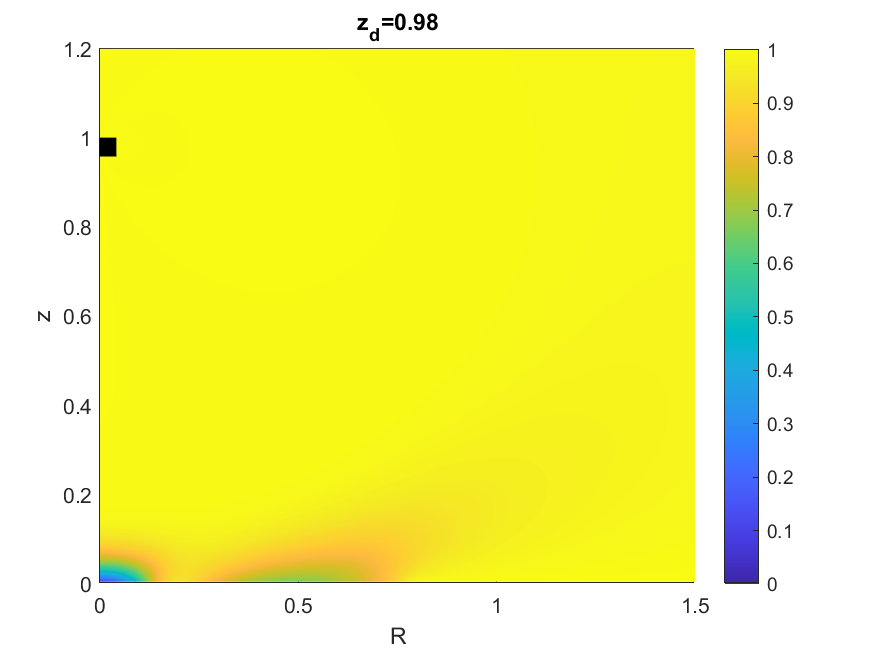}
	 \includegraphics[width=0.32\linewidth]{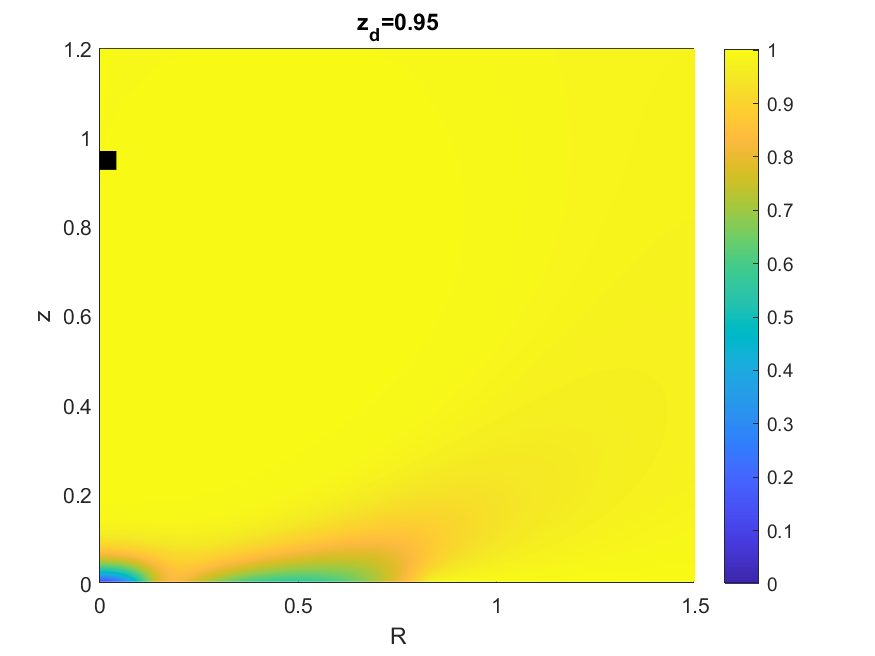}
	 \includegraphics[width=0.32\linewidth]{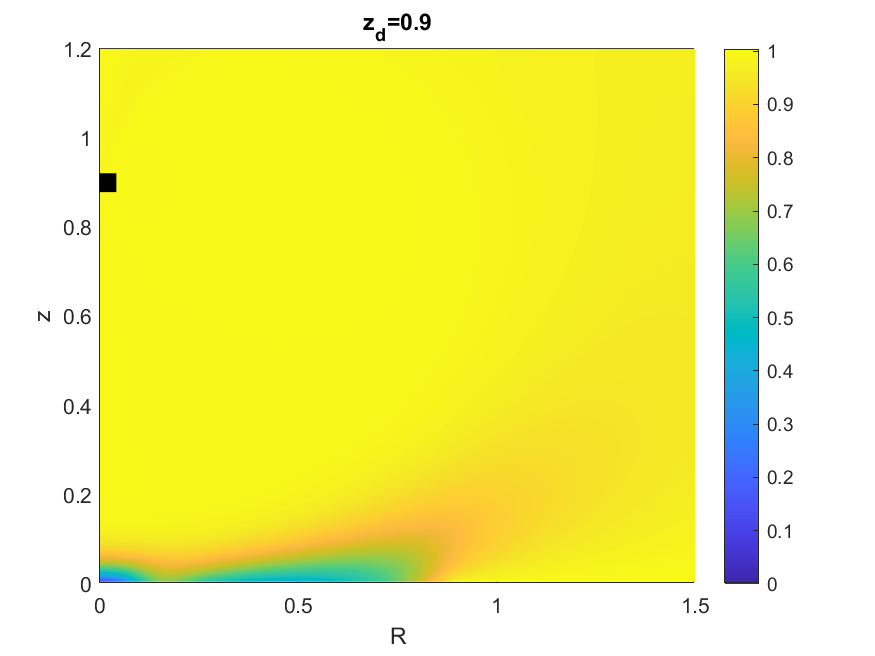}
	 \includegraphics[width=0.32\linewidth]{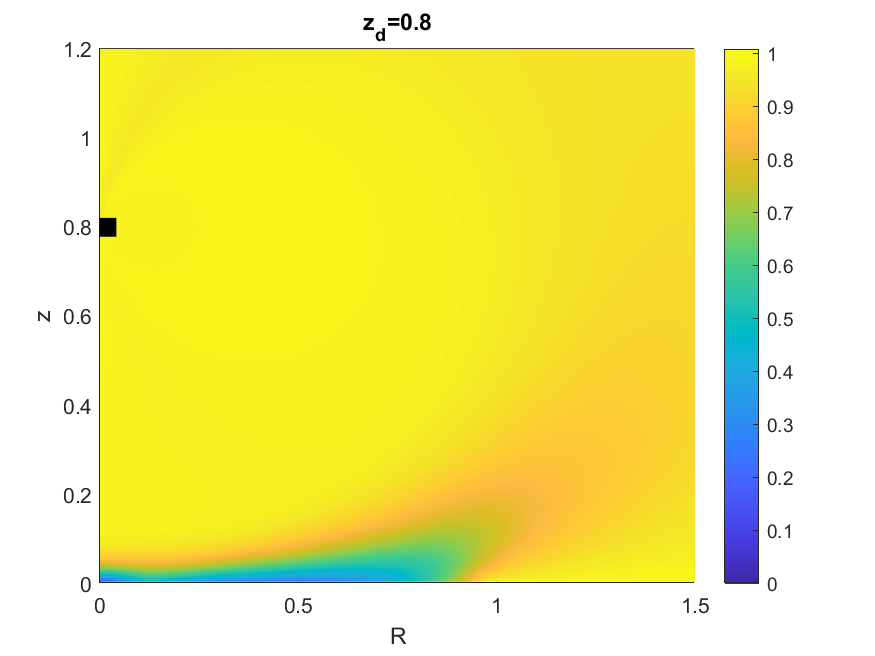}
	 \includegraphics[width=0.32\linewidth]{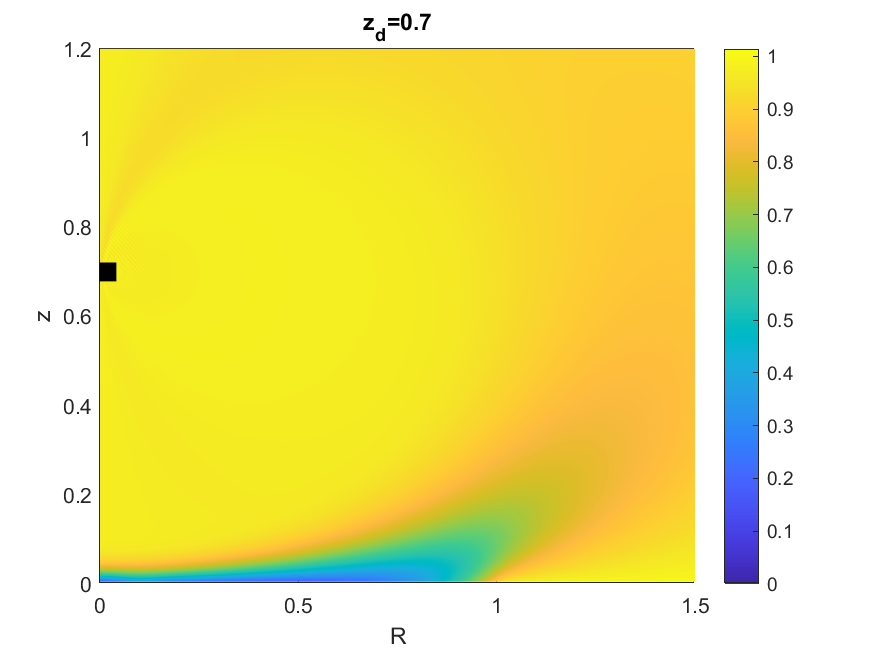}
	 \includegraphics[width=0.32\linewidth]{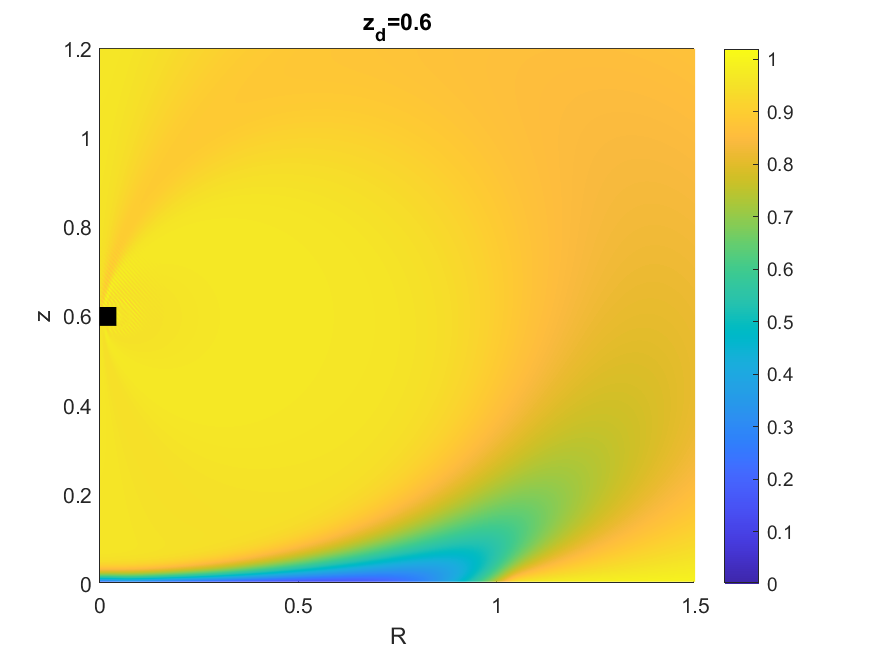}
	 \includegraphics[width=0.32\linewidth]{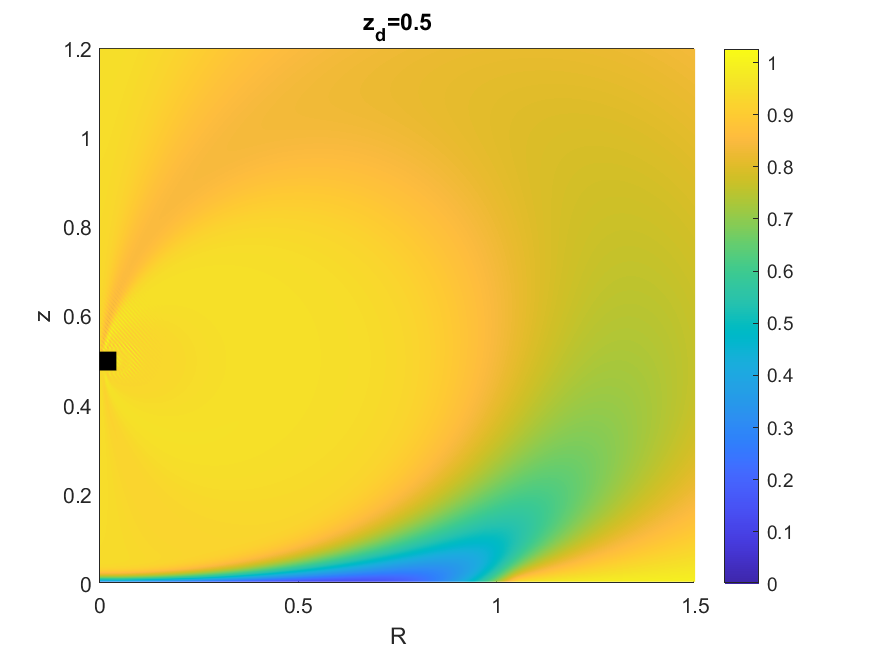}
	 \includegraphics[width=0.32\linewidth]{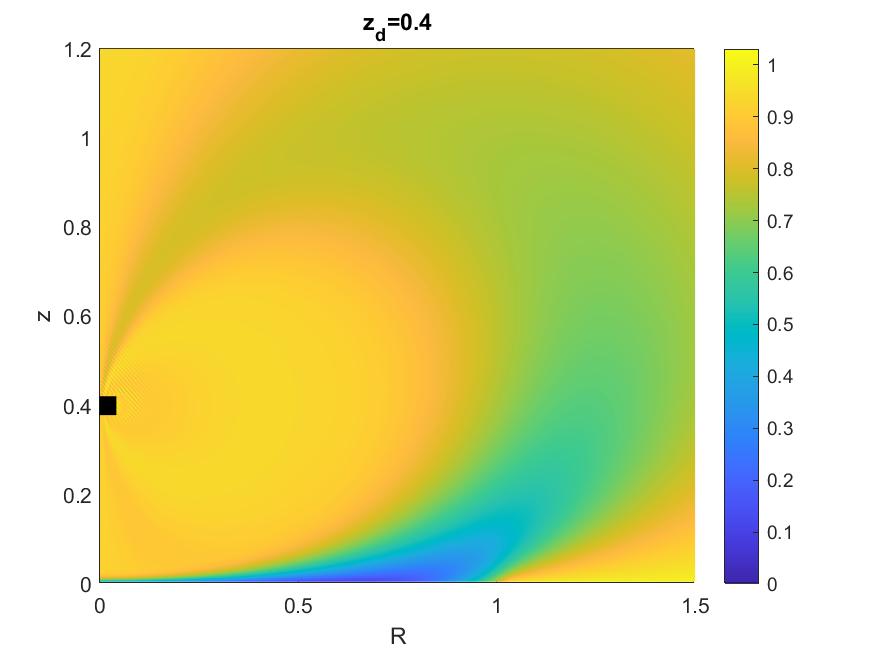}
     \caption{Voidage field $E$, example~\ref{eg5}, $\alpha=0.02$, $z_d\in[0.4,1.1]$.}
   \label{fig:eg5c}
   \end{figure}
\end{example}

\section{Conclusion}
In this paper we have developed a model for the formulation and analysis of problems relating to the flow of an incompressible fluid above an otherwise static particle bed, and the consequent development and evolution of a fluidized particle cloud in the fluid flow. With the flow in the bulk fluidized region modelled as a two-phase flow, the principal contribution of the paper has been to develop a rational theory for the key processes of particle entrainment and detrainment from the otherwise static particle bed into the fully fluidized region above. This leads to a natural additional macroscopic boundary condition at the interface of the fully fluidized region and the particle bed, which renders a closed problem in the fully fluidized region, for the voidage field and a fluid velocity potential.

To illustrate the applicability of this theory we have formulated it in relation to a simple model of the helicopter cloud problem, leading to a fully determined nonlinear elliptic boundary value problem for the voidage field throughout the fluidized region.  This provides a rational and efficient approach to computing the voidage field for this application.  Careful numerical solution via a finite difference approximation scheme has demonstrated through a series of experiments how the qualitative behaviour of the voidage field varies as key parameters change, in a way that is entirely consistent with our understanding of the underlying physical processes.  It may be of interest to look at this example in relation to controlled scale model experiments in order to specifically determine the qualitative and quantitative accuracy of the rational model that has been developed in this paper, but such considerations are left to future work.  However, we do note that qualitative structural agreement between the high speed photography presented in \citet{thesis}, and particularly relating to Figure 2.6 therein, and the theoretical figures presented here in \S7, is very encouraging.

\backsection[Funding]{This research received no specific grant from any funding agency, commercial or not-for-profit sectors.}

\backsection[Declaration of interests]{The authors report no conflict of interest.}

\backsection[Data availability statement]{The authors confirm that the data supporting the findings of this study are available within the article.}

\backsection[Author ORCIDs]{D. J. Needham, https://orcid.org/0000-0002-4958-6976; S. Langdon, https://orcid.org/0000-0002-0572-5137}

\bibliographystyle{jfm}  
\bibliography{refs}        
\begin{appendix}
\section{The problem [BVP] in cylindrical polar coordinates $(R,\theta,z)$}
  On adopting~(\ref{eqn:h33}), the problem [BVP]  becomes, in terms of the cylindrical polar coordinates $(R,\theta,z)$,
\begin{eqnarray}
  & E\left(E_{RR} + \frac{1}{R} E_R + E_{zz} \right) - \bar{a}(R,z) E_R - \bar{b}(R,z) E_z = 0, \quad (R,z)\in \omega,& \label{eqn:h34} \\
  & E=g(R) \mbox{ on } z=0, \quad R\geq 0, & \label{eqn:h35} \\
  & E_R = 0 \mbox{ on } \displaystyle{\left\{ \begin{array}{ll} R=0, & z\in(0,z_d^-) \cup (z_d^+,\infty) \\ R=\Delta, & z\in[z_d^-,z_d^+], \end{array} \right.}& \label{eqn:h36} \\
  & E_z = 0 \mbox{ on } z= z_d^+,z_d^-, \quad R\in(0,\Delta), & \label{eqn:h37} \\
  & E\rightarrow 1 \mbox{ as } (R^2 + z^2)\rightarrow\infty \mbox{ uniformly in } \omega. & \label{eqn:h38}
\end{eqnarray}
Here
\[
  \bar{a}(R,z) = \frac{9}{2\alpha} a(R,z), \quad \bar{b}(R,z) = \frac{9}{2\alpha} b(R,z),
\]
for $(R,z)\in \omega$, whilst
\[
  z_d^+ = z_d + \frac{\Delta}{2}, \quad z_d^- = z_d - \frac{\Delta}{2},
\]
and
\[
  \omega = \{ (R,z): \, z>0, \, R>0 \} \backslash \{ (R,z): \, z_d^- \leq z \leq z_d^+, \, 0 < R\leq \Delta \}.
\]

\section{The numerical method, convergence and error estimates}
To solve the nonlinear boundary value problem [BVP], given by~(\ref{eqn:h34})--(\ref{eqn:h38}), we proceed in an iterative fashion. First, we set
\[ E^{(0)} = 1, \quad\mbox{for all }(R,z)\in\omega. \]
Then, for $n\geq 1$, we choose a tolerance level $\epsilon$, and seek $E^{(n)}$ solving
\begin{eqnarray}
  & E^{(n-1)}\left(E^{(n)}_{RR} + \frac{1}{R} E^{(n)}_R + E^{(n)}_{zz} \right) - \bar{a}(R,z) E^{(n)}_R - \bar{b}(R,z) E^{(n)}_z = 0, \quad (R,z)\in \omega,& \label{eqn:h34it} \\
  & E^{(n)}=g(R) \mbox{ on } z=0, \quad R\geq 0, & \label{eqn:h35it} \\
  & E^{(n)}_R = 0 \mbox{ on } \displaystyle{\left\{ \begin{array}{ll} R=0, & z\in(0,z_d^-) \cup (z_d^+,\infty) \\ R=\Delta, & z\in[z_d^-,z_d^+], \end{array} \right.}& \label{eqn:h36it} \\
  & E^{(n)}_z = 0 \mbox{ on } z= z_d^+,z_d^-, \quad R\in(0,\Delta), & \label{eqn:h37it} \\
  & E^{(n)}\rightarrow 1 \mbox{ as } (R^2 + z^2)\rightarrow\infty \mbox{ uniformly in } \omega. & \label{eqn:h38it}
\end{eqnarray}
We continue until we have
\begin{equation}
  \max_{(R,z)\in \omega}|E^{(n)}-E^{(n-1)}| < \epsilon.
  \label{eqn:tol}
\end{equation}
Given $E^{(n-1)}$, we solve the linear [BVP]~(\ref{eqn:h34it})--(\ref{eqn:h38it}) for $E^{(n)}$ numerically, using a finite difference approximation scheme.

We begin by truncating the unbounded domain $\omega$.  We choose truncation parameters $L_R, L_z\geq 1$, and truncate $\omega$ at $R=R_{\infty}:=(L_R+1)\Delta$ and at $z=z_{\infty}:=z_d+(L_z+1/2)\Delta$, i.e., $R_{\infty}$ is $L_R\Delta$ to the right of $R=\Delta$, and $z_{\infty}$ is $L_z\Delta$ above $z=z_d^+$, so that the truncated domain is
\[ \omega_L := \{ (R,z): \, z\in(0,z_{\infty}), \, R\in(0,R_{\infty}) \} \backslash \{ (R,z): \, z_d^- \leq z \leq z_d^+, \, 0 < R\leq \Delta \}. \]

To discretise the domain $\omega_L$, we choose a discretisation parameter $N\geq1$, and set the mesh size $h:=\Delta/N$.  This allows for a uniform mesh, in both $R$ and $z$ directions, for $z>z_d^-$.  For $z\leq z_d^-$, in order to ensure that our mesh aligns with the boundaries of $\omega_L$, we choose $N_z\in\Z$ to be the smallest integer such that
\[ N_z \geq N\left(\frac{z_d}{\Delta}-\frac{1}{2}\right), \]
and set the mesh size $h_z := z_d^-/N_z \leq h$, with equality (i.e., $h_z=h$, and hence a uniform mesh throughout $\omega_L$) if $N(z_d/\Delta-1/2)\in\Z$, since in that case
\[ h_z = \frac{z_d^-}{N(z_d/\Delta-1/2)} = \frac{z_d^-h}{\Delta(z_d/\Delta-1/2)} = h. \]

The mesh we use for our finite difference scheme then consists of the points $(x_i,y_j)\cap\bar{\omega_L}$, where
\[ x_i := ih, \quad i=0, \ldots, N(L_R+1)-1, \]
and
\[ y_j := \left\{ \begin{array}{ll} jh_z, & j=1,\ldots,N_z, \\ z_d^- + (j-N_z)h, & j=N_z+1, \ldots, N_z+N(L_z+1)-1. \end{array} \right. \]
We approximate the solution $E^{(n)}$ of~(\ref{eqn:h34it})--(\ref{eqn:h38it}) at the mesh points $(x_i,y_j)\cap\bar{\omega_L}$ by $E_N$, where
\[ E^{i,j} := E_N(x_i,y_j) \approx E^{(n)}(x_i,y_j). \]
The mesh points $(x_i,y_j)$, $i=0,\ldots,N-1$, $j=N_z+1,\ldots,N_z+N$, are excluded from our computational domain, as they lie outside $\omega_L$, and hence
the total number of degrees of freedom in our numerical solution is given by
\[ \mbox{DOF}:=N(N_z(L_R+1) + (N-1)L_R + NL_z(L_R+1)). \]

At the boundary $z=0$, recalling~(\ref{eqn:h35it}) we set
\[ E^{i,0} := g(x_i), \quad i=0, \ldots, N(L_R+1)-1. \]
At the boundary $R=R_{\infty}$, recalling~(\ref{eqn:h38it}), we set
\[ E^{N(L_R+1),j} = 1, \quad j=1,\ldots, N_z+N(L_z+1)-1, \]
and at the boundary $z=z_{\infty}$, again recalling~(\ref{eqn:h38it}), we set
\[ E^{i,N_z+N(L_z+1)} = 1, \quad i=0, \ldots, N(L_R+1)-1. \]

We next approximate the $R$-derivatives in~(\ref{eqn:h34it}) and the boundary condition~(\ref{eqn:h36it}).  Away from the boundaries of $\omega_L$, we approximate the second order derivative in the $R$-direction to \textit{O}$(h^2)$ accuracy by
\begin{equation}
  E^{(n)}_{RR}(x_i,y_j) \approx \frac{E^{i+1,j}-2E^{i,j}+E^{i-1,j}}{h^2},
  \label{eqn:ERR}
\end{equation}
and we approximate the first order derivative in the $R$-direction to \textit{O}$(h^2)$ accuracy by
\begin{equation}
  E^{(n)}_R(x_i,y_j) \approx \frac{E^{i+1,j}-E^{i-1,j}}{2h}.
  \label{eqn:ER}
\end{equation}
On the left boundary of $\omega_L$, i.e.\ at the points
\begin{align*}
  (x_0,y_j), & \quad j = 1,\ldots,N_z, \\
  (x_N,y_j), & \quad j = N_z+1,\ldots,N_z+N, \\
  (x_0,y_j), & \quad j = N_z+N+1, \ldots, N_z+N(L_z+1)-1,
\end{align*}
the boundary condition~(\ref{eqn:h36it}) implies, using again the approximation~(\ref{eqn:ER}),
\begin{align*}
  E^{-1,j} = E^{1,j}, & \quad j = 1,\ldots,N_z, \\
  E^{N-1,j} = E^{N+1,j}, & \quad j = N_z+1,\ldots,N_z+N, \\
  E^{-1,j} = E^{1,j}, & \quad j = N_z+N+1, \ldots, N_z+N(L_z+1)-1.
\end{align*}
These values are inserted into the formula~(\ref{eqn:ERR}) near the boundaries of $\omega_L$, as required.

Finally, we approximate the $z$-derivatives in~(\ref{eqn:h34it}) and the boundary condition~(\ref{eqn:h37it}).  We define
\[ h_{z,j} := y_{j+1}-y_j= \left\{ \begin{array}{ll} h_z & j=1,\ldots,N_z, \\ h & j = N_z+1, \ldots, N_z+N(L_z+1)-1, \end{array}\right. \]
and then, away from the boundaries of $\omega_L$, and for $y_j\neq z_d^-$, we approximate the second order derivative in the $z$-direction to \textit{O}$(h^2)$ accuracy (recalling that $h_z\leq h$) by
\begin{eqnarray}
  E^{(n)}_{zz}(x_i,y_j) & \approx & \frac{E^{i+1,j}-2E^{i,j}+E^{i-1,j}}{h_{z,j}^2}, \quad i=0, \ldots, N(L_R+1)-1, \nonumber \\
  & & \qquad j=1, \ldots, N_z-1, N_z+1, \ldots, N_z+N(L_z+1)-1.
  \label{eqn:Ezz}
\end{eqnarray}
For $y_j = z_d^-$, i.e.\ for $j=N_z$, we need a different formula, to account for the fact that the mesh spacing above and below $y_{N_z}$ are not equal.  In this case, we  approximate the second order derivative in the $z$-direction to \textit{O}$(h^2)$ accuracy (this can be shown through simple Taylor Series expansions) by
\begin{eqnarray}
  E^{(n)}_{zz}(x_i,y_{N_z}) & \approx & \left(\frac{h_z}{h^2(h+h_z)}\right)(E^{i,N_z+2}-2E^{i,N_z+1}) + \frac{1}{h+h_z}\left(\frac{h_z}{h^2}+\frac{h}{h_z^2}\right)E^{i,N_z} \nonumber \\
  & & \qquad + \left(\frac{h}{h_z^2(h+h_z)}\right)(-2E^{i,N_z-1}+E^{i,N_z-2}), \quad i=0, \ldots, N(L_R+1)-1. \nonumber \\
  & & \label{eqn:Ezz_Nz}
\end{eqnarray}
Similarly, we approximate the first order derivative in the $z$-direction to \textit{O}$(h^2)$ accuracy by
\begin{equation}
  E^{(n)}_z(x_i,y_j) \!\approx\! \left(\frac{h_{z,j-1}}{h_{z,j}(h_{z,j}\!+\!h_{z,j-1})}\right)E^{i,j+1} - \left(\frac{h_{z,j-1}\!-\!h_{z,j}}{h_{z,j-1}h_{z,j}}\right)E^{i,j}
    - \left(\frac{h_{z,j}}{h_{z,j-1}(h_{z,j}\!+\!h_{z,j-1})}\right)E^{i,j-1}.
  \label{eqn:Ez}
\end{equation}
In the case that $j\neq N_z$, in which case $y_j\neq z_d^-$, we have $h_{z,j}=h_{z,j-1}$ in~(\ref{eqn:Ez}), which then becomes
\[
  E^{(n)}_z(x_i,y_j) \approx \left\{ \begin{array}{ll} \displaystyle{\frac{E^{i,j+1}-E^{i,j-1}}{2h_z}}, & j=1, \ldots, N_z-1, \\
  \displaystyle{\frac{E^{i,j+1}-E^{i,j-1}}{2h}}, & j=N_z+1, \ldots, N_z+N(L_z+1)-1. \end{array} \right.
\]
The formula~(\ref{eqn:Ez}) is applied on the upper and lower boundaries of $\omega_L$ to approximate the boundary condition~(\ref{eqn:h37it}), in an identical fashion to the approximation of~(\ref{eqn:h36it}) described above, with the values attained inserted into the formulae~(\ref{eqn:Ezz}) and~(\ref{eqn:Ezz_Nz}) as required.

If we choose $L_R,L_z\propto N$, then, recalling~(\ref{eqn:h27}), we truncate at $R_{\infty},z_{\infty}=\mbox{\textit{O}}(h^{-1})$, and in~(\ref{eqn:tol}) take $\epsilon=\mbox{\textit{O}}(h^{2})$.  With mesh size $h$, and each step of our finite difference approximation accurate to \textit{O}$(h^2)$, we thus anticipate that as we decrease $h$ a reasonable expectation would be that the overall error in our numerical scheme is \textit{O}$(h^{2})$.  We test this in the following numerical example.
\begin{example}[Numerical convergence]
   \label{eg1}
   \textnormal{In this example, we choose the dimensionless parameters as follows:
   \[  E_s = 0.01,\quad \alpha = 0.1,\quad \gamma= 500,\quad \Omega = 0.5,\quad \Delta = 0.04,\]
   and we consider the three cases $z_d = 0.5$, $z_d=0.9$, and $z_d=1.3$ (note that for these values $(z_d/\Delta-1/2)\in\Z$ and hence $h_z=h$).
   We fix $L_R=12N$ and $L_z=6N$, take $\epsilon=1/(10N^2)$, and then successively increase $N=2,4,8,16$, to see if our numerical scheme is converging, taking the solution computed with $N=16$ as the reference solution for the purpose of computing errors.  We denote the number of iterations required before~(\ref{eqn:tol}) is satisfied by ``iterations'', and we calculate the expected order of convergence (EOC) as
   \[ \mbox{EOC} = \log_2\left|\frac{\|(E_N-E_{16})/E_{16}\|_{\infty}}{\|(E_{2N}-E_{16})/E_{16}\|_{\infty}}\right|, \]
   where, to enable comparison, we compute the relative errors for each value of $N$ at the mesh points from the example with $N=2$, noting that the size of the computational domain grows as $N$ increases.  Under the hypothesis that the overall error in our numerical scheme is \textit{O}$(h^{2})$, we might expect to see EOC$\sim 2$, as $N$ increases.  Results are shown in table~\ref{table:eg1a}, with the last column having values only where EOC is well-defined.}
   \begin{table}
     \begin{center}
       \begin{tabular}{crcrrcc}
         $z_d$ & $N$ & $h$ & DOF & iterations & $\|(E_N-E_{16})/E_{16}\|_{\infty}$ & EOC \\[3pt]
         0.5 &   2 & 2.0$\times10^{-2}$ &      2448 &  4 & 4.6$\times10^{+0}$ & 3.9 \\
             &   4 & 1.0$\times10^{-2}$ &     28800 &  6 & 3.1$\times10^{-1}$ & 1.0 \\
             &   8 & 5.0$\times10^{-3}$ &    377856 &  9 & 1.5$\times10^{-1}$ & \\
             &  16 & 2.5$\times10^{-3}$ &   5382144 & 11 & & \\[3pt]
         0.9 &   2 & 2.0$\times10^{-2}$ &      3448 &  4 & 2.9$\times10^{+0}$ & 4.1 \\
             &   4 & 1.0$\times10^{-2}$ &     36640 &  6 & 1.7$\times10^{-1}$ & 1.1 \\
             &   8 & 5.0$\times10^{-3}$ &    439936 &  8 & 7.7$\times10^{-2}$ &  \\
             &  16 & 2.5$\times10^{-3}$ &   5876224 &  9 & & \\[3pt]
         1.3 &   2 & 2.0$\times10^{-2}$ &      4448 &  4 & 1.9$\times10^{+0}$ & 4.3 \\
             &   4 & 1.0$\times10^{-2}$ &     44480 &  5 & 9.6$\times10^{-2}$ & 1.2 \\
             &   8 & 5.0$\times10^{-3}$ &    502016 &  7 & 4.1$\times10^{-2}$ &  \\
             &  16 & 2.5$\times10^{-3}$ &   6370304 &  8 & & \\
       \end{tabular}
       \caption{Convergence of our numerical approximation scheme as $N$ increases, example~\ref{eg1}.}
       \label{table:eg1a}
     \end{center}
   \end{table}

   \textnormal{As $N$ increases, we see the error decrease, but the rate of convergence is somewhat erratic.  To see more clearly how the solution converges, in figure~\ref{fig:eg1c} we plot the solution for $z_d=0.9$, for $N=2$, $N=4$, $N=8$ and $N=16$.  These are plotted on the computational domain, which grows with increasing $N$.  To ease comparison, we additionally plot (on the same figure) the $N=8$ and $N=16$ solutions over the same range as the $N=4$ solution.  In figure~\ref{fig:eg1d} we plot the relative errors $|(E_N-E_{16})/E_{16}|$, each over the same range as the $N=2$ solution.}
   \begin{figure}
     \centering
     \includegraphics[width=0.45\linewidth]{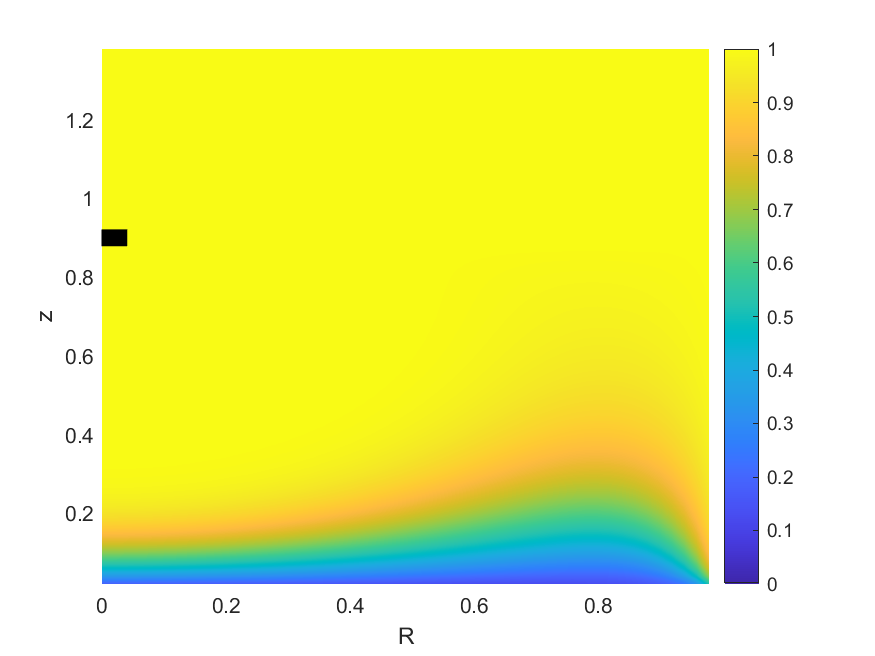}
	 \includegraphics[width=0.45\linewidth]{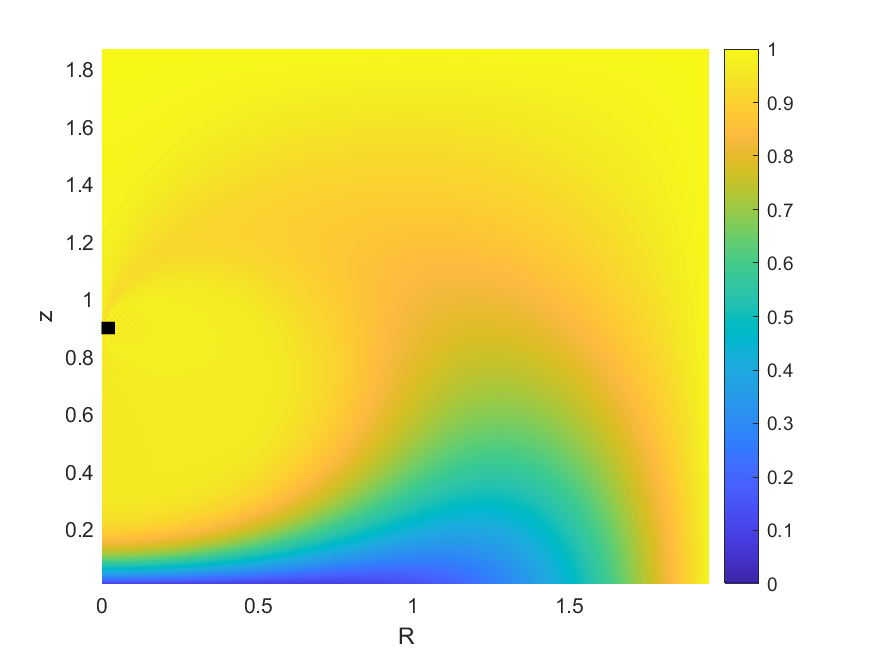}
	 \includegraphics[width=0.45\linewidth]{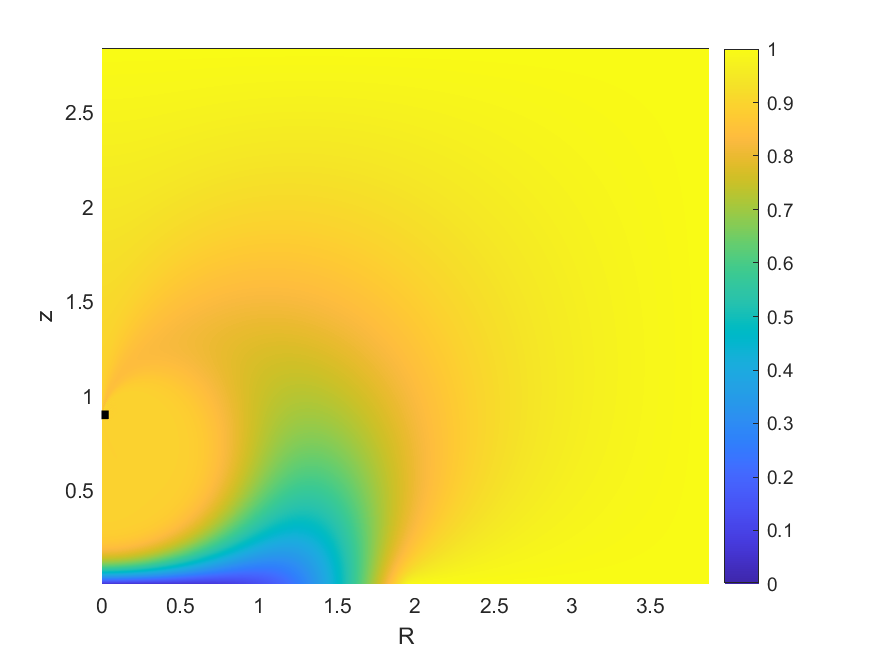}
	 \includegraphics[width=0.45\linewidth]{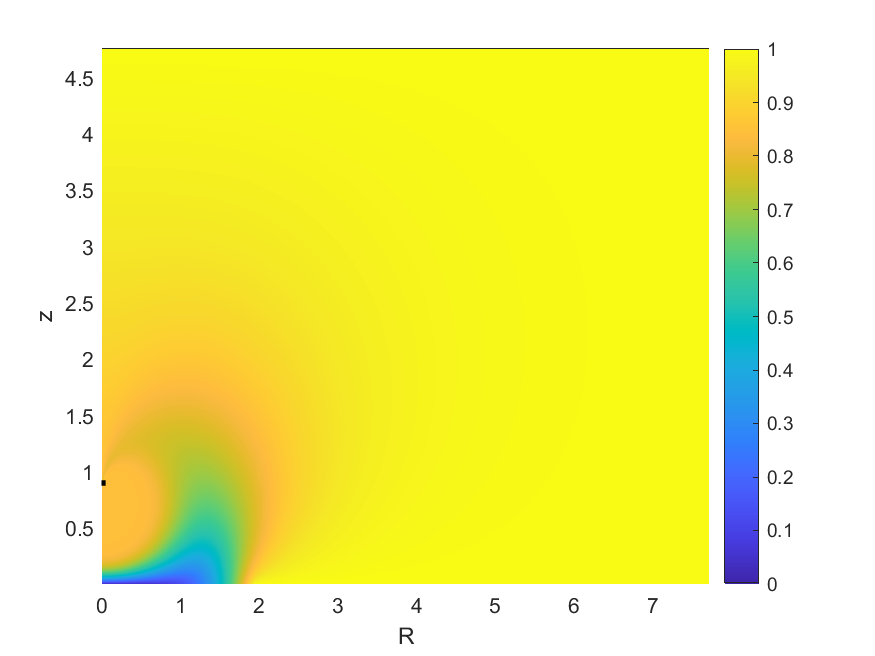}
	 \includegraphics[width=0.45\linewidth]{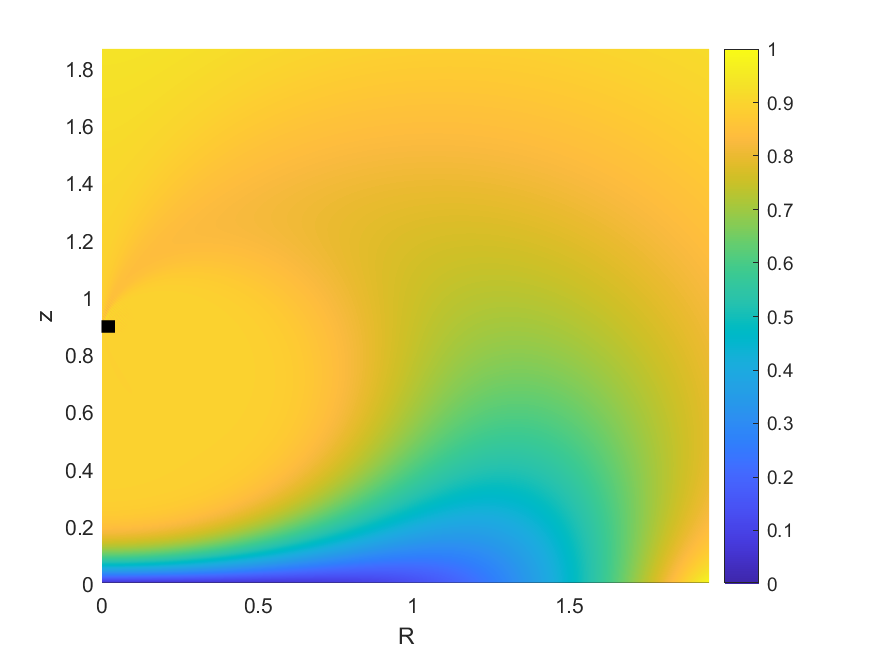}
	 \includegraphics[width=0.45\linewidth]{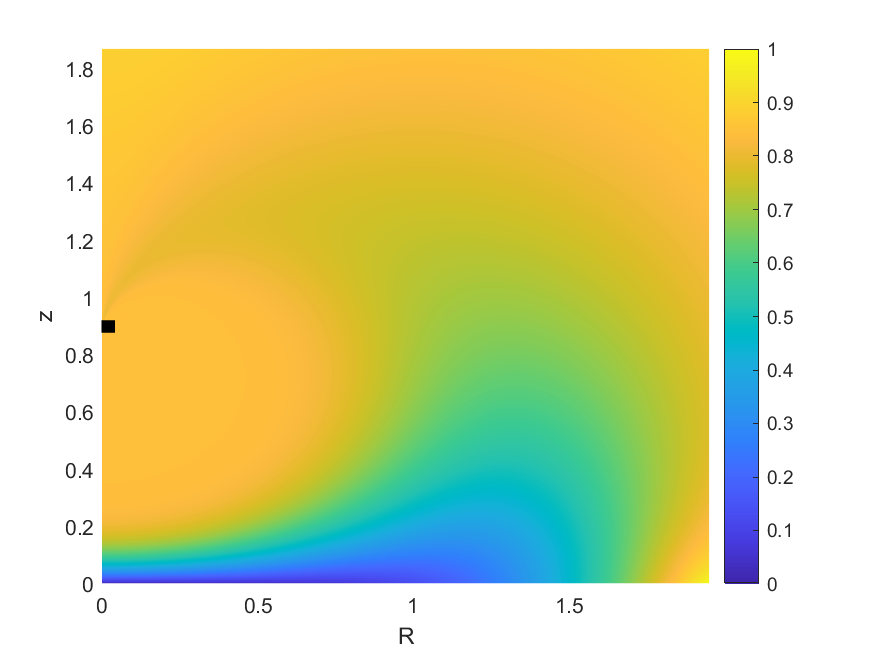}
     \caption{Numerical solution $E_N$, $z_d=0.9$, example~\ref{eg1}, computed for $N=2$ (top left), $N=4$ (top right), $N=8$ (middle left) and $N=16$ (middle right), each shown on the full computational domain.  The $N=8$ (bottom left) and $N=16$ (bottom right) solutions are also plotted over the same range as for the $N=4$ solution, for easier comparison.}
   \label{fig:eg1c}
   \end{figure}
   \begin{figure}
     \centering
     \includegraphics[width=0.45\linewidth]{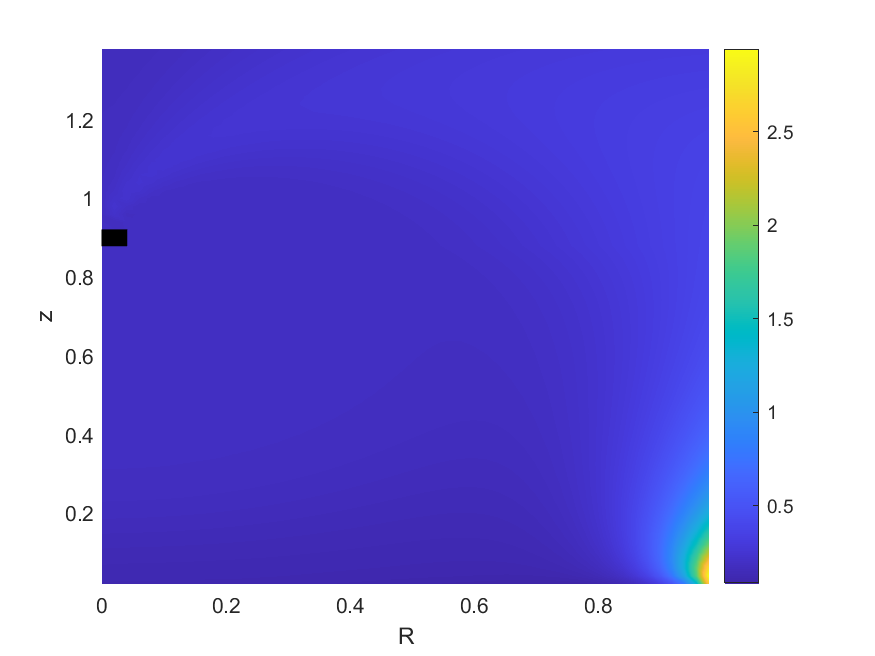}
	 \includegraphics[width=0.45\linewidth]{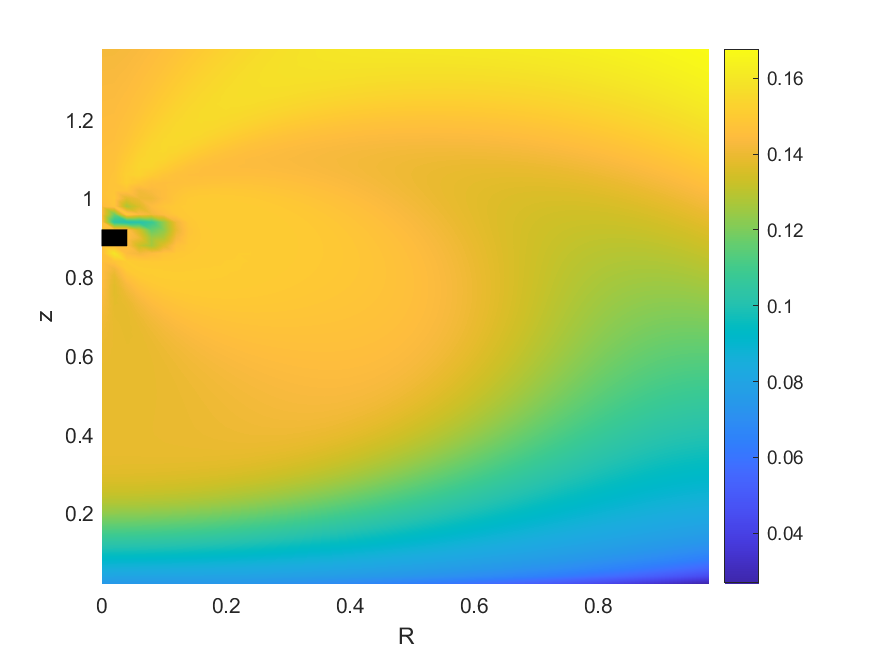}
	 \includegraphics[width=0.45\linewidth]{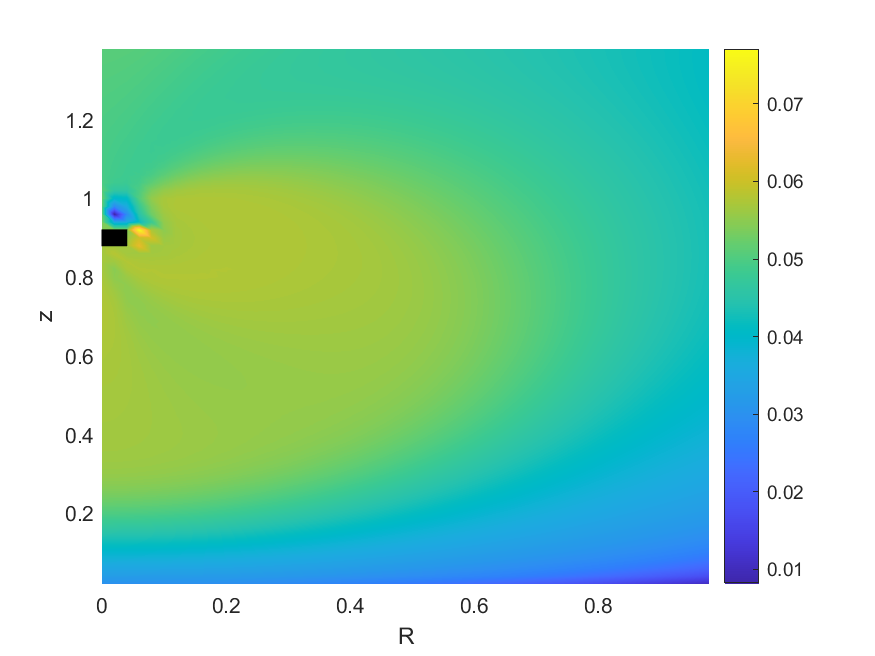}
     \caption{Relative error $|(E_N-E_{16})/E_{16}|$, for $z_d=0.9$, example~\ref{eg1}, computed for $N=2$ (top left), $N=4$ (top right), $N=8$ (bottom).  Note the different scales on the color bar for each plot.}
   \label{fig:eg1d}
   \end{figure}

   \textnormal{For each value of $z_d$, we see broadly comparable results.  In each case, for $N=2$ the maximum relative error (as calculated in table~\ref{table:eg1a}) is where $R$ is largest and $z$ is smallest (lower right corner of plot) - this is because the domain is not large enough in this case to adequately capture the complete boundary data.  As we increase $N$, the maximum relative error moves closer to the location of the helicopter rotor.  The norm we have used in table~\ref{table:eg1a} to measure error, namely the maximum pointwise relative error, solely measures the relative error at the worst point in the domain.  To present a more balanced picture of the effectiveness of our method, illustrating numerically what we have seen qualitatively in figure~\ref{fig:eg1d}, in table~\ref{table:eg1b} we focus on three specific points in the domain, to see how our numerical solution converges at those points.  The values of $h$, DOF and iterations are the same in table~\ref{table:eg1b} and~\ref{table:eg1a}.}
   \begin{table}
     \begin{center}
       \begin{tabular}{cccccccc}
         $z_d$ & $N$ & $\mbox{error}_{1}$ & EOC & $\mbox{error}_{2}$ & EOC & $\mbox{error}_{3}$ & EOC\\[3pt]
         0.5 &   2 & 6.1$\times10^{-1}$ & 1.2 & 4.1$\times10^{-1}$ & 0.7 & 2.9$\times10^{-1}$ & 0.9 \\
             &   4 & 2.7$\times10^{-1}$ & 1.8 & 2.6$\times10^{-1}$ & 1.5 & 1.6$\times10^{-1}$ & 1.4 \\
             &   8 & 7.9$\times10^{-2}$ &     & 9.3$\times10^{-2}$ &     & 5.8$\times10^{-2}$ &     \\[3pt]
         0.9 &   2 & 1.8$\times10^{-1}$ & 0.3 & 1.8$\times10^{-1}$ & 0.4 & 2.4$\times10^{-1}$ & 1.4 \\
             &   4 & 1.5$\times10^{-1}$ & 1.4 & 1.4$\times10^{-1}$ & 1.4 & 8.8$\times10^{-2}$ & 1.3 \\
             &   8 & 5.8$\times10^{-2}$ &     & 5.4$\times10^{-2}$ &     & 3.5$\times10^{-2}$ &     \\[3pt]
         1.3 &   2 & 9.2$\times10^{-2}$ & 0.2 & 9.3$\times10^{-2}$ & 0.5 & 3.3$\times10^{-1}$ & 2.9  \\
             &   4 & 7.7$\times10^{-2}$ & 1.2 & 6.6$\times10^{-2}$ & 1.2 & 4.4$\times10^{-2}$ & 1.3  \\
             &   8 & 3.3$\times10^{-2}$ &     & 2.8$\times10^{-2}$ &     & 1.8$\times10^{-2}$ &      \\
       \end{tabular}
       \caption{Convergence of our numerical approximation scheme at three fixed points as $N$ increases, example~\ref{eg1}.  Here, $\mbox{error}_{m} = |(E_N(\bx_m)-E_{16}(\bx_m))/E_{16}(\bx_m)|$, where $\bx_1=(0.2,0.8)$, $\bx_2=(0.5,0.5)$, $\bx_3=(0.8,0.2)$.}
       \label{table:eg1b}
     \end{center}
   \end{table}

   \textnormal{Though the convergence rate varies somewhat across these examples, we typically see a relative error of ~2--10\% for $N=8$, with the error (for fixed $N$) decreasing as $z_d$ increases - this is not surprising, noting (from table~\ref{table:eg1a}) that we have more degrees of freedom for larger $z_d$, with all other discretisation parameters fixed.  We can see from the results above the importance of making the computational domain sufficiently large, and of having a sufficiently fine mesh near the helicopter.  The optimal way to achieve this might be through using a nonuniform mesh, with smaller mesh width nearer the helicopter and larger mesh width further away from the helicopter.  Noting though that we can achieve sufficiently accurate results with our fixed mesh approach to demonstrate key qualitative features of the solution, we leave such considerations to future work.}
\end{example}

\end{appendix}
\end{document}